\documentclass[12pt]{article}
\usepackage{mheck}
\usepackage{amsmath}
\usepackage{amssymb,amsfonts}
\usepackage{amsthm}
\usepackage[all]{xy}
\usepackage{graphicx}
\usepackage[parfill]{parskip}
\usepackage{tabu}
\usepackage{float}
\usepackage[utf8]{inputenc}

\numberwithin{equation}{section}
\numberwithin{table}{section}

\newcommand{\be}{\begin{equation}}
\newcommand{\ee}{\end{equation}}
\def\beq{\begin{eqnarray}}
\def\eeq{\end{eqnarray}}
\def\ba{\begin{eqnarray}}
\def\ea{\end{eqnarray}}

\def\ep1{\epsilon_1}
\def\eps2{\epsilon_2}

\newcommand{\IZ}{\mathbb{Z}}
\newcommand{\IC}{\mathbb{C}}
\newcommand{\IP}{\mathbb{P}}
\newcommand{\IN}{\mathbb{N}}
\newcommand{\IR}{\mathbb{R}}
\newcommand{\IQ}{\mathbb{Q}}

\newcommand{\IF}{\mathbb{F}}

\newcommand{\bz}{\boldsymbol{z}}
\newcommand{\bn}{\boldsymbol{n}}
\newcommand{\brho}{\boldsymbol{\rho}}
\newcommand{\bzero}{\mathbf{0}}

\newcommand{\bX}{\boldsymbol{X}}
\newcommand{\bt}{\boldsymbol{t}}
\newcommand{\bk}{\boldsymbol{k}}
\newcommand{\bnu}{\boldsymbol{\nu}}

\newcommand{\NefY}{\Nef(Y_\Sigma)}
\newcommand{\Lroot}{\Lambda^{root}}
\newcommand{\Lweight}{\Lambda^{weight}}

\newcommand{\rk}{\mathrm{rk\, }}

\newcommand{\nn}{\nonumber}

\newcommand{\cM}{{\cal M}}

\newcommand{\cE}{{\cal E}}
\newcommand{\cO}{{\cal O}}
\newcommand{\cC}{{\cal C}}

\newcommand{\cK}{{\cal K}}
\newcommand{\cR}{{\cal R}}

\newcommand{\mg}{\mathfrak{g}}
\newcommand{\mR}{\mathfrak{R}}

\newcommand{\DB}{D^{\IF_n}}
\newcommand{\DX}{D^X}
\newcommand{\DY}{D^Y}

\newcommand{\LP}{P}

\newcommand{\blue}{\textcolor{blue}}
\newcommand{\red}{\textcolor{red}}
\newcommand{\green}{\textcolor{green}}

\newcommand{\Ccontr}{C_{contr}}
\newcommand{\tCcontr}{{C}_{contr}'}

\newcommand{\KYn}{\cK_{Y_n}}
\newcommand{\KDY}{K_{Y}}
\newcommand{\KDYn}{K_{Y_n}}
\newcommand{\KB}{\cK_{\IF_n}}
\newcommand{\Pp}{P^{\circ}}
\newcommand{\PKY}{P_{-\KDY}}
\newcommand{\Psing}{P_{sing}}
\newcommand{\Msing}{S_{sing}}

\newcommand{\mult}{\mathrm{mult}}
\newcommand{\MC}{\overline{\mathrm{NE}}}
\newcommand{\Nef}{\mathrm{nef}}
\newcommand{\ISR}{I_{\mathrm Stanley-Reisner}}

\newcommand{\Mcplx}{\cM_{cplx}}
\newcommand{\Mkaehler}{\cM_{J}}

\preprint{}

\begin{document}
	\begin{titlepage}
		{}~ \hfill\vbox{ \hbox{} }\break
		
		
		\vskip 1 cm

		\begin{center}
			\Large \bf Determining F-theory matter via Gromov-Witten invariants
		\end{center}
		
		\vskip 0.8 cm
		
		\centerline{Amir-Kian Kashani-Poor}

		\vskip 0.2in
		\begin{center}{\footnotesize
				\begin{tabular}{c}
					{\em LPENS, CNRS, PSL Research University, Sorbonne Universit\'{e}s, UPMC, 75005 Paris, France}\\[0ex]
				\end{tabular}
		}\end{center}

		\setcounter{footnote}{0}
		\renewcommand{\thefootnote}{\arabic{footnote}}
		\vskip 60pt
		\begin{abstract} 
			{}{\bf Abstract:} We show how to use Gromov-Witten invariants to determine the matter content of F-theory compactifications on elliptically fibered Calabi-Yau manifolds $X$ over Hirzebruch surfaces. To determine the representations of these matter multiplets under the gauge algebra~$\mg$, we use toric methods to embed the weight lattice of $\mg$ into the integer homology lattice of $X$. We then apply mirror symmetry to determine whether classes in this lattice which correspond to weights of given representations are represented by irreducible curves. Applying mirror symmetry efficiently to such geometries requires obtaining good approximations to their Mori cones. We propose an algorithm for obtaining such approximations. When the algorithm yields a smooth cone, we find that the latter in fact coincides with the Mori cone of $X$ and already contains information on the matter content of compactifications on $X$. Our algorithm relies on studying toric ambient spaces for the Calabi-Yau hypersurface $X$ which are merely birationally equivalent to fibrations over Hirzebruch surfaces. We study the flops relating such varieties in detail.
			
		\end{abstract}
		
		{\let\thefootnote\relax
			\footnotetext{\tiny amir-kian.kashani-poor@phys.ens.fr}}
		
	\end{titlepage}

	\pagebreak
	\hspace{0pt}
	\vfill
	{\it To Noah Kian, tenaciously, adorably, pursuing his sleep deprivation research agenda.}
	\vfill
	\hspace{0pt}
	\pagebreak

	\vfill \eject

	\tableofcontents
	
	\newpage
	
	\section{Introduction}
	
	String theory has an intimate connection to complex geometries that serve as supersymmetry preserving compactification manifolds. This relation is particularly elaborate in F-theory \cite{Vafa:1996xn}, where also the vacuum expectation value of the axio-dilaton is geometrized: it is identified with the complex structure modulus of a torus fibered over the compactification manifold. The understanding of F-theory compactifications on elliptic fibrations over Hirzebruch surfaces $\IF_n$ \cite{Morrison:1996na,Morrison:1996pp} was substantially advanced in the seminal paper \cite{Bershadsky:1996nh} in the context of heterotic/F-theory duality. In particular, it was found that while the gauge symmetry of the corresponding six dimensional theories follows easily from the Kodaira classification of the singularities of the elliptic fibration, the matter content is more difficult to extract from the geometry.
	
	There has been a tremendous amount of work on F-theory compactifications on elliptically fibered geometries in the intervening twenty plus years. We touch upon a very few developments in this paragraph, and refer to the excellent recent review \cite{Weigand:2018rez} for a more complete list of references. The geometries discussed in \cite{Bershadsky:1996nh} were introduced in the context of toric geometry slightly earlier in \cite{Candelas:1996su}. Some  observations made in that work regarding the interplay of toric data and the gauge symmetry of the compactification were explained in \cite{Perevalov:1997vw}. The question of identifying matter in F-theory compactifications, the topic of this paper, has received much attention, with some important developments being \cite{Katz:1996xe,Morrison:2011mb,Grassi:2011hq}. A decompactification limit of the Hirzebruch base of the compactification Calabi-Yau manifold yields six dimensional superconformal field theories. Interest in the study of these elusive theories was revived by classification proposals \cite{Heckman:2013pva,Heckman:2015bfa}, based on work classifying all bases appropriate for F-theory compactifications \cite{Morrison:2012np}. The observation that the elliptic genus of tensionless strings in these theories is captured by the topological string with target space the compactification manifold \cite{Lockhart:2012vp} has led to an improved understanding both of the tensionless strings that arise upon compactification of F-theory on elliptically fibered Calabi-Yau spaces \cite{Kim:2014, Haghighat:2014vxa, Kim:2016foj, DelZotto:2016pvm, Kim:2018gak, DelZotto:2018tcj, Lee:2018urn} and on the topological string with such target spaces \cite{Huang:2015sta, Gu:2017ccq, DelZotto:2017mee, Cota:2019cjx}. In particular, the connection has led to the computation of all genus modular results for the topological string partition function, order by order in base wrapping number.
	
	In this work, we will put some of the enumerative invariants extracted from the topological string to work to determine the field content of 6d theories systematically and computationally effectively. We descend from the lofty heights of all genus results, as all we require are genus 0 invariants (though in the case of E-strings, these in fact uniquely specify all invariants \cite{Duan:2018sqe}), which we obtain via mirror symmetry computations \cite{Candelas:1990rm,Hosono:1993qy}.\footnote{We refer to these invariants as  ``genus 0 Gromov-Witten invariants''  throughout the paper  to emphasize that the only tool involved in their computation is mirror symmetry. These invariants of course coincide with genus 0 Gopakumar-Vafa invariants \cite{Gopakumar:1998ii,Gopakumar:1998jq}. The higher genus Gopakumar-Vafa invariants reflect the BPS spectrum of the theory that remains massive also at the singular point of the fibration.}
	
	The backbone of this work consists in extracting the matter content of F-theory compactifications from the Gromov-Witten invariants of the compactification manifold $X$. As we review in section \ref{s:F-matter}, this requires knowledge of the curves $C$ occurring in $X$ and their intersection numbers with a set of distinguished divisors $\{D_i\}$ which arise when constructing $X$ by resolving a singular starting point $X_{sing}$. In sections \ref{ss:anti_canonical_hypersurface} to \ref{ss:g_n_varieties}, we construct $X_{sing}$ and its resolution $X$ as hypersurfaces in ambient toric varieties $Y_{sing}$ and $Y$. We then identify, in subsection \ref{ss:distinguished_curves}, the curves in $X$ giving rise to vector multiplets associated to a gauge symmetry $\mg$ in terms of intersections of $X$ with torus invariant surfaces of $Y$. This provides a map $\phi$ from the root lattice $\Lroot(\mg)$  to the space $N_1(X)$ spanned by classes of curves of $X$. We then use mirror symmetry in section \ref{s:mirror_sym} to get a handle on all irreducible curves in $X$ (up to a certain degree, depending on the computer time invested). Finally, extending $\phi$ over $\IQ$ to access the weight lattice $\Lweight(\mg)$, we identify the representations $\mR_i$ of all light hypermultiplets present in the spectrum (i.e. those which become massless in the singular limit) in section \ref{s:formalism_applied}. In the appendix, we gather data regarding the various varieties that arise from the construction outlined in section \ref{s:construction}.
	
	Beyond the primary focus of this paper, we demonstrate in section \ref{s:formalism_applied} that information about the matter content of the theory can already be extracted from toric geometry, before invoking mirror symmetry. To this end, using an idea which goes back to Sheldon Katz, as cited in \cite{Braun1998}, we compute an approximation to the Mori cone of the Calabi-Yau manifold $X$ -- which we call the toric Mori cone -- which improves upon that derived from the Mori cone of the ambient toric space $Y$: roughly speaking, the toric Mori cone is obtained by taking all possible toric ambient spaces into account. This idea is developed in section \ref{ss:mori_cone}. 
	
	Having computed the toric Mori cone, we use mirror symmetry to determine when it coincides with the actual Mori cone of the Calabi-Yau manifold. We find, for the many examples that we consider, that this is the case whenever the toric Mori cone is smooth. 
	
	In section \ref{ss:flops_II}, we study in some detail the flops relating the birationally equivalent varieties which resolve the singularities of a given fibration, allowing us to explain in section \ref{ss:distinguished_curves} why the intersection numbers determining gauge symmetry and matter representations are (largely) independent of these.\footnote{This provides an explicit demonstration of a fact proved in a more abstract setting in \cite{Grassi:2018rva}.} This study also lays the geometrical groundwork for extending the study in \cite{DelZotto:2017pti} of phases of five dimensional theories obtained from compactifying six dimensional theories on a circle beyond the maximally Higgsed case.
	
	As the novelty in the mirror symmetry computations that we perform, compared e.g. to \cite{Haghighat:2014vxa}, is the use of the toric Mori cone, we review in detail in section \ref{s:mirror_sym} how the relation between the Mori cone (or an approximation thereof) of a Calabi-Yau manifold $X$ and distinguished coordinates on the complex structure moduli space of its mirror $X'$ arises, and we discuss how to proceed if the approximate Mori cone is not smooth.
	
	For all things toric, we follow the notation of the wonderful book \cite{CoxLittleSchenck}, where, unless otherwise noted, the proofs of all toric facts that we cite in this paper can be found. Extensive use of the mathematics software system SageMath \cite{sagemath} was made to perform the toric computations in this work. The mirror symmetry computations were performed in Mathematica \cite{Mathematica}. The Mathematica package LieART \cite{Feger:2012bs} proved very useful for all computations involving Lie algebras and their representations.

	As this paper was in the final phase of completion, we learned of the paper \cite{Paul-KonstantinOehlmann:2019jgr}, which has some overlap with this work.

	\section{F-theory matter via F-theory/M-theory duality} \label{s:F-matter}
	In this section, we briefly review M-theory compactifications on Calabi-Yau manifolds $X$ and how they related to F-theory compactifications on $X \times S^1$ when $X$ is elliptically fibered \cite{Vafa:1996xn}. These matters have recently been discussed in great detail in the review \cite{Weigand:2018rez}.
	
	Perturbative gauge fields arise in M-theory compactifications on a Calabi-Yau manifold $X$ via the expansion of the supergravity field $C_3$ in harmonic two forms,
	\be \label{eq:perturbative_vector_fields}
	C_3 = \sum_i A_i \omega^i  \,.
	\ee
	Perturbative states are not charged under the gauge fields $A_i$. In particular, the perturbative gauge symmetry is abelian. Non-perturbatively, the story is much richer. $C_3$ is sourced electrically by M2 branes. This coupling is described by an interaction term
	\be
	I_{int} = \int C_3
	\ee
	on the worldvolume of M2 branes. An M2 brane wrapping a holomorphic curve $C$ in $X$ gives rise to a BPS multiplet of particles. Their worldline action contains the coupling
	\be \label{eq:non_perturbative_charges}
	\int dt \,A_i \int_C \omega^i = C \cdot D^i \, \int dt \, A_i\,,
	\ee
	where we have introduced divisors $D_i$, representatives of the homology classes in $H_4(X,\IZ)$ which are Poincar\'e dual to the cohomology classes $[\omega_i]$. From \eqref{eq:non_perturbative_charges}, we can read off the charge of these particles under the gauge field $A_i$: it is given by the intersection product $C \cdot D^i$. The spins of these particles depend on the moduli space of the M2 brane: an isolated curve gives rise to a hypermultiplet, a curve with a genus $g$ Riemann surface as moduli space gives rise to a vector multiplet and 2g hypermultiplets \cite{Katz:1996ht, Witten:1996qb}.
	
	M-theory on an elliptically fibered Calabi-Yau manifold $X$ is dual to F-theory on $X \times S^1$. The elliptic fiber of $X$ becomes fully physical in the M-theory picture; its size maps to the inverse radius of the $S^1$ in the F-theory frame. The 5d theory obtained by compactification of M-theory on $X$ lifts to a 6d theory by mapping to the F-theory picture and decompactifying the $S^1$. In this paper, we will be interested in particular in vector fields that lift to 6d vector fields (rather than lifting to a component of the metric or a tensor field). All but one of the perturbative vector fields introduced in \eqref{eq:perturbative_vector_fields} belong to this class; in the following, the notation $A_i$ will exclude the Kaluza-Klein gauge field which does not. M2 branes wrapping curves $C$ which arise from resolving singularities in the elliptic fiber of $X$ give rise to further such vector fields. The associated vector multiplets are charged under a subset $\{A_i\}_{i \in J}$ of the perturbative gauge fields. This mechanism results in enhanced gauge symmetry with gauge group of rank $|J|$ at the singular point of the geometry.
	
	To pinpoint the Lie algebra $\mathfrak{g}$ underlying the gauge symmetry (restricting for simplicity to the case of simple $\mathfrak{g}$), we identify the $A_i$ with the Cartan generators $H_{\alpha_i}$ of $\mathfrak{g}$ in the Chevalley basis. Recall that in this basis, Cartan generators are labeled by simple roots $\alpha_i$ ($\alpha_i \in \Delta$), and completed to a basis of $\mg$ via elements $E_\alpha$ labeled by the roots $\alpha$ ($\alpha \in \Phi)$ (matching the count $\dim \mg = |\Phi| + |\Delta|$). The structure constants of the Lie algebra are then determined by the bilinear form $( \cdot , \cdot )$ induced on $\Phi$ via the Killing form of $\mathfrak{g}$. In particular,
	\be
	[ H_{\alpha_i}, E_\alpha ] = ( \alpha, \alpha_i^\vee ) E_\alpha   = \alpha(H_{\alpha_i}) E_\alpha\,.
	\ee
	A curve $C$ giving rise to a BPS multiplet containing the vector field associated to the Lie algebra generator $E_\alpha$ thus must exhibit intersection number with the divisor $D^i$ associated to the perturbative gauge fields $A_i$ equal to the negative of the $i^{th}$ coefficient of the root $\alpha$ in an expansion in fundamental weights,\footnote{That a minus sign must be present in \eqref{eq:intersection_no_and_Cartan} is  perhaps clearest in the case of elliptic surfaces, where rational curves $C$ with $C \cdot C = -2$ map to simple roots $\alpha$ with $\alpha(H_{\alpha}) = 2$.}
	\be \label{eq:intersection_no_and_Cartan}
	C \cdot D^i = -\alpha(H_{\alpha_i}) \,.
	\ee
	Note that the charge of a field associated to a curve $C$ is fixed entirely by its homology class (or more precisely by its class in the space $N_1$, which we introduce in section \ref{ss:mori_cone} below). Conversely, in order for a field with the charges associated to a given homology class to be part of the spectrum of the theory compactified on $X$, that class must be represented by an irreducible curve.
	
	Note further that the Cartan matrix underlying the Lie algebra can be determined by identifying the curves $C_i$ corresponding to generators $E_{\alpha_i}$ of $\mathfrak{g}$,
	\begin{equation}
		(\alpha_i, \alpha_j^\vee) = -C_i \cdot D^j \,.
	\end{equation}
	This matrix is generically not symmetric. Indeed, compactifications on Calabi-Yau 3-folds can lead to non-simply laced gauge groups.
	
	Isolated curves in the fiber will give rise to fields belonging to charged hypermultiplets, transforming in a representation $\mR$ of $\mg$. Such curves are called {\it matter curves}. To each $\lambda \in \Pi(\mR)$,  $\Pi(\mR)$ denoting the set of weights of the representation $\mR$, are associated one or multiple basis vectors of the representation space $V_{\mR}$. The charges of the corresponding fields under the gauge fields $A_i$ are given by $\lambda(H_{\alpha_i})$. A curve $C$ giving rise to such a multiplet must hence exhibit intersection numbers satisfying
	\be \label{eq:intersection_no_and_weights}
	C \cdot D^i =- \lambda(H_{\alpha_i}) \,.
	\ee
	The two relations \eqref{eq:intersection_no_and_Cartan} and \eqref{eq:intersection_no_and_weights} suggest identifying the divisors $D^i$ with elements of the coroot lattice $(\Lambda^{root})^\vee(\mathfrak{g})$, and irreducible curves with elements of the weight lattice $\Lambda^{weight}(\mathfrak{g})$.
	
	We have seen that determining the field content of an F-theory compactification on a variety $X$ requires identifying the divisors $D^i$ and the irreducible curves $C$ occurring in $X$ and computing their intersection numbers. We will turn to these questions in the following sections.
	
	Note that studying the intersection theory of curves and divisors will teach us about the Lie algebra $\mathfrak{g}$ underlying the gauge group $G$ of the compactification, as well as the matter representations present. Unless the latter are only compatible with the simply connected Lie group $G_0$ associated to $\mathfrak{g}$, this does not uniquely fix $G$.\footnote{E.g., both the Lie groups $SU(2)$ and $SO(3)$ have $\mathfrak{a}_1$ as their Lie algebra; the presence of half-integer spin representations would  identify the former as the correct gauge group.} The global structure of  $G$ in such cases can be read off from the torsional part of the Mordell-Weil group of the fibration \cite{Aspinwall:1998xj,Guralnik:2001jh}. The authors of  \cite{Mayrhofer:2014opa} provide a direct link between  non-trivial torsion in the Mordell-Weil group and excluded representations: they show that the former leads to the existence of  divisors which are non-integer linear combinations of the divisors $D^i$ introduced above; candidate matter curves of non-integer intersection with this divisor are ruled out. The study of the existence of such divisors is amenable to toric methods \cite{Mayrhofer:2014opa}, but we will not pursue such questions further in this paper.
	
	\section{The geometries and their curves}\label{s:construction}
	
	\subsection{Anti-canonical hypersurfaces in projective bundles over Hirzebruch surfaces} \label{ss:anti_canonical_hypersurface}
	The starting point of our considerations is the projective bundle $Y_{\Sigma_n}=\IP^{2,3,1}(2\KB \oplus 3\KB \oplus \cO)$ over a Hirzebruch surface $\IF_n$. $\KB$ here indicates the canonical line bundle of $\IF_n$, and we follow the convention that $\IP(\cE)$ indicates the projectivization of the dual bundle $-\cE$. The total space of the projective bundle is a (singular) toric variety. The notation indicates that this geometry is associated to the toric fan $\Sigma_n$. For brevity, we will also write $Y_n$. The generators $u_\rho \in N$ of the 1-cones $\rho \in \Sigma_n(1)$ of this geometry are given in table \ref{table:one_cones_for_Y}.
	\begin{table}[h]
		\centering
		\begin{tabular}{c|*{4}{>{$}c<{$}}|*{3}{>{$}c<{$}}}
			&		&		&		&		&(\IC^*)_1	&	(\IC^*)_2	&	(\IC^*)_3\\
			\hline
			$u_{\rho_x}$	&	1	&	0	&	0	&	0	&	2	&	0	&	0\\
			$u_{\rho_y}$	&	0	&	1	&	0	&	0	&	3	&	0	&	0\\
			$u_{\rho_z}$	&	-2	&	-3	&	0	&	0	&	1	&	n-2	&	-2\\
			$u_{\rho_s}$	&	-2	&	-3	&	0	&	-1	&	0	&	-n	&	1\\
			$u_{\rho_t}$ 	&   -2  &   -3  &   0   &   1 	&	0	&	0	&	1\\
			$u_{\rho_u}$	&	-2	&	-3	&	-1	&	-n	&	0	&	1	&	0\\
			$u_{\rho_v}$	&	-2	&	-3	&	1	&	0	&	0	&	1	&	0\\
		\end{tabular}
		\caption{Toric data for $\IP^{2,3,1}(2\KB \oplus 3\KB \oplus \cO) \rightarrow \IF_n$.} \label{table:one_cones_for_Y}
	\end{table}

	The top dimensional cones of the fan $\Sigma_n$ are given by the set of pairwise sums of an element of $\{\sigma_{x y}, \sigma_{x z}, \sigma_{y z}\}$ with a top dimensional cone of the fan of the Hirzebruch surface. Here and in the following, we use the notation
	\be
	\sigma_{i \ldots j} = \langle u_{\rho_i}, \ldots, u_{\rho_j} \rangle \,.
	\ee

	Recall that one useful way of thinking about toric varieties associated to a fan $\Sigma$ is in terms of the homogeneous coordinate ring (or Cox ring) \cite{CoxHomCord,CoxHomCordERR}: each 1-cone generator $u_{\rho_i}$ is assigned a $\IC$ valued coordinate $x_i$, called a homogeneous coordinate. Coordinates whose associated 1-cone generators do not jointly belong to any cone in the fan cannot vanish simultaneously. The monomials formed from the products of the elements of each such set generate the so-called Stanley-Reisner ideal $\ISR$, with vanishing locus $Z$. The toric variety is obtained as the almost geometric quotient 
	\be
	Y_\Sigma = (\IC^{|\Sigma(1)|} - Z) /\!/ G \,,
	\ee
	with $G \cong (\IC^*)^{|\Sigma(1)|-d}$ the group of relations among the 1-cone generators. 
	
	On the right in table \ref{table:one_cones_for_Y}, we give a set of generators for the group $G$ of relations of the toric variety $Y_{\Sigma_n}$. Note that we have indexed the rays $\rho \in \Sigma(1)$ by the variables we shall assign to them.
	
	The projective bundle $Y_{\Sigma_n}$ is chosen such that the zero set of the generic section of the anti-canonical bundle $-\KYn$ of its total space defines an elliptically fibered Calabi-Yau manifold.\footnote{Note that as $Y_{\Sigma_n}$ is singular, we need to argue that it is Gorenstein such that its dualizing sheaf is indeed a line bundle. We will do so in section \ref{ss:desingularization}.} A basis for $\Gamma(Y,-\KYn)$, the vector space of global holomorphic sections of $-\KYn$, can be obtained from the polyhedron
	\be \label{eq:PY}
	P_{-\KDYn} = \{ m \in M_{\IR}\, |\, \langle m, u_\rho \rangle \ge -1 \,\, \forall \rho \in \Sigma(1) \} \,,
	\ee
	as we shall now review. The relation \eqref{eq:PY} is a specialization of the polyhedron
		\be \label{eq:polyhedron_D}
		P_{D} = \{ m \in M_{\IR}\, |\, \langle m, u_\rho \rangle \ge - a_\rho \,\, \forall \rho \in \Sigma(1) \} 
		\ee
	associated to a torus invariant Weil divisor $D = \sum_\rho a_\rho D_\rho$ to the case of the anti-canonical divisor, which for any toric variety is given by $-K = \sum D_\rho$. $M$ in these formulae is the dual lattice to the lattice $N$ introduced above (with $M_{\IR} = M \otimes \IR$). It can be identified with the lattice of characters of the torus $T \subset Y$ underlying the toric variety $Y$. We write $\chi^m$ for the character associated to $m \in M$. Upon choosing coordinates $\bt = (t_1 , \ldots, t_4)$ on $T$, it can be explicitly written as $\chi^m = \prod t_i^{(m)_i}$. The elements of the lattice $M$ lying in the intersection with $P_{-\KDYn}$ yield the desired basis of holomorphic sections of $-\KYn$:
	\be \label{eq:holomorphic_sections}
	\Gamma(Y,-\KYn) = \bigoplus_{m \in P_{-\KDY} \cap M} \IC \cdot \chi^m \,.
	\ee
	In section \ref{s:mirror_sym}, we will use the notation
	\be \label{eq:Laurent_polynomials_to_polytope}
	L(P_{-\KDY}) 
	\ee
	for $\Gamma(Y,-\KYn)$ expressed in these coordinates. Note that $\bt$ is not a good global coordinate on $Y$; the characters $\chi^m$ appearing on the RHS of \eqref{eq:holomorphic_sections} are generically not holomorphic in these coordinates. In homogeneous coordinates, which do provide good coordinates patchwise, one obtains the holomorphic sections of $-\KYn$ via the map  
	\be \label{eq:mTOmonom}
	\chi^m \mapsto \prod_\rho x_\rho^{\langle m, u_\rho \rangle +1} \,,
	\ee
	which is the specialization of the so-called $D$-homogenization of $\chi^m$, $\chi^m \mapsto \prod_\rho x_\rho^{\langle m, u_\rho \rangle +a_\rho}$, for $D = \sum_{\rho} a_{\rho} D_\rho$, to the anti-canonical divisor.
	
	It now follows from the 1-cone generators given in table \ref{table:one_cones_for_Y} that a general section of $-\KYn$ in homogeneous coordinates is of the form
	\be  \label{eq:generic_anti_canonical_section}
	s_{-\KYn} = \alpha y^2 + a_1 x y z + a_3 y z^3 - (\beta x^3 +a_2 x^2 z^2+a_4 x z^4+a_6 z^6) \,,
	\ee
	with the coefficients $a_i$ functions of the remaining coordinates $u,v,s,t$. $\alpha$ and $\beta$ are constants that can be absorbed in the coordinates $x$ and $y$. The minus sign is to match the conventions of \cite{Bershadsky:1996nh}, see also \cite{Katz:2011qp}. The anti-canonical hypersurface $X_n$ of $Y_n$ is given by
	\be
	X_n = \{p \in Y_n \, | \, s_{-\KYn}(p) = 0\} \,.
	\ee
	It is Calabi-Yau by construction. From \eqref{eq:generic_anti_canonical_section}, it is evident that it is given by an elliptic fibration over the Hirzebruch base $\IF_n$. This fibration will generically be singular at $y = s =0$. The vanishing degree $n_i$ in the variable $s$ of the coefficients $a_i$ determines the singularity type of the elliptic fiber over the corresponding points of the base surface $\IF_n$.
	 
	 Note that in patches permitting $z=0$, the equation $s_{-\KYn} = 0$ has a solution $(x:y:z) \sim (1 : \sqrt{\tfrac{\beta}{\alpha}} : 0)$ (the two points corresponding to the two signs of the square root are identified by the weighted projective action) which is independent of the coordinates on $\IF_n$. This defines a global holomorphic section (called the zero section) of the elliptic fibration, allowing the embedding of the base $\IF_n$ into $X_n$. We will use the term `zero section' also to refer to the image $Z$ of this section. The point in each fiber defined by the intersection with the divisor $z=0$ yields the distinguished point (the zero element of the additive group) of the elliptic fiber.

	\subsection{Desingularization} \label{ss:desingularization}
	The strategy for desingularizing $X_n$ will be to find a $Y$ birationally equivalent to the ambient space $Y_n$, $\phi: Y \rightarrow Y_n$, such that $\phi^{-1}X_n$ is smooth. Note that $\phi$ is generically not a resolution of singularities of $Y_n$. In fact, our considerations here can equally well be applied to the projective bundle $\IP^{1,1,1}(2\KB \oplus 3\KB \oplus \cO)$, which is smooth. The procedure we will review \cite{Batyrev:1994hm} relies on constructing a toric variety $Y$ which is a Gorenstein orbifold with terminal singularities (we will explain the term `Gorenstein' and how to diagnose terminal singularities in the toric setting in the ensuing discussion). It can be shown that the zero set of a generic section of a base-point free line bundle on such spaces share all three properties, at least in the case of toroidal singularities (see theorem 2.6 and proposition 4.3 of \cite{Batyrev:1994pg}).\footnote{Thanks to Antonella Grassi for requesting a reference, and to Andrew Harder for providing one.} As Gorenstein orbifolds with terminal singularities have singularities in codimension 4 and higher, the three dimensional anti-canonical hypersurfaces in $Y$ that will be the objects of our interest will be smooth. 
	
	As we are considering toric orbifolds $Y$, i.e. ambient spaces that are generically not smooth, it is not guaranteed that their anti-canonical divisor is Cartier, hence describes a line bundle over $Y$. $Y$ is called Gorenstein when this is the case. There is a simple criterion for when a toric Weil divisor is Cartier: the polyhedron \eqref{eq:polyhedron_D} must be a lattice polytope. A Gorenstein variety is hence characterized by the fact that the polyhedron $P_{-\KDY}$ defined in \eqref{eq:PY} be a lattice polytope. 
	
	Via the correspondence
	\begin{gather*}
	\{(Y_{\Sigma},D)\,| \text{$\Sigma$ a complete fan in $N_\IR$, $D$ a torus invariant ample divisor on $Y_\Sigma$} \} \\ 
	\updownarrow \\
	\{\text{lattice polytopes $P \subset M_\IR$}\}
	\end{gather*}
	we can recover the fan $\Sigma$ together with the ample line bundle $D$ on $X_{\Sigma}$ from a lattice polytope: $\Sigma$ coincides with the normal fan $\Sigma_P$ of $P$, and the facet presentation \eqref{eq:polyhedron_D} of the polytope uniquely determines $D$ on $Y_{\Sigma_P}$. As we will discuss presently, desingularization will involve considering fans $\Sigma'$ which are so-called maximal projective subdivisions of the polytope $P$, with $\Sigma_P(1) \subset \Sigma'(1)$. When $\Sigma'$ refines $\Sigma_P$, a toric morphism $\phi : Y_{\Sigma'} \rightarrow Y_{\Sigma_P}$ exists; the pullback $D' = \phi^* D$ of $D$ to $Y_{\Sigma'}$ via this morphism will generically no longer be ample. It will however satisfy the weaker property of base-point freeness: a divisor $D$ is base-point free if the points $m_\sigma$, $\sigma \in \Sigma(4)$, constituting the Cartier data of $D$ lie in $P_D$. As $D'$ is the pullback of $D$, the set of points $\{m_\sigma | \sigma \in \Sigma(4)\}$ and $\{m_{\sigma'} | \sigma' \in \Sigma'(4)\}$ coincide. As furthermore $P_D = P_{D'}$ by the definition of maximal projective subdivision, the claim follows. 
	
	Ampleness and base-point freeness can be read off from the intersection ring of the toric variety. By the toric Kleiman criterion, ampleness of a Cartier divisor $D$ is equivalent to the positivity of the intersection of $D$ with any torus invariant irreducible curve $C$, $D\cdot C > 0$, while base-point freeness is equivalent to the weaker condition $D\cdot C \ge 0$ on complete\footnote{We use the algebraic geometric language; for those more familiar with the analytic setting, `complete' can be replaced with `compact' in the following, with little loss of precision. Completeness of the toric variety expresses itself in the support of its fan coinciding with the lattice $N$. Note that here and in the following, we will often replace the weaker condition ``fan of convex support of full dimension'' with completeness, which suffices for our purposes.}  toric varieties, hence to $D$ being nef, a notion we will introduce below.

	A lattice polytope of the form \eqref{eq:PY} is called reflexive. Its unique interior lattice point is the origin. The concept of dual (or polar) polytopes is particularly useful in the case of reflexive polytopes. The general definition
	\be
	\Pp = \{ u \in N_{\IR} \, | \, \langle m,u \rangle \ge -1 \,\, \forall m \in P \}
	\ee
	applied to the polytope $P$ with facet presentation
	\be 
	P = \{ m \in M_{\IR}\, | \,\langle m, u_F \rangle \ge - a_F \,\, \text{ for all facets } F \} \,,
	\ee
	all $a_F>0$, yields
	\be \label{eq:polar_facet_presentation}
	P^\circ = \mathrm{Conv}(\frac{1}{a_F} u_F \, | \, F \text{ a facet of P}) \,.
	\ee 
	As facet normals $u_F$ are lattice vectors of $N$ by definition, it follows easily that the dual of a reflexive polytope is again a lattice polytope. It is in fact also reflexive. The normal fan of $P$ is the face fan of $P^\circ$ and vice versa.
	
	As singularities are local properties of a variety, we can study them by considering the variety patch by patch. Given an affine patch $U_\sigma$ associated to a cone $\sigma \in \Sigma$ of a fan, we can determine the type of singularities it contains by considering the lattice points contained in the polytope 
	\be
	P_\sigma = \mathrm{Conv}(0, u_\rho \, | \, \rho \in \sigma(1))
	\ee
	given by the convex hull of the origin and the tips of the 1-cone generators $u_\rho$ of $\sigma$: if all of the $u_\rho$ lie on a hyperplane at a distance 1 from the origin, i.e.
	\be
	\exists m \in M : \langle m, u_\rho \rangle = 1 \quad \forall \rho \in \sigma(1)\,,
	\ee
	then $U_\sigma$ has terminal singularities iff the only lattice points of $P_\sigma$ are given by its vertices.
	
	It is now clear that to obtain a fan $\Sigma$, with $\Sigma(1) \supset \Sigma_P(1)$, whose associated toric variety has terminal singularities, we must introduce additional 1-cones through all lattice points $\Pp \cap N$. This leads to the following definition \cite{Batyrev:1994hm, CoxKatz}: a maximal projective subdivision of a reflexive polytope $P$ is a regular simplicial fan $\Sigma$  such that $\Sigma(1) = \{ \langle u_\rho \rangle \,|\, u_\rho \in \Pp \cap N \}$. Simplicial fans lead to toric varieties with at worst orbifold singularities. For smooth fans, regularity ensures that the corresponding toric variety is projective. Note that in \cite{CoxKatz}, maximal projective subdivisions are required to refine the normal fan of $P$. We will not do so, as subdivisions which do not refine the normal fan will also prove of interest to us.
	
	Consider now a maximal projective subdivision $\Sigma$ of $P$. As $\Sigma$ is simplicial by definition, $Y_\Sigma$ is an orbifold. The polyhedron associated to the anti-canonical divisor of $Y_\Sigma$ is the lattice polytope $P$, hence $Y_\Sigma$ is Gorenstein. If $\Sigma$ refines the normal fan $\Sigma_P$ of $P$, we have argued above that the anti-canonical divisor $-K_{Y_{\Sigma}}$ is base-point free. If not, we have found computationally (with one exception, see below) that it can be related to a fan that does refine $\Sigma_P$ by a series of flops, allowing us to conclude base-point freeness of its anti-canonical divisor (flops will be discussed at some length in section \ref{ss:flops_II}). Finally, as $P^\circ$ has the origin as unique interior point and by maximality of the subdivision, all patches $U_\sigma$ of $\Sigma$ have terminal singularities. We thus conclude that $Y_\Sigma$ is a Gorenstein orbifold with terminal singularities.
	
	Let us now relate this discussion to the fans $\Sigma_n$ introduced in section \ref{ss:anti_canonical_hypersurface}. To use maximal projective subdivisions of the dual polytope to desingularize  $X_n$, two conditions need to be met:
	\begin{enumerate}
		\item $P_{-\KDYn}$ must be a lattice polytope, thus reflexive. This is the case for $n = 0 ,1,2,3,4,6,12$. By the above, this is the statement that for these values of $n$, a birationally equivalent variety to $Y_n$ exists, given by the normal fan to $P_{-\KDYn}$, that is Fano, i.e. whose anti-canonical bundle is ample. We will discuss how to treat some other values of $n$ below.
		\item For $P_{-\KDYn}$ reflexive, any maximal projective subdivision $\Sigma$  satisfies 
		\be
		\Sigma_n(1) \subset \Sigma(1)  \,.
		\ee
		However, as already mentioned above, the subdivisions need not be refinements of $\Sigma_n$. In fact, for $n=3$ (and only for this value among the 7 for which $P_{-\KDYn}$ is reflexive), non of the maximal projective subdivisions of $P_{-K_{Y_3}}$ refine $\Sigma_3$. In section \ref{s:type_II}, we will see that they nevertheless give rise to the expected gauge symmetry upon F-theory compactification.
	\end{enumerate}

	\subsection{Enhancing the singularity (before desingularizing anew)} \label{ss:enhancing}
	The discussion up to this point was concerned with the resolution of the generic singularity of the zero set of an anti-canonical hypersurface of $Y_n$. In the local context, the resulting geometries give rise to the maximally Higgsed theories discussed in \cite{Morrison:2012js}.
	To enhance the singularities of the zero set of  the generic anti-canonical section \eqref{eq:generic_anti_canonical_section}, and thus unHiggs the gauge symmetry of the F-theory compactification, we need to define a new lattice polytope $\Psing$ in $M_{\IR}$  by excluding those points of $P_{-\KDY}$ which map to sections with a vanishing degree in $s$ too low for the singularity desired. To obtain $\Psing$, first determine the subset $\Msing$ of $\PKY \cap M$ corresponding to all sections compatible with the desired singularity. Next, define the polyhedron 
	\be \label{singConstraint}
	P_{aux} = \{ u \in N_{\IR}\, |\, \langle m, u \rangle \ge - 1 \,\, \forall m \in \Msing \} \,.
	\ee
	$\Msing$ needs to be sufficiently rich, i.e. the constraints on the vanishing degrees in $s$ sufficiently lax, for $P_{aux}$ to be a compact polyhedron. Its dual is then the lattice polytope we are after,
	\be
	\Psing = P_{aux}^\circ =  \{ m \in M_{\IR} \,|\, \langle m, u \rangle \ge - 1 \,\, \forall u \in P_{aux} \} \,.
	\ee	
	To argue that, for $P_{aux}$ compact, the polyhedron $\Psing$ is a lattice polytope, and hence reflexive, note that a selection $\{ m_F \} \subset M$ among the elements of $\Msing$ is possible such that
	\be
	P_{aux} =  \{ u \in N_{\IR} \,|\, \langle m_F, u \rangle \ge - 1 \,\,  \text{for all facets $F$ of $P_{aux}$} \} \,.
	\ee	
	By the discussion around \eqref{eq:polar_facet_presentation}, the dual of $P_{aux}$ is given by the convex hull of the facet normals $m_F$ of $P_{aux}$, is hence a lattice polytope.
	
	Clearly, $\Msing \subset \Psing$, and $\Psing$ is the smallest reflexive polytope containing $\Msing$. If $\Msing \neq \Psing \cap M$, the points of $\Msing$ cannot be described as all lattice points of a lattice polytope, and the singularity desired cannot be realized torically over the given base. 
	
	Let us apply these considerations to the set $\Msing = P_n \cap M$ with $P_n$ the lattice polytopes defined in the previous subsection. For all $n$ for which $P_n$ is a lattice polytope (recall that these are $n= 0,\ldots,4$ and $n=6,12$), we necessarily find $P_n = \Psing$. For all other  $n$ in the range $0 \le n \le 12$, $\Psing$ can replace $P_n$ as the starting point of the desingularization process described in subsection \ref{ss:desingularization}.  Beyond $n=12$, $\Psing$ is non-compact for any choice of singularity.
	
	In all cases where $\Psing$ is a lattice polytope, we observe by computation that with one exception to be discussed below, at least one of its maximal projective subdivisions $\Sigma$ is a refinement of $\Sigma_n$. This implies that in terms of the coordinates in which the 1-cone generators of $\Sigma_n$ are expressed in \ref{table:one_cones_for_Y}, the projection $\bar{\pi} : N \rightarrow \langle e_3, e_4 \rangle$ gives rise to a toric morphism $Y_\Sigma \rightarrow \IF_n$ which induces the fibration structure $X_\Sigma \rightarrow \IF_n$ on the anti-canonical hypersurface $X_\Sigma$ of $Y_\Sigma$. As we will discuss further in section \ref{ss:g_n_varieties}, many of the hypersurfaces $X$ we will study are fibered over $\IF_n$ even though the corresponding ambient spaces $Y$ are not, yet we shall also encounter examples of  hypersurfaces which do not exhibit this fibration structure.
	
	In some cases, merely studying the vanishing orders of sections in the variable $s$ can be misleading, as variable definitions are possible which increase some vanishing orders (cf. the discussion of split, semi-split, and non-split singularities in \cite{Bershadsky:1996nh}, to which we will return in section \ref{ss:constructing_Higgsing_trees}). In such cases, computing $h^{1,1}(X)$ of the resolved geometry can serve as an indication for which singularity among the two or three possibilities was realized before desingularization.

	\subsection{$h^{1,1}(X)$} \label{ss:h11}
	Mirror symmetry requires knowledge of the K\"ahler cone of the anti-canonical hypersurface $X \subset Y$. This will be the topic of subsection \ref{ss:mori_cone}. A cruder question is that of the homology in complex codimension 1. For the ambient space $Y$, given by a maximal projective subdivision $\Sigma$ of the lattice polytope $P$, this space is generated by the linear equivalence classes of torus invariant divisors, hence
	\be
	h^{1,1}(Y) = |\Sigma(1)| - 4 \,.
	\ee
	The intersections of the toric divisors of $Y$ with the anti-canonical hypersurface $X$ are either empty or yield reducible or irreducible divisors of $X$. It can be shown \cite{Batyrev:1994hm} that a divisor generically misses $X$ iff the corresponding cone generator lies on a facet of $\Pp$. Such divisors will thus be of no interest to us in the following and the corresponding  1-cones will be discarded (rendering the ambient space associated to the fan built from the retained 1-cones more singular). The divisors that potentially lead to reducible intersections are those whose associated 1-cone generators lie in the interior of codimension 2-faces $\Theta^\circ$ of $\Pp$ \cite{Batyrev:1994hm}. The number of irreducible components of such intersections is captured by the number of interior points of the dual face to $\Theta^\circ$, the 1-face $\Theta$ of $P$: it is given by  $ l^*(\Theta)+1$, where we have introduced the notation $l^*(\Theta)$ for the number of interior points lying on a face $\Theta$. This yields the following expression for the (1,1) Hodge number of $X$:
	\be
	h^{1,1}(X) = |\Sigma(1)| - 4 - \sum_{\Gamma^\circ \in F_1(\Pp)} l^*(\Gamma^\circ) + \sum_{\Theta^\circ \in F_2(\Pp)} l^*(\Theta^\circ) l^*(\Theta)  \,,
	\ee
	with $F_n(P)$ indicating the set of codimension $n$ faces of the polytope $P$. 
	
	To have toric methods capture the geometry of the variety $X$ as accurately as possible, we will hence need to describe it as an anti-canonical hypersurface in a toric ambient space with fan $\Sigma$ which maximizes 
	\be
	|\Sigma(1)| - 4 - \sum_{\Gamma^\circ \in F_1(\Pp)} l^*(\Gamma^\circ) \,,  \label{eq:good_cones}
	\ee
	thus reducing the correction term. We conclude that the construction from section \ref{ss:enhancing} can be useful even for $\Msing = P_n \cap M$: the polytopes $\Psing$ we construct in these cases contain the same interior points as $P_n$. As they are the smallest lattice polytopes to contain these points, they will generically have more facets than $P_n$, thus have duals $\Pp$ with more vertices, leading to a maximization of \eqref{eq:good_cones}. It can happen however that even upon considering these smallest polytopes, the intersection of some divisors of $Y$ with $X$ is reducible. We discuss this further in subsection \ref{ss:constructing_Higgsing_trees} in the context of constructing Higgsing trees.

	\subsection{The class $(\mg)_n$ of hypersurfaces over $\IF_n$} \label{ss:g_n_varieties}

	Our interest lies not with the ambient toric variety $Y$, but with its anti-canonical hypersurface $X$. We can thus identify all maximal projective subdivisions of the polytope $P$ introduced in section \ref{ss:enhancing} which differ only in a locus which does not intersect the associated anti-canonical hypersurface. Removing all 1-cones whose generators lie on facets of $P^\circ$ was a first step in this direction, as explained in section \ref{ss:h11}. 
	
	To determine when two fans $\Sigma_1$, $\Sigma_2$ have isomorphic anti-canonical hypersurfaces, we will make use of the fact that a toric variety can be decomposed as a union of orbits of the torus action. It  hence suffices to consider the set of orbits by which the toric varieties	$Y_{\Sigma_1}$ and $Y_{\Sigma_2}$ differ, and determine whether these  intersect the respective anti-canonical divisor.  To perform this analysis, recall that by the orbit-cone correspondence, the set $\Sigma(k)$ of $k$-dimensional cones of $\Sigma$ is in 1:1 relation to the set of $n-k$ dimensional orbits of the torus action, with $n$ the dimension of the toric variety. For future reference, we introduce the notation $V(\sigma)$ for the closure of the orbit associated to the cone $\sigma$. The set of torus invariant divisors of $Y_\Sigma$ is consequently given by $\{ V(\rho) \,|\, \rho \in \Sigma(1)\}$. We will use the notation $D_\rho = V(\rho)$ in this case. A fact we will use repeatedly below is that for $Y$ complete and simplicial, 
	\be
	[D_{\rho_1}] \cdot \ldots \cdot [D_{\rho_k}] = \frac{1}{\mult(\sigma_{1,\ldots,k})} [V(\sigma_{1,\ldots,k})]
	\ee
	for $\rho_1\,, \ldots, \rho_k \in \Sigma(1)$ distinct and such that $\sigma_{1,\ldots,k} \in \Sigma(k)$. For $\sigma$ simplicial, its multiplicity  $\mult(\sigma)$ is given by the number of interior lattice points contained in the parallelotope spanned by its generators, plus 1. A smooth cone has multiplicity 1.

	Let us now assume that the point corresponding to the top-dimensional cone $\sigma_{ijkl} \in \Sigma_1(4) \setminus \Sigma_2(4)$ lies in $-K_{\Sigma_1}$. As $\Sigma_1$ and $\Sigma_2$ are assumed complete, at least one of the facets of $\sigma_{ijkl}$, say $\sigma_{ijk}$,  will not lie in $\Sigma_2(3)$, and satisfy $V(\sigma_{ijk}) \cdot (-K_{\Sigma_1}) \not = 0$. A facet of this cone, say $\sigma_{ij}$, will not lie in $\Sigma(2)$, and satisfy $V(\sigma_{ij}) \cdot D  \cdot (-K_{\Sigma_1}) \not = 0$ for at least one torus invariant divisor of $Y_\Sigma$ (notably $D_k$). We can conclude that to determine that the anti-canonical hypersurfaces of $Y_{\Sigma_1}$ and $Y_{\Sigma_2}$ are isomorphic, it suffices to show that
	\be \nn
	\forall \sigma \in \Sigma_1(2) \setminus \Sigma_2(2) \, : \,  V(\sigma)  \cdot D \cdot (-K_{\Sigma_1}) \not = 0 \quad \textrm{for at least one torus invariant divisor }D
	\ee	
	together with the reciprocal condition arising from interchanging $\Sigma_1$ and $\Sigma_2$.
	 
	Not surprisingly, we find that the number of isomorphism classes among the hypersurfaces $X$ is far smaller than the corresponding number for the ambient space $Y$. E.g., the polytope $P_n$ associated to an $E_6$ singularity over $\IF_n$ exhibits 200 maximal projective subdivisions for all $n$, but merely four (for $n$ even) or eight (for $n$ odd) isomorphism classes of hypersurfaces. We will denote the set of all isomorphism classes of hypersurfaces associated to a given singularity $\mg$ over the base $\IF_n$ as $(\mg)_n$. Each class corresponds to one variety; different members of a class differ by the embedding of this variety into different (but birationally equivalent, see section \ref{ss:flops_II}) ambient spaces. We will refer to the elements of $(\mg)_n$ as $(\mg)_n$ varieties. 
	
	$(\mg)_n$ varieties can be divided into two classes, determined by whether the Stanley-Reisner ideal $\ISR$ of any of the associated ambient spaces $Y$ contains the monomial $u*v$.  We will call varieties for which this condition does or does not hold of type~I , type~II respectively. $u*v \in I_{\mathrm Stanley-Reisner}$ implies that no cone of $Y$ exists containing both $\rho_u$ and $\rho_v$ as faces. This will be the case for all refinements of $\Sigma_n$, the fans introduced in section \ref{ss:anti_canonical_hypersurface} as the starting point of our considerations. We find that all varieties of type I exhibit at least one ambient space $Y$ whose fan is a refinement of $\Sigma_n$. This in particular implies that these varieties are fibrations
	over a Hirzebruch surface. On the other hand, as discussed in section \ref{ss:desingularization}, the existence of cones containing both $\rho_u$ and $\rho_v$ as faces is equivalent to the absence of the toric projection map $\pi : Y \rightarrow \IP^1$, with $\IP^1$ identified with the base of the Hirzebruch surface $\IF_n$. Hence, type~II varieties do not exhibit this fibration structure.\footnote{It would be interesting to determine whether this fact has repercussions for the physics on these spaces, in particular for the case in which no birationally equivalent geometries exist which exhibit the projection map $\pi$.}
	
	We find that $(A_2)_3$ is the only non-empty class of varieties containing no variety of type~I. As for type~II varieties, no such varieties of class $(\mg)_n$ exist for $n$ even. On the other hand, all non-empty classes $(\mg)_n$ with $n$ odd do exhibit such varieties.

	As we will discuss in detail in subsection \ref{ss:flops_II}, any two $(\mg)_n$ varieties are related by a sequence of flops in the ambient space.

	\subsection{The Mori cone} \label{ss:mori_cone}
	By the discussion in section \ref{s:F-matter}, BPS states in M-theory compactified on a Calabi-Yau manifold $X$ arise from M2 branes wrapping curves in $X$. The charges of the associated states are determined by the intersection numbers of these curves with divisors of $X$. A first step in studying these curves will hence consist in studying them modulo numerical equivalence $\equiv$ (i.e. identifying curves that have identical intersection numbers with all divisors). The free abelian group generated by the classes of irreducible complete curves up to numerical equivalence is denoted as $N_1(X)$. The cone within $N_1(X)$ generated by these classes is called $\mathrm{NE}(X)$. Its closure, $\MC(X)$, is called the Mori cone and will play an important role in the following: this is where the representatives of curves we wish to identify sit. The edges of the Mori cone are called extremal rays; they will play an important role when we discuss flipping transitions in section \ref{ss:flops_II}.
	
	The Mori cone of the complete $n$-dimensional toric variety $Y$ with at worst orbifold singularities (i.e. with simplicial fan $\Sigma$) is easy to determine, as it is spanned by the classes of complete torus invariant curves. As stated above, the latter are in one-to-one relation to $(n-1)$-cones $\tau$ of $\Sigma_Y$ and denoted as $V(\tau)$. 
	
	In the following, we will in particular be interested in identifying elements of the Mori cone via their intersection numbers with divisors. For $Y$ complete, an $(n-1)$-cone $\tau$ always lies at the intersection of two $n$-cones $\sigma_1$ and $\sigma_2$, $\tau = \sigma_1 \cap \sigma_2$, and is hence called a wall. We will call the tuple of intersection numbers of such a curve with all torus invariant divisors $D_\rho$ the (toric) Mori vector of the curve. Its entries are computed as follows: 
	\begin{enumerate} 
		\item \label{computeMori1} If $\rho \not \in \sigma(1)$ for either $\sigma \in \{\sigma_1, \sigma_2\}$, the intersection number is zero. 
		\item \label{computeMori2} If $\{\rho\} = \sigma(1) \setminus \tau(1)$ for one $\sigma \in \{\sigma_1, \sigma_2\}$, the intersection number is
			\be \label{eq:intNumSimp}
				D_\rho \cdot V(\tau) = \frac{\mult(\tau)}{\mult(\sigma)} \,.
			\ee
		\item If $\rho \in \tau(1)$,  invariance of the intersection number under linear equivalence can be used to choose a representative of the class $[D_\rho]$ which is a linear combination of torus invariant divisors that does not involve $D_{\rho'}$ for any $\rho' \in \tau(1)$. The intersection product of $D_\rho$ with $V(\tau)$ can then be evaluated term by term by invoking \ref{computeMori1}. and \ref{computeMori2}.
	\end{enumerate}
	We can sidestep the above procedure and obtain the intersection number $D_\rho \cdot V(\tau)$ for all $\rho \in \Sigma(1)$ in one go via the following observation. Any character $\chi^m$ for $m\in M$ defines a principal divisor, which can be decomposed in terms of the torus invariant divisors $D_\rho$ as 
	\be \label{eq:principal_divisor}
	0 \sim (\chi^m) = \sum_{\rho \in \Sigma(1)} \langle m, u_\rho \rangle D_\rho \,.
	\ee
	Considering the intersection number with a curve $C$ yields
	\be \label{eq:intNumAsRel}
	 \forall m \in M  : \quad 0 = \langle m , \sum_{\rho \in \Sigma(1)} (C \cdot D_\rho) u_\rho \rangle \quad   \Rightarrow  \quad \sum_{\rho \in \Sigma(1)} (C \cdot D_\rho) u_\rho   = 0 \,.
	\ee
	Equation \eqref{eq:intNumAsRel} implies that Mori vectors encode relations between the cone generators $u_\rho$. The relation that follows from taking the intersection product of \eqref{eq:principal_divisor} with $V(\tau)$ is called the wall relation induced by $\tau$. From the discussion of intersection numbers, we know that for a wall $\tau = \sigma_1 \cap \sigma_2$, the only non-vanishing coefficients arise for $\rho \in \sigma_1(1) \cup \sigma_2(1)$. Thus, by normalizing the coefficients of the relation among these $n+1$ cone generators, e.g. by using \eqref{eq:intNumSimp}, we obtain the Mori vector of the curve $V(\tau)$.  As stated above, these wall relations generate the Mori cone of the toric variety $Y$.
	
	The computation of the Mori cone of the Calabi-Yau hypersurface $X$ is substantially more difficult. Note that the curves on $Y$ generically intersect $X$ in points or not at all. It would hence seem that $\MC(Y)$ contains little information with regard to $\MC(X)$. To see that this is not the case, consider the dual cone to the Mori cone, the so-called $\Nef$ cone. The $\Nef$ cone sits inside $N^1$, the abelian group generated by Cartier divisors up to numerical equivalence. Modding out by numerical equivalence to define both $N_1$ and $N^1$ guarantees that the intersection product defines a non-degenerate pairing between these two spaces, and allows for the definition $\Nef = \MC^\vee$. From the definition of $\MC$, we see that the $\Nef$ cone is spanned by Cartier divisor classes $[D]$ that satisfy $D \cdot C \ge 0$ for all complete irreducible curves $C$ (indeed, such Cartier divisors are called {\bf n}umerically {\bf ef}fective). We already encountered this condition in section \ref{ss:desingularization}, where we stated it to be equivalent to base-point freeness on complete toric varieties (we are using that the Mori cone on complete toric varieties is spanned by the set of torus invariant curves). Dealing with divisors rather than curves has the advantage that the non-empty intersections of divisors of $Y$ with $X$ yield (not necessarily irreducible) divisors of $X$. Indeed, 
	\be \label{eq:nefX_to_nefY}
	\Nef(Y)|_X \subset \Nef(X) \,.
	\ee
	While the torus invariant divisors of $Y$ are in one-to-one relation with the elements of $\Sigma(1)$, the $\Nef$ condition depends on how these 1-cones are assembled into fans. 
	
	As we discussed in section \ref{ss:g_n_varieties}, many different maximal projective subdivisions of $P$ can lead to the same hypersurface $X$. We hence obtain a better approximation of the $\Nef$ cone of $X$ than the inclusion \eqref{eq:nefX_to_nefY} by considering 
	\be \label{eq:toric_Kaehler_cone}
	\bigcup_{\substack{ \{  \text{maximal projective} \\ \text{subdivisions $\Sigma$ of $P$} \\ \text{yielding isomorphic $X$} \}}} \Nef(Y_{\Sigma})|_X \subset \Nef(X) \,.
	\ee
	The dual of this relation,
	\be \label{eq:toric_Mori_cone}
	\bigcap_{\substack{ \{  \text{maximal projective} \\ \text{subdivisions $\Sigma$ of $P$} \\ \text{yielding isomorphic $X$} \}}} \left(\Nef(Y_{\Sigma})|_X\right)^\vee \supset \MC(X) \,,
	\ee
	allows us to determine an approximation to the Mori cone of $X$. Note that this approximation is substantially improved by the fact that we take all maximal projective subdivisions of $P$ into account which yield isomorphic hypersurfaces $X$, not just those that refine the normal fan of $P$. We will refer to the cone defined in \eqref{eq:toric_Kaehler_cone} as {\it the toric K\"ahler cone}, and the cone in  \eqref{eq:toric_Mori_cone} as {\it the toric Mori cone}.

	Once we have computed the Gromov-Witten invariants for $X$, we can determine whether the toric Mori cone coincides with $\MC(X)$ by determining the Gromov-Witten invariants of its generators. If an invariant is non-vanishing, we can conclude that the corresponding class is represented by a curve in $X$.\footnote{This strategy was also pursued in \cite{Anderson:2017aux} in the context of studying the fibration structure of complete intersection Calabi-Yau manifolds.} As $\MC(X)$ is contained in the toric Mori cone, all generators of the latter having non-vanishing invariants suffices to conclude that the two coincide. In practice, we have found that this is the case whenever the toric Mori cone is smooth.
	
	We will use the terminology {\it toric Mori vector} also for curves in $X$, to indicate the tuple of intersection numbers of the curve with the divisors of $X$ descending from the torus invariant divisors of $Y$.
	
	\paragraph{Example: $(E_6)_3$  } 
	The generators of the toric Mori cone of one of the eight $(E_6)_3$ varieties (of type I, in the terminology to be introduced in section \ref{ss:distinguished_curves}) are given in table  \ref{table:toric_mori_cone_E_6_3}. $(E_6)_n$ for $n=1,2,4, 5$ each contain a variety with nearly identical Mori cone, differing only in the last generator $C_9$. This is also true for the other three type I varieties in $(E_6)_3$. Upon providing a useful basis of torus invariant curves in the next subsection, we will return to this example and the interpretation of the generators of the toric Mori cone in section \ref{sss:uv_in_ISR}.
	\begin{table}[!]
		\begin{center}
			\begin{tabular}{c|c|c|c|c|c|c|c|c|c|c|c|c|c} 
			& $D^X_{r_1}$ & $D^X_{r_2}$ & $D^X_{r_3}$ & $D^X_{r_4}$ & $D^X_{r_5}$ & $D^X_{r_6}$ & $D^X_{s}$ & $D^X_{t}$ & $D^X_{u}$ & $D^X_{v}$ & $D^X_{x}$ & $D^X_{y}$ & $D^X_{z}$ \\ \hline
			$C_{1}$ & $-1$ & $0$ & $0$ & $0$ & $1$ & $0$ & $0$ & $0$ & $0$ & $0$ & $0$ & $1$ & $0$ \\ 
			$C_{2}$ & $-1$ & $1$ & $0$ & $0$ & $-1$ & $0$ & $0$ & $0$ & $0$ & $0$ & $1$ & $0$ & $0$ \\ 
			$C_{3}$ & $0$ & $0$ & $0$ & $0$ & $0$ & $-1$ & $-1$ & $0$ & $1$ & $1$ & $0$ & $0$ & $0$ \\ 
			$C_{4}$ & $0$ & $0$ & $0$ & $0$ & $0$ & $0$ & $1$ & $1$ & $0$ & $0$ & $0$ & $0$ & $-2$ \\ 
			$C_{5}$ & $0$ & $0$ & $0$ & $0$ & $0$ & $1$ & $-2$ & $0$ & $0$ & $0$ & $0$ & $0$ & $1$ \\ 
			$C_{6}$ & $0$ & $0$ & $1$ & $-2$ & $1$ & $0$ & $0$ & $0$ & $0$ & $0$ & $0$ & $0$ & $0$ \\ 
			$C_{7}$ & $0$ & $0$ & $1$ & $0$ & $0$ & $-2$ & $1$ & $0$ & $0$ & $0$ & $0$ & $0$ & $0$ \\ 
			$C_{8}$ & $0$ & $1$ & $-2$ & $1$ & $0$ & $1$ & $0$ & $0$ & $0$ & $0$ & $0$ & $0$ & $0$ \\ 
			$C_{9}$ & $1$ & $-1$ & $0$ & $1$ & $-1$ & $0$ & $0$ & $0$ & $0$ & $0$ & $0$ & $0$ & $0$ \\ 
			\end{tabular}
		\caption{Toric Mori cone generators for an $(E_6)_3$ variety of type I.} \label{table:toric_mori_cone_E_6_3}
		\end{center}
	\end{table}

	\subsection{Relating the Mori cones of geometries related by a flop} \label{ss:flops_II}
	In the following subsection, we will be interested in the Mori vectors of a set of distinguished curves in the hypersurface $X$ which are cut out from torus invariant surfaces of the ambient space $Y$: the curves that give rise to the gauge fields and charged matter hypermultiplets of the F-theory compactification, as discussed in section \ref{s:F-matter}. In this subsection, we will study how curves in $X$ cut out from torus invariant surfaces in $Y$ behave under flop transitions. 
	
	All elements of $(\mg)_n$ are related by flops which are induced by flops of the toric ambient space. More precisely, we can choose representatives $Y_{\Sigma_i}$ among the ambient spaces with anti-canonical hypersurface $X_i \in (\mg)_n$ such that any two fans among the $\Sigma_i$ are related by a finite sequence of elementary flops and with each elementary flop resulting in a fan contained among the $\Sigma_i$. We shall thus begin by describing toric elementary flops on the ambient space, and then consider the flop at the level of the anti-canonical hypersurfaces. 
	
	Recall that a {\it flip} or {\it flipping transition} is the composition $(\phi')^{-1} \circ \phi$ of  two birational maps $\phi, \phi'$ which are a special type of {\it extremal contractions}: $\phi$ is a birational map between two normal projective varieties  $X$ and $X_0$ which is associated to an extremal ray $\cR$ of the Mori cone of $X$ such that 
	\begin{itemize}
		\item $X$ and $X_0$ are isomorphic in codimension 1,
		\item a curve $C$ in $X$ is mapped to a point by $\phi$ iff $[C] \in \cR$.
	\end{itemize}
	We say that $\phi$ contracts $\cR$. Likewise, the map $\phi': X' \rightarrow X_0$ contracts $\cR' = -\cR$. We will call $\cR$ the flipping extremal ray of the flipping transition. Note that to compare rays $\cR$ and $\cR'$ in the Mori cones of $X$ and $X'$ respectively, we rely on the isomorphism 
	\be \label{eq:iso_N1}
	N_1(X) \cong N_1(X') \,.
	\ee
	This isomorphism is the dual of the isomorphism 
	\be \label{eq:iso_Pic}
	N^1(X) \cong N^1(X')
	\ee
	induced by the flipping transition.	 In concrete terms, we can represent elements of $N_1(X)$, $N_1(X')$ in terms of $r$-dimensional vectors with integer (or, in the case of orbifold singularities, rational) entries, where $r$ is the rank of the Picard group.	
	
	 In the toric context, if the toric variety $Y_{\Sigma}$ is complete and quasi-projective  with at worst orbifold singularities, any of its extremal rays can be contracted, resulting in a toric variety $Y_{\Sigma_0}$ whose (perhaps generalized\footnote{A fan is a degenerate generalized fan if it contains convex polyhedral cones which are not strongly convex. In this case, the origin is not an element of the fan. One can nevertheless associate a toric variety to such a fan, by modding out by the intersection of all of its cones.}) fan $\Sigma_0$ is refined by $\Sigma$.
	
	In the case of  complete  toric varieties, all extremal rays are generated by torus invariant curves, each associated to a wall $\tau \in \Sigma(n-1)$. As we reviewed in section \ref{ss:mori_cone}, each wall $\tau = \sigma_1 \cap \sigma_2$ leads to a relation among the 1-cone generators $\{u_{\rho_i} : \rho_i \in \sigma_1(1) \cup \sigma_2(1) \}$,
	\be
	\sum_{i=1}^{\dim{Y_\Sigma}+1} b_i u_{\rho_i} = 0 \,.
	\ee
	Such a wall relation also specifies $\cR$. The nature of $Y_{\Sigma_0}$ depends sensitively on the distribution of signs among the coefficients $b_i$. A contraction featuring in a flipping transition requires at least two positive and two negative coefficients. Introducing the three subsets
	\be
	J_- = \{i:b_i <0\} \,,\,\, J_0 = \{i:b_i =0\} \,, \,\,J_+ = \{i:b_i >0\} \,,
	\ee
	we have that $\sigma_{J_+} \in \Sigma$, the non-simplicial cone $\sigma_{J_- \cup J_+} \in \Sigma_0$, and subdividing $\sigma_{J_- \cup J_+}$ yields $\sigma_{J_-} \in \Sigma'$. If $\Sigma$ is smooth to begin with, then any non-simplicial top-dimensional cone $\sigma_0$ of $\Sigma_0$ is spanned by $|J_- \cup J_+|+1$ generators and exhibits $\sigma_{J_- \cup J_+}$ as a face. The extremal contractions $\phi, \phi'$ define birational morphisms to $Y_0$ with exceptional locus $V(\sigma_{J_+})$, $V(\sigma_{J_-})$ respectively; all torus invariant curves of $Y$ whose associated 3-cones $\tau$ meet the interior of a non-simplicial top-dimensional cone $\sigma_0$ satisfy $[V(\tau)] \in \cR$ and are contracted by $\phi$; likewise, all torus invariant curves of $Y'$ whose associated 3-cones $\tau'$ meet the interior of a non-simplicial top-dimensional cone $\sigma_0$ satisfy $[V(\tau')] \in \cR'$ and are contracted by $\phi'$.  
	
	The flipping extremal rays $\cR$ we shall be considering will have exactly four non-zero coefficients $b_i$ in their wall relation, twice 1 and twice -1. Upon reordering the 1-cones, we can thus take the toric Mori vector spanning $\cR$ to be
	\be \label{eq:extremal_ray}
	\cR = \langle (1,1,-1,-1, \mathbf{0}) \rangle\,.
	\ee
	As the sum of the coefficients $b_i$ vanishes,
	\be
	-K \cdot \cR = 0 = -K' \cdot \cR' \,,
	\ee
	i.e. the degree of the respective anti-canonical divisor on all contracted curves is zero. Flipping transitions with this property are called {\it flop transitions} or simply {\it flops}. Note in particular that flop transitions preserves the nefness of the anti-canonical bundle. As the fans we are considering come from triangulations of polytopes, the anti-canonical divisors of the varieties they define are always nef, as discussed above.  The only flipping transitions that can occur between them are hence flops.

	Flop transitions in four complex dimensions with flipping extremal ray of the form \eqref{eq:extremal_ray} are very closely related to the familiar flop transition of non-compact Calabi-Yau 3-folds:  let us call the five 1-cone generators involved in the wall relation underlying $\cR$ $u_1, \ldots, u_5$, such that the wall relation is given by
	\be \label{eq:extremal_wall_relation}
	u_1 + u_2 - u_3 - u_4 = 0 
	\ee
	(the coefficient of $u_5$ is zero). In the notation introduced above, we hence have
	\be
	\sigma_{J_+} = \sigma_{12} \,, \quad \sigma_{J_- \cup J_+} = \sigma_{1234} \,, \quad \sigma_{J_-} = \sigma_{34} \,.
	\ee

\begin{figure}
		\centering
			\includegraphics[width=1\linewidth]{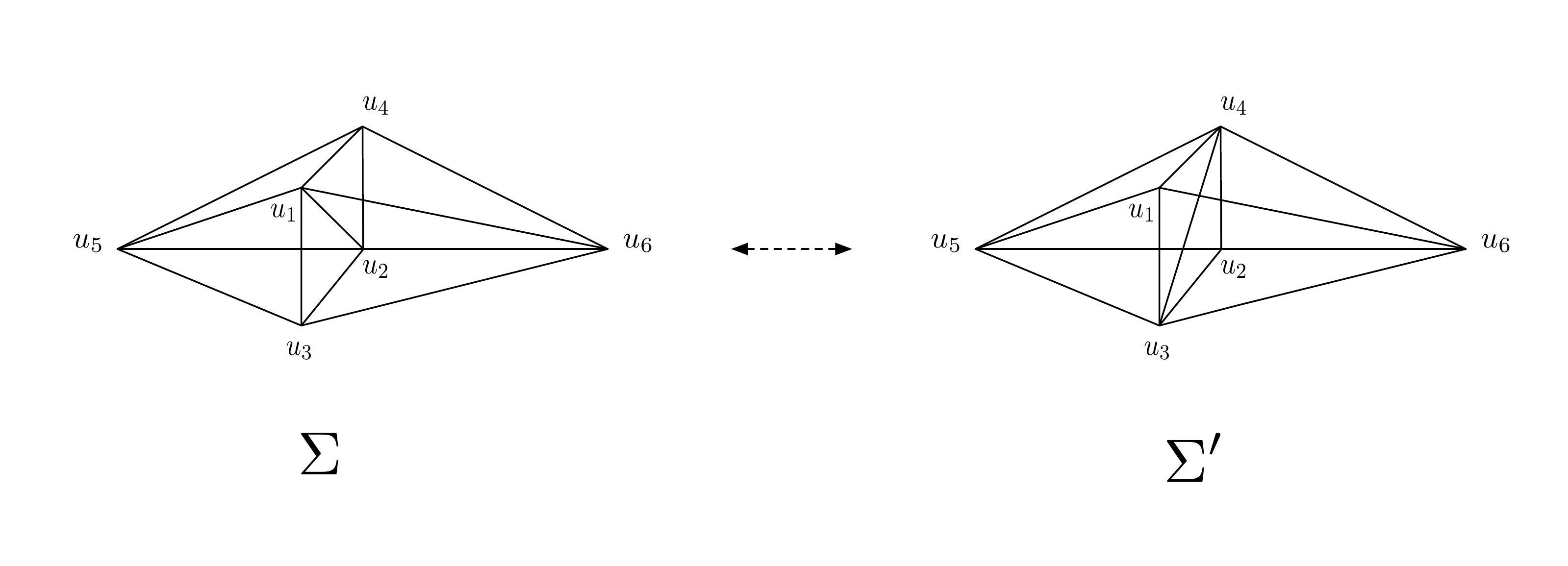}
		\caption{The cones involved in a flipping transition.}
		\label{fig:generic_flop}
	\end{figure}

	To clarify the local geometry, consider the $GL(4,\IZ)$ mapping 
	\begin{eqnarray} \label{eq:local_presentation}
	u_1 &\rightarrow& e_1 + e_2 + e_3 \,, \\
	u_2 &\rightarrow& - e_1 \,, \nn\\
	u_3 &\rightarrow& e_2 \,, \nn\\
	u_4 &\rightarrow& e_3 \,, \nn\\ 
	u_5 &\rightarrow& e_4 \,.\nn 
	\end{eqnarray}
	The intersection of the corresponding cones with the hyperplane orthogonal to the $e_1$ direction are depicted on the LHS of figure \ref{fig:generic_flop} (the generator $u_6$ will be discussed momentarily).
	We recognize a sum of three line bundles over a rational curve $\IP^1$ whose fan extends in the $e_1$ direction, and can read off the corresponding divisors in terms of the class $D = \langle e_1 \rangle$ from the coefficients of $e_2$, $e_3$, and $e_4$ in the image of $u_1$, yielding
	\be \label{eq:5d_conifold}
	\cO(-1) \oplus \cO(-1) \oplus \cO \,,
	\ee
	i.e. the familiar conifold, up to a factor of $\IC$. 
	
	Given the existence of a flipping extremal ray of the form \eqref{eq:extremal_ray} in the Mori cone of $Y_{\Sigma}$, we can draw further conclusions regarding the local form of $\Sigma$. We will argue in terms of the coordinate system \eqref{eq:local_presentation}. 
	
	To obtain the list of all torus invariant curves whose class lies in $\cR$, we need to identify all top-dimensional cones in the fan $\Sigma$ which meet the interior of any non-simplicial top-dimensional cone $\sigma_0 \in \Sigma_0$ with face $\sigma_{1234}$. Exactly two such cones exist: $\sigma_{12345}$, and a cone $\sigma_{12346}$, with $u_6$ a 1-cone generator extending in the negative $e_4$ direction. Imposing the smoothness of $\sigma_{1236}$ fixes the $e_4$ coordinate of $u_6$ to be $\pm1$. It must hence have the form
	\be \label{eq:u6}
	u_6 \rightarrow \alpha e_1 + \beta e_2 + \gamma e_3 - e_4 \,.
	\ee
	As the wall $\sigma_{126}$ meets the interior of $\sigma_{12346}$, $[V(\sigma_{126})] \in \cR$. 
	
	We have thus arrived at the following local structure of the fans $\Sigma$ and $\Sigma'$: they exhibit 1-cone generators $u_1, \ldots, u_6$ which assemble to cones as indicated in figure \ref{fig:generic_flop}.
	In particular, taking $u_6$ as given in \eqref{eq:u6} into consideration allows us further insight into the local geometry \eqref{eq:5d_conifold}: the summand $\cO$ is in fact compactified to a rational curve $\IP^1$ whose fan extends in the $e_4$ direction of the lattice and over which the local geometry is fibered,
	\be
	(\cO(-1) \oplus \cO(-1)   \rightarrow \IP^1) \rightarrow \IP^1 \,.	
	\ee
	 The fan $\Sigma_0$ is obtained from $\Sigma$ by removing the walls $\sigma_{125}$ and $\sigma_{126}$. Triangulating the resulting non-simplicial cones of $\Sigma_0$ by adding the walls $\sigma_{345}$ and $\sigma_{346}$ yields the fan $\Sigma'$. 
	 
	 All type~I varieties are related via flops induced by the contraction of curves contained in the fiber of the elliptic fibration, such that $\{u_5, u_6\} = \{u_{\rho_u},u_{\rho_v} \}$. The transition between type~I and type~II varieties occurs via a flop induced by the contraction of a curve with the base curve of the Hirzebruch surface as a component. In terms of the ambient space, such flops are characterized by $\{u_3, u_4\} = \{u_{\rho_u},u_{\rho_v} \}$. The flop thus induces a top-dimensional cone containing both $\rho_u$ and $\rho_v$ as faces: the monomial $u*v$ is removed from the Stanley-Reisner ideal. Type~II varieties amongst themselves are again related via flops such that $\{u_5, u_6\} = \{u_{\rho_u},u_{\rho_v} \}$. 
	
\begin{figure}
		\centering
		\includegraphics[width=1\linewidth]{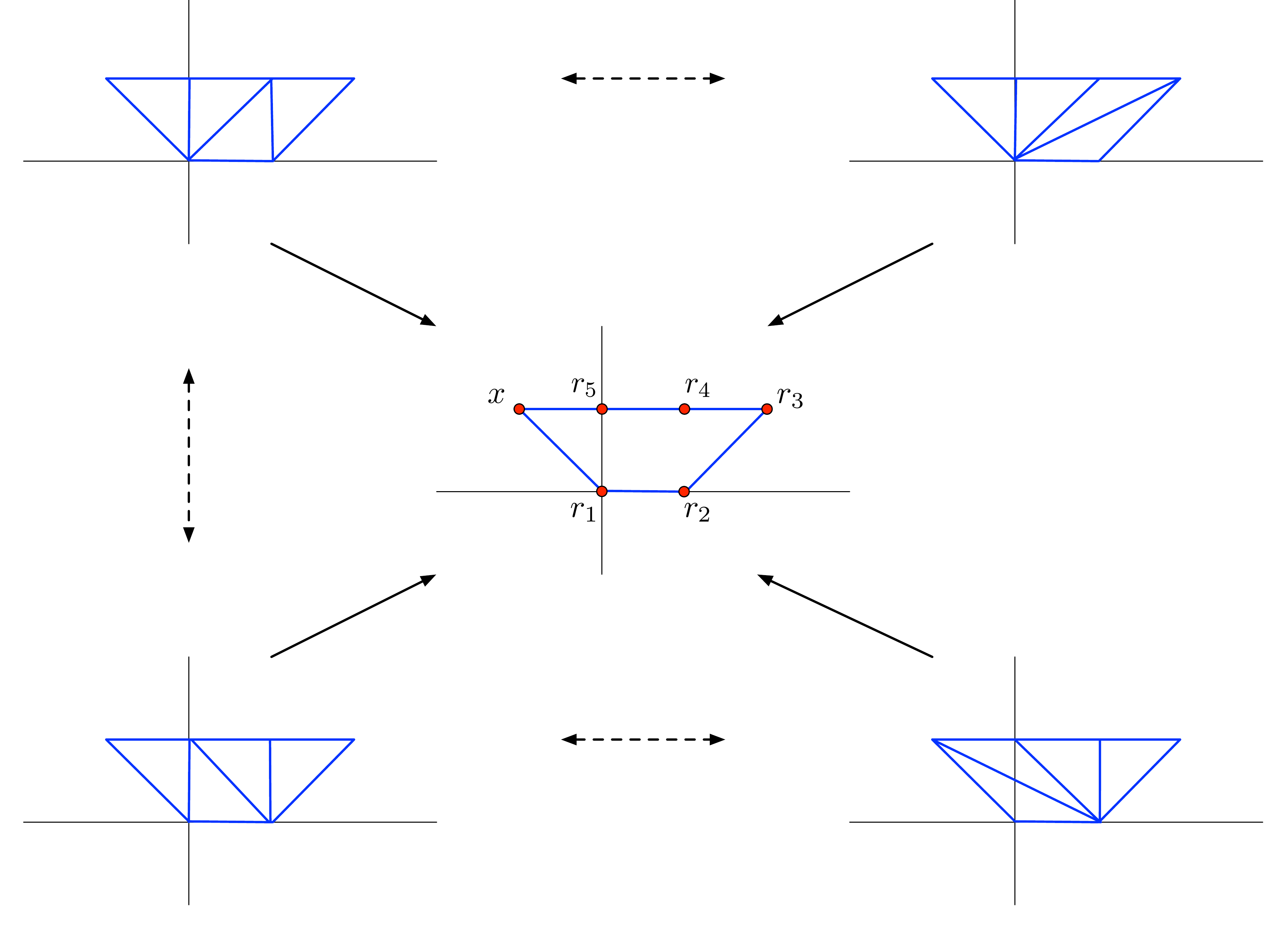}
		\caption{All flops relating $(E_6)_n$ varieties of type~I.}
		\label{fig:E6_flops_without_central_node}
	\end{figure}

\begin{figure}
		\centering
		\includegraphics[width=1\linewidth]{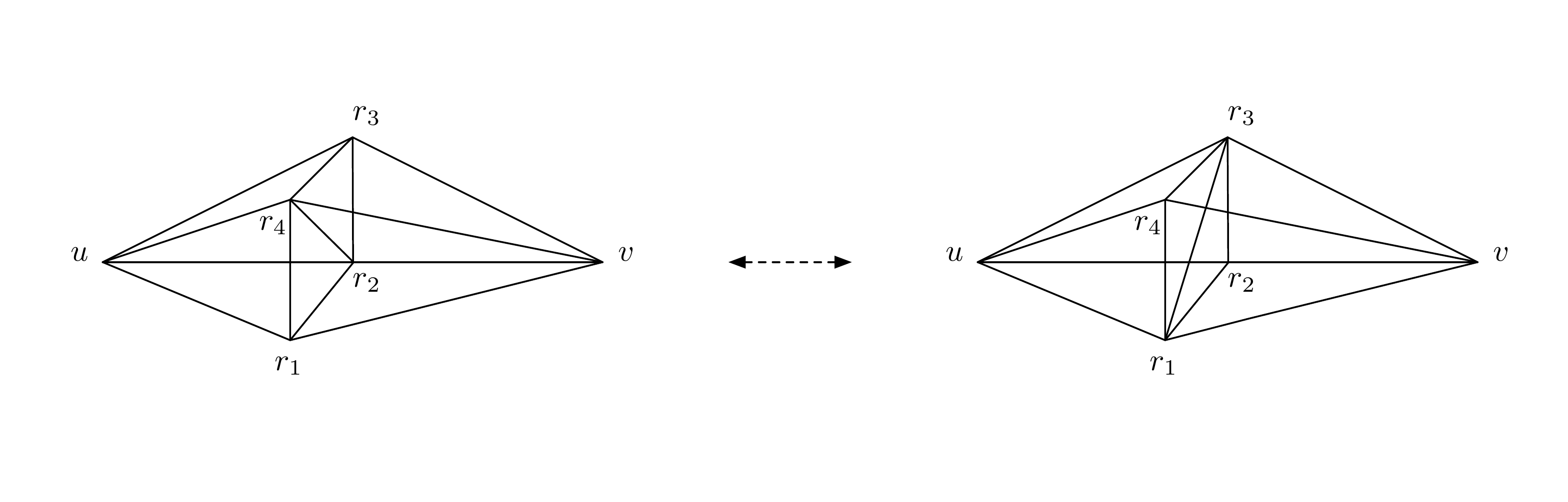}
		\caption{The top flop in figure \ref{fig:E6_flops_without_central_node} in detail.}
		\label{fig:a_complete_E6_flop}
	\end{figure}

	\paragraph{Example:}
	The classes $(E_6)_n$ for $n=1, \ldots, 5$ each contain four  varieties of type~I, related via flops such that $\{ u_5, u_6\} = \{u_{\rho_u},u_{\rho_v}\}$. All further 1-cone generators involved in the various flops end on a common 2-plane, as depicted in figure \ref{fig:E6_flops_without_central_node}. 
	In more detail, the flop relating the top two partial fans in figure \ref{fig:E6_flops_without_central_node} is given in figure \ref{fig:a_complete_E6_flop}.
	The classes $(E_6)_n$ for $n=1,3,5$ each also contain four varieties of type~II.
	\hspace{1cm}
	
	We now turn to comparing the Mori cone of two varieties $Y_{\Sigma}$ and $Y_{\Sigma'}$ related by a flipping transition induced by the contraction of an extremal ray $\cR$ of the form \eqref{eq:extremal_ray}. Only the curve classes $[V(\tau_i)]$, $\tau_i$ a facet of one of the top-dimensional cones modified by the flop, will be affected. We will call this set of cones $S_{flop}$, $S_{flop}'$ respectively:
	\be
	S_{flop} = \{\sigma_{1235},\sigma_{1245},\sigma_{1236}, \sigma_{1246}\} \,,
	\ee
	\be
	S_{flop}' = \{\sigma_{1345},\sigma_{2345},\sigma_{1346}, \sigma_{2346}\} \,.
	\ee
	We will perform the following analysis for curves $V(\sigma)$ with $\sigma \prec \sigma_{1235}$ or $\sigma \prec \sigma_{1245}$. The discussion for curves $V(\sigma)$ with $\sigma$ contained in the other two cones $\sigma_{1236}$, $\sigma_{1246}$ proceeds by replacing $5 \rightarrow 6$ in all ensuing steps. 
	
	By the fundamental property of a flipping transition, we have that
	\be
	[V(\sigma_{125})] = - [V(\sigma_{345})]' \,,
	\ee
	where we have indicated the classes in $Y_\Sigma$, $Y_{\Sigma'}$ by $[\cdot]$, $[\cdot]'$ respectively, and the comparison of curve classes in the Mori cones of $Y$ and $Y'$ proceeds via the identification \eqref{eq:iso_N1}. 
	
	To study a facet different from the wall, recall that the relations among the generators of the Chow group $A_k(Y_{\Sigma})$ are determined by the exact sequence
	\be
	\bigoplus_{\tau \in \Sigma(n-k-1)} M(\tau)_{\IQ} \overset{\alpha}{\longrightarrow} \bigoplus_{\sigma \in \Sigma(n-k)} \IQ[V(\sigma)] \overset{\beta}{\longrightarrow} A_k(Y_\Sigma)_{\IQ} \longrightarrow 0 \,,
	\ee
	where the image of the map $\alpha$ on an element $m \in M(\tau) = \tau^\perp \cap M$ is given by
	\be  \label{eq:trivial_chow_class}
	\alpha(m) = \sum_{\sigma \in \Sigma(n-k), \tau \prec \sigma} \langle m, u_{\rho,\tau} \rangle [V(\sigma)] 
	\ee
	and maps to a trivial class in the Chow group.	Here, $u_{\rho,\tau}$ is the image of $u_\rho \in \sigma$ in the quotient $N/N_\tau$, with $N_\tau = \IZ(\tau \cap N)$ the sublattice of $N$ spanned by $\tau$, and is uniquely determined by the fact that $N_\sigma/N_\tau \cong \IZ$.
	
	Let us determine  e.g. how  the class of the curve $V(\sigma_{135})$ changes due to the flop. If we choose the cone $\sigma_{15}$ as the $\tau$ in \eqref{eq:trivial_chow_class}, we have
	\be
	\{ \sigma \in \Sigma(3) \,|\, \tau \prec \sigma \prec \sigma_{top} \in S_{flop}\} = \{ \sigma_{135}, \sigma_{125}, \sigma_{145} \} \,.
	\ee
	Next, we choose $m \in M(\sigma_{15}) = (\sigma_{15})^\perp \cap M$ to also be orthogonal to $u_4$. By \eqref{eq:extremal_wall_relation}, it follows that $\langle m, u_{2,\tau} \rangle = \langle m, u_{3,\tau} \rangle$. Therefore,
	\be \label{eq:curve_relations}
	[V(\sigma_{135})] = - [V(\sigma_{125})] + [C]\,,
	\ee
	where $C$ is a sum involving curves $V(\sigma)$, $\tau \prec \sigma$ but $\sigma$ not a face of $\sigma_{1235}$ or $\sigma_{1245}$. The same considerations for $Y'$ yield
	\be
	\{ \sigma \in \Sigma'(3) \,|\, \tau \prec \sigma' \prec \sigma_{top} \in S_{flop}'\} = \{ \sigma_{135}, \sigma_{145} \} \,.
	\ee
	The cone $\sigma_{125}$ is missing on the RHS, as $\sigma_{125} \not \in \Sigma'(3)$. 
	Hence
	\be
	[V(\sigma_{135})]'=  [C]' \,,
	\ee
	with the same $C$ as in \eqref{eq:curve_relations}. As $[C] = [C]'$, we conclude that
	\be
	[V(\sigma_{135})] + [V(\sigma_{125})] = [V(\sigma_{135})]'  \,.
	\ee
	This relation remains true upon replacing $\sigma_{135}$ by $\sigma_{145},\sigma_{235},\sigma_{245}$ respectively. The remaining two facets $\sigma_{123}$ and $\sigma_{124}$ of $\sigma_{1235}$ and $\sigma_{1235}$ do not feature in the fan of $\Sigma'$, hence do not give rise to torus invariant curves of $Y_{\Sigma'}$.
	
	We have now laid the groundwork for performing the analogous analysis at the level of the hypersurfaces $X$ and $X'$. Here, we consider the surface classes in the respective ambient spaces which restrict to the curves of interest upon intersection with the anti-canonical class.	In particular, the curve
	\be
	\Ccontr = V(\sigma_{12}) \cdot (-K) \,,
	\ee
	considered as a curve of $X$, generates the extremal ray in the Mori cone of $X$ that is contracted in the flipping transition, as
	\be \label{eq:identifying_flopped_curve_in_hypersurface}
	[\Ccontr] = k[V(\sigma_{125})] \,, \quad k>0 \,.
	\ee
	This almost follows from the local structure of $\Sigma$ derived above: the relation between the divisor classes associated to the 1-cones generated by $u_1, \ldots, u_6$ is  read off from the presentation  \eqref{eq:local_presentation} and \eqref{eq:u6}:
	\begin{eqnarray} \label{eq:divisor_relations}
	\lbrack D_{u_2}]  &=& [D_{u_1}] + \alpha [D_{u_6}] +\ldots \,, \nn\\
	\lbrack D_{u_3} \rbrack &=& -[D_{u_1}] - \beta [D_{u_6}] +\ldots\,, \nn\\
	\lbrack D_{u_4}] &=& -[D_{u_1}] - \gamma [D_{u_6}] +\ldots\,, \nn\\
	\lbrack D_{u_5}] &=&  [D_{u_6}] +\ldots\,,
	\end{eqnarray}
	where $\ldots$ indicate divisor classes not involving $[D_{u_i}]$, $i=1, \ldots, 6$. Hence,
	\be \label{eq:anti_canonical_specialized}
	-[K] = \sum_{\{i: \langle u_i \rangle \in \Sigma(1)\}} [D_{u_i}] = (\alpha -\beta-\gamma+2) [D_{u_5}] + \ldots \,,
	\ee
	with the dots having the same significance as before. Therefore,
	\be
	[\Ccontr] = (\alpha - \beta - \gamma +2) [V(\sigma_{125})] \,.
	\ee
	We can obtain the coefficient of $[V(\sigma_{125})]$ as the intersection number
	\be
	V(\sigma_{123}) \cdot (-K) = (\alpha - \beta - \gamma +2)  \,.
	\ee
	As the maximal projective subdivisions we are considering can have merely nef (rather than ample) anti-canonical divisors, this only allows us to conclude  $(\alpha - \beta - \gamma +2) \ge 0$. However, in all instances studied, strict inequality holds. Given \eqref{eq:identifying_flopped_curve_in_hypersurface}, we can easily relate the Mori vectors of the curves of interest in $X$ and $X'$:
	
	\begin{itemize}
		\item The Mori vectors of curves of the form $V(\sigma_{i5}) \cdot (-K)$, $V(\sigma_{i6}) \cdot (-K)$ for $i = 1, \ldots, 4$ are invariant under the flop as by \eqref{eq:identifying_flopped_curve_in_hypersurface} and \eqref{eq:divisor_relations}, $D_5 \cdot \Ccontr = D_6 \cdot \Ccontr = 0$.
		\item To work out the behavior of the Mori vectors of the the curves of the form $V(\sigma_{ij})\cdot(-K)$, $i \in \{1,2\}$, $j \in \{3,4\}$, we can apply the same line of reasoning to the surfaces $V(\sigma_{ij})$ as used above for curves of the form $V(\sigma_{ij5})$, $i \in \{1,2\}$, $j \in \{3,4\}$ . Choosing $\tau = \sigma_1$, we have
		\be
		\{ \sigma \in \Sigma(2) \,|\, \tau \prec \sigma \prec \sigma_{top} \in S_{flop}\} = \{ \sigma_{12}, \sigma_{13}, \sigma_{14},\sigma_{15},\sigma_{16}\} \,.
		\ee
		Choosing $m \in M(\sigma_1)$ to also be orthogonal to $u_4$, we obtain e.g.
		\be \label{eq:surface_relations}
		[V(\sigma_{13})] = - [V(\sigma_{12})] + [S]\,,
		\ee
		with $S$ a sum of surfaces $V(\sigma)$ such that $\tau \prec \sigma$ but $\sigma$ does not contain the 1-cones $\langle u_i \rangle$, $i = 1, \ldots 4$ as a face. The same consideration for $Y'$ yield
		\be
		\{ \sigma \in \Sigma'(2) \,|\, \tau \prec \sigma \prec \sigma_{top} \in S_{flop}'\} = \{  \sigma_{13}, \sigma_{14},\sigma_{15},\sigma_{16} \} \,.
		\ee
		Hence,
		\be
		[V(\sigma_{13})]'=  [S]' \,,
		\ee
		with the same $S$ as in \eqref{eq:surface_relations}. Note that $S$ can contain surface components of the form $V(\sigma_{i5}), V(\sigma_{i6})$ which are affected by the flop, we hence cannot conclude $[S] = [S]'$. By the previous bullet point however, we can equate the class of the intersection of $S$ with $-K$, i.e. $[S\cdot(-K)] = [S\cdot(-K)]'$, allowing us to conclude
		\be
		[V(\sigma_{13})\cdot(-K)] + [\Ccontr] = [V(\sigma_{13}) \cdot(-K)]'  \,.
		\ee
		The relation remains true upon replacing $\sigma_{13}$ by $\sigma_{14},\sigma_{23},\sigma_{24}$ respectively.
		
	\end{itemize}

	\subsection{Constructing torus invariant curves in the Mori cone of $X$} \label{ss:distinguished_curves}
	Distinguished curve and divisor classes in elliptically fibered Calabi-Yau manifolds $X$ in the context of F-theory models have been discussed in \cite{Intriligator:1997pq,Aspinwall:2000kf,Park:2011ji} and reviewed in the recent lecture notes \cite{Weigand:2018rez}. Aside from the embedding of the base and fiber curve of the Hirzebruch surface $\IF_n$  into $X$ via the zero section of the fibration, called $C_B$ and $C_F$ below, the curves of interest are fibral curves:  the generic elliptic fiber $C_E$, the rational curves $C_{r_i}$, $i \in 1, \ldots, \rk(\mg)$, resolving the singularities of $X$, and the rational curve $C_{r_0}$. Unlike the $C_{r_i}$, $i >0$, $C_{r_0}$ intersects the zero section. The curve classes $[C_E]$, $[C_{r_0}]$ and $[C_{r_i}]$ are related by
	\be \label{eq:generic_fiber_in_terms_of_marks}
	[C_E] = \sum_{i=0}^{\rk{\mg}} a_i [C_{r_i}] \,,
	\ee
	with $a_0 =1$ and $a_i$, $i>0$, the marks of $\mg$, i.e. the coefficients of the highest root in an expansion in simple roots.
	
	The distinguished divisors $D^i$ associated to coroots of $\mg$ in section \ref{s:F-matter} arise, for $\mg$ simply-laced, from fibrations of the curves $C_{r_i}$, $i>0$, over the discriminant locus of the fibration. Unlike the case for elliptically fibered surfaces, elliptically fibered threefolds can have singularities associated to non-simply laced Lie algebras (see e.g. \cite{Esole:2017qeh,Esole:2017rgz} for an in depth study of this phenomenon). This happens when the fiber of the distinguished divisor $D^i$ over a generic point of the discriminant locus is reducible, but the components are not invariant under monodromy. In this case, $D^i$ has the description of a rational fibration over a branched cover of the discriminant locus. In both cases, we will refer to the distinguished divisors $D^i$ as {\it resolution divisors}.
	
	In this subsection, we will identify the resolution divisors of the anti-canonical hypersurface $X$ with intersections of appropriate toric invariant divisors of the ambient space $Y$ with the anti-canonical divisor $-\KDY$. Likewise, we will
	obtain the distinguished curves $C_i$ as intersections of two torus invariant divisors of the ambient space $Y$ with $-\KDY$. We will use the notation 
	\be
	\DX_{x_i} = - \DY_{x_i} \cdot \KDY \,;
	\ee 
	which intersection product is meant can be read off from the superscript of the divisors. 
	
	We note that when the resolution divisor $D^i$ has reducible fiber over a generic point of the base, the appropriate intersection of toric invariant divisors to obtain the fiber of $D^i$  yields a curve with Mori vector divisible by an integer $n^{comp}$. In this case, we identify the correct Mori vector for the rational curve $C_{r_i}$ by dividing by $n^{comp}$ .
	
	For the following discussion, it will be useful to identify a torus invariant divisor $D_\rho$ with the zero set of the corresponding homogeneous coordinate $x_\rho$. To see that the latter describes the former, note that the polyhedron $P_{D_\rho}$ associated to the divisor $ D_\rho$, as given in \eqref{eq:polyhedron_D}, always contains $m = \boldsymbol{0}$ as an interior point. From the $D_\rho$-homogenization of $\chi^m$, we conclude that the associated line bundle to $D_\rho$ exhibits a global meromorphic section which is given by $x_\rho$ in homogeneous coordinates.

	The proper identification of the distinguished curves $C_i$ in $X$ depends on whether $X$ is a variety of type ~I or type~II, in the terminology introduced in section \ref{ss:g_n_varieties}. 
	We will discuss these two cases in turn.
	
	\subsubsection{Type I: $u*v \in I_{\mathrm Stanley-Reisner}$} \label{sss:uv_in_ISR}
	As discussed in section \ref{ss:g_n_varieties}, when $u*v \in I_{\mathrm Stanley-Reisner}$, $X$ is fibered over the base $\IF_n$,
	\be
	\pi: X \rightarrow \IF_n \,,
	\ee
	with $\IF_n$ embedded in $X$ as the image $Z$ of the zero section induced by that of $X_n$. $\pi |_Z$ induces a push-forward map (we drop the $|_Z$ in the notation) $\pi_*: N^1(Z) \rightarrow N^1(\IF_n)$. We will denote divisor classes both in $N^1(X)$ and $N^1(\IF_n)$ by $[\cdot]$. In particular, we write $[\DB_B]$ and $[\DB_F]$ for the base and fiber class of $\IF_n$. 
	
	The divisors that will be relevant in our discussion are the following:
	
	\paragraph{$\DX_z$: }$[\DX_z] = [Z]$
	
	\vspace{-0.5cm}
	
	\paragraph{$\DX_u$: } $\pi_*(\DX_u\cdot [Z]) = [\DB_F]$
		
	\vspace{-0.5cm}
	
	\paragraph{$\DX_s$: } $\pi_*([\DX_s]\cdot [Z]) = [\DB_B]$
	
	\vspace{-0.5cm}
	
	\paragraph{$\DX_{r_i}$: } $[\DX_{r_i}]\cdot [Z] = 0$
	
	\vspace{0.5cm}
	
	By studying the intersection properties of these divisors, we can identify the distinguished curves in the geometry with the following torically invariant intersections:
	
	\paragraph{$C_B$, the base curve of the zero section $Z$}
	\begin{equation} \label{eq:CB}
	C_B = \DX_z \cdot \DX_s
	\end{equation}
	
	\paragraph{$C_F$, the fiber curve of the zero section $Z$}
	\begin{equation} \label{eq:CF}
	C_F = \DX_z \cdot \DX_u
	\end{equation}
	
	\paragraph{$C_{r_0}$, the fibral rational curve intersecting the zero section}
	\begin{equation} \label{eq:Cr0}
	C_{r_0} = \DX_s \cdot \DX_u
	\end{equation}
	
	\paragraph{$C_E$, the generic elliptic fiber}
	\begin{equation} \label{eq:CE}
		C_E = \DX_t \cdot \DX_u
	\end{equation}
	
	\paragraph{$C_{r_i}$, the exceptional curves}
	\begin{equation}  
	C_{r_i} = \frac{1}{n_i^{comp}}\DX_{r_i} \cdot \DX_u  \label{eq:exceptional_curves}
	\end{equation}
	
	To justify the identifications \eqref{eq:Cr0} and \eqref{eq:CE} of $C_{r_0}$ and $C_E$ respectively, recall from the discussion in section \ref{ss:anti_canonical_hypersurface} that the singularity of the anti-canonical hypersurface of $Y_n$ lies, in terms of homogeneous coordinates, at $s=0$. Furthermore, as $s*t \in I_{Stanley-Reisner}$ of $Y_n$ and hence any of its refinements, the loci $s=0$ and $t=0$ are disjoint.
	
	The identification of the curves \eqref{eq:CB} to \eqref{eq:exceptional_curves} is valid for all varieties of type~I, as the classes of these curves   are invariant under flipping transitions for which, in  the notation of section \ref{ss:flops_II},  $\{u_5,u_6\} = \{u,v\}$, i.e. for which the flipping extremal ray $\cR$ has no component in the $C_B$ direction. For the curves $C_F$, $C_{r_0}$, and $C_{r_i}$, this is immediate, as by the discussion of  \ref{ss:flops_II}, all curves of the form $V(\sigma_{u_i, u_5}) \cdot (-K)$ are invariant under such flops. To establish the invariance for the curve $C_B$, we need to argue that the difference $u_s - u_z$ cannot be extended to yield a wall relation of the form \eqref{eq:extremal_wall_relation}. This is difficult to argue for in general, but can easily be checked case by case (e.g. against the 1-cone generators cited in the appendix).\footnote{Over the base $\IF_1$, the wall relation $u_s + u_v - u_z - u_u = 0$ exists, but the corresponding flop is not of the form we are discussing in this subsection, as $\{u_5,u_6\} \not = \{u,v\}$.}
	
	With regard to the discussion in section \ref{s:F-matter}, we identify the  $\DX_{r_i}$ with the resolution divisors associated to the perturbative gauge fields $A_i$; the curves $C_{r_i}$, when wrapped by M2 branes, give rise to the vector multiplets associated to the Lie algebra element $E_{\alpha_i}$, for $\alpha_i \in \Delta$, i.e. a simple root. Indeed, one can check that the intersection numbers between these divisor and curve classes  reproduce the negative Cartan matrix of a Lie algebra $\mg$.\footnote{Note that  the identification of divisors with simple coroots and curves with simple roots in section \ref{s:F-matter} implies that in the case of the gauge algebra $\mathfrak{b}_n$, which exhibits $n-1$ long roots and one short root, exactly one of the constants $n_i^{comp}$ is equal to 2, whereas the gauge algebra $\mathfrak{c}_n$ should exhibit $n-1$ constants $n_i^{comp}$ equal to 2.}.	
		  In fact, more is true: including $\DX_{r_0} := \DX_{s}$ and $C_{r_0}$ in our considerations, we have
	\be \label{eq:affine_cartan}
	[C_{r_i}] \cdot [\DX_{r_j}]   =  -\widehat{A}_{ij} \,, \quad i,j = 0, \ldots, \rk \mg \,,
	\ee 
	with $\widehat{A}$ the affine Cartan matrix of $\mg$.

	As the curves $C_B$ and $C_F$ arise via intersections with the divisor $\DX_z$, the computation of their Mori vectors reduces to a computation of intersection numbers in $\IF_n$. Thus, for $C_B$, the only non-vanishing intersection numbers are
	\ba
	[\DX_s] \cdot [C_B] &=& [\DX_s] \cdot [\DX_z] \cdot [\DX_s] = [\DB_B]^2 = -n  \,, \\[0.5cm]
	 [\DX_u] \cdot [C_B] &=& [\DX_u] \cdot [\DX_z] \cdot [\DX_s] =  [\DB_F]\cdot [\DB_B] = 1 \\
	&=&[\DX_v] \cdot [C_B]\,, \nn\\[0.5cm]
	[\DX_z]\cdot [C_B] &=& [\DX_z] \cdot [\DX_z] \cdot [\DX_s] = [K^{\IF_n}] \cdot [\DB_B] \\
	&=& -(2[\DB_B] +(n+2) [\DB_F]) \cdot [\DB_B] \nn \\
	&=& 2n - n - 2 = n-2 \,,  \nn
	\ea 
	while for $C_F$,
	\ba
	[\DX_s] \cdot [C_F] &=& [\DX_s] \cdot [\DX_z] \cdot [\DX_u] =  [\DB_B]\cdot [\DB_F] = 1   \,, \\[0.5cm]
	[\DX_z]\cdot [C_F] &=& [\DX_z] \cdot [\DX_z] \cdot [\DX_u] = [K^{\IF_n}] \cdot [\DB_F]  \\
	&=& -(2[\DB_B] +(n+2) [\DB_F]) \cdot [\DB_F]  \nn \\
	&=& -2 \,. \nn 
	\ea 
	The intersection numbers with the divisor $\DX_v$ follow from the linear equivalence
	\be
	\DY_v \sim \DY_u \,. 
	\ee
	
	Note that the relation \eqref{eq:generic_fiber_in_terms_of_marks} between the class of the generic and the exceptional fibers allows us to identify the fourth component of the 1-cone generators $u_{r_i}$, in the coordinate system \eqref{table:one_cones_for_Y}, with the negative comarks $a_i^\vee$ of the Lie algebra $\mathfrak{g}$, as follows. The divisor $\DY_t$ enjoys the linear equivalence
	\be
	\DY_t \sim  n \DY_u + \DY_s - \sum_{i=1}^{\rk \mathfrak{g}} (u_{r_i})_4 \DY_{r_i} \,.
	\ee
	Intersecting with $-\KDY$ and $\DY_u$, we obtain
	\be
	[C_E] =   [C_{r_0}] - \sum_{i=1}^{\rk \mathfrak{g}} n_i^{comp} (u_{r_i})_4 [C_{r_i}] \,.
	\ee
	We have used the fact that $[\DY_u] \cdot [\DY_u] = [\DY_u] \cdot [\DY_v] = 0$, as $u*v \in I_{Stanely-Reisner}$. Comparing with \eqref{eq:generic_fiber_in_terms_of_marks}, we obtain
	\be
	(u_{r_i})_4 = - \frac{1}{n_i^{comp}} a_i \,, \quad i = 1, \ldots, \rk \mg
	\ee
	and the claim follows.
	
	The Mori vector of the generic fiber $C_E$ can also be computed universally: by \eqref{eq:generic_fiber_in_terms_of_marks} and \eqref{eq:affine_cartan},
	\be
	 [\DX_{r_j}] \cdot [C_E] = -\sum_{i=0}^{\rk \mg} a_i (\widehat{A})_{ij} = 0 \,, \quad j = 0, \ldots, \rk \mg \,.
	\ee
	The only non-vanishing intersections are with $D_x$, $D_y$, $D_z$: the generic fiber is a degree 2, 3 curve respectively in the homogeneous coordinates $x$, $y$, hence
	\be
	[\DX_x ]\cdot [C_E ]= 3 \,, \quad [\DX_y] \cdot [C_E] =2 \,.
	\ee
	Finally,
	\be
	[\DX_z] \cdot [C_E] = [\DB_B + n \DB_F] \cdot [\DB_F] = 1 \,.
	\ee

	\paragraph{Example: $(E_6)_n$}
	 The intersection numbers of the curves $C_\bullet$ just constructed with the divisors $\DX_{x_i}$ are given in table \ref{table:mori_vectors_E6_n}, for general $\IF_n$.
	\begin{table}[!]
		\begin{center}
			\begin{tabular}{c|c|c|c|c|c|c|c|c|c|c|c|c|c} 
					& $D^X_{r_1}$ & $D^X_{r_2}$ & $D^X_{r_3}$ & $D^X_{r_4}$ & $D^X_{r_5}$ & $D^X_{r_6}$ & $D^X_{s}$ & $D^X_{t}$ & $D^X_{u}$ & $D^X_{v}$ & $D^X_{x}$ & $D^X_{y}$ & $D^X_{z}$ \\ \hline
					$C_{E}$ & $0$ & $0$ & $0$ & $0$ & $0$ & $0$ & $0$ & $0$ & $0$ & $0$ & $2$ & $3$ & $1$ \\
					$C_{r_1}$ & $-2$ & $1$ & $0$ & $0$ & $0$ & $0$ & $0$ & $0$ & $0$ & $0$ & $1$ & $1$ & $0$ \\
					$C_{r_2}$ & $1$ & $-2$ & $1$ & $0$ & $0$ & $0$ & $0$ & $0$ & $0$ & $0$ & $0$ & $1$ & $0$ \\
					$C_{r_3}$ & $0$ & $1$ & $-2$ & $1$ & $0$ & $1$ & $0$ & $0$ & $0$ & $0$ & $0$ & $0$ & $0$ \\
					$C_{r_4}$ & $0$ & $0$ & $1$ & $-2$ & $1$ & $0$ & $0$ & $0$ & $0$ & $0$ & $0$ & $0$ & $0$ \\
					$C_{r_5}$ & $0$ & $0$ & $0$ & $1$ & $-2$ & $0$ & $0$ & $0$ & $0$ & $0$ & $1$ & $0$ & $0$ \\
					$C_{r_6}$ & $0$ & $0$ & $1$ & $0$ & $0$ & $-2$ & $1$ & $0$ & $0$ & $0$ & $0$ & $0$ & $0$ \\
					$C_{r_0}$ & $0$ & $0$ & $0$ & $0$ & $0$ & $1$ & $-2$ & $0$ & $0$ & $0$ & $0$ & $0$ & $1$ \\
					$C_F$ & $0$ & $0$ & $0$ & $0$ & $0$ & $0$ & $1$ & $1$ & $0$ & $0$ & $0$ & $0$ & $-2$ \\
					$C_B$ & $0$ & $0$ & $0$ & $0$ & $0$ & $0$ & $-n$ & $0$ & $1$ & $1$ & $0$ & $0$ & $n-2$ \\
			\end{tabular}	
		\caption{Intersection numbers for an $(E_6)_n$ variety of type~I.} \label{table:mori_vectors_E6_n}
		\end{center}
	\end{table}
	We recognize the negative affine Cartan matrix of $E_6$ in the upper left block of the table.

	\paragraph{Example: $(F_4)_5$}
	The intersection numbers of the distinguished curves $C_\bullet$ with the divisors $\DX_{x_i}$ are given in table \ref{table:mori_vectors_F4_5}. We have used $n_3^{comp}=n_4^{comp}=2$ in the definition of $C_{r_3}$ and $C_{r_4}$.
	\begin{table}[!]
		\begin{center}
		\begin{tabular}{c|c|c|c|c|c|c|c|c|c|c|c|c|c|c} 
			& $D^X_{r_1}$ & $D^X_{r_2}$ & $D^X_{r_3}$ & $D^X_{r_4}$ & $D^X_{s}$ & $D^X_{t}$ & $D^X_{u}$ & $D^X_{v}$ & $D^X_{x}$ & $D^X_{y}$ & $D^X_{z}$ \\ \hline
			$C_{E}$ & $0$ & $0$ & $0$ & $0$ & $0$ & $0$ & $0$ & $0$ & $2$ & $3$ & $1$ \\
			$C_{r_1}$ & $-2$ & $1$ & $0$ & $0$ & $1$ & $0$ & $0$ & $0$ & $0$ & $0$ & $0$ \\
			$C_{r_2}$ & $1$ & $-2$ & $2$ & $0$ & $0$ & $0$ & $0$ & $0$ & $0$ & $1$ & $0$ \\
			$C_{r_3}$ & $0$ & $1$ & $-2$ & $1$ & $0$ & $0$ & $0$ & $0$ & $0$ & $0$ & $0$ \\
			$C_{r_4}$ & $0$ & $0$ & $1$ & $-2$ & $0$ & $0$ & $0$ & $0$ & $1$ & $0$ & $0$ \\
			$C_{r_0}$ & $1$ & $0$ & $0$ & $0$ & $-2$ & $0$ & $0$ & $0$ & $0$ & $0$ & $1$ \\
			$C_{F}$ & $0$ & $0$ & $0$ & $0$ & $1$ & $1$ & $0$ & $0$ & $0$ & $0$ & $-2$ \\
			$C_{B}$ & $0$ & $0$ & $0$ & $0$ & $-n$ & $0$ & $1$ & $1$ & $0$ & $0$ & $n-2$ \\
		\end{tabular}
			\caption{Intersection numbers for an $(F_4)_n$ variety of type~I.} \label{table:mori_vectors_F4_5}
		\end{center}
	\end{table}

	\subsubsection{Type II: $u*v \not \in I_{\mathrm Stanley-Reisner}$} \label{s:type_II}
	A number of the type I varieties $X$ we are considering contain contractible curves $\Ccontr$ whose class have a component $[C_B]$, $C_B$ being the base curve of the zero section $Z$ as introduced in \eqref{eq:CB}. Other than the case $n=1$, for which this base curve itself generates an extremal ray $\cR$ of the form \eqref{eq:extremal_ray}, the class of $\Ccontr$ takes the form
	\be
	[\Ccontr] = [C_B] - \sum_i k_i [C_{f_i}] \,, \quad C_{f_i} \in \{C_{r_0}, C_{r_1}, \ldots, C_{r_k}\} \,, \quad k_i \in \IN \,.
	\ee
	The wall relations of such curves are of the form
	\be
	u_1 + u_2 - u_{\rho_u} - u_{\rho_v} = 0 \,.
	\ee
	Performing a flipping transition along the ray $\cR$ generated by such a wall relation yields a birationally equivalent hypersurface $X'$ of type II with a contractible curve $\tCcontr = \DX_u \cdot \DX_v$.\footnote{We are again using the isomorphism \eqref{eq:iso_Pic} to speak of  the  same set of divisor classes on $X$ and $X'$.} As explained above, this intersection being non-trivial implies that in the flopped fan, $u_{\rho_u}$ and $u_{\rho_v}$ figure as 1-faces of common cones. Upon such flops, the fibration structure over the base $\IF_n$ we began with is therefore lost. In terms of homogeneous coordinates, the variables $u$ and $v$ are now permitted to simultaneously vanish, i.e. $u *v \not \in I_{\mathrm Stanley-Reisner}$, hence no longer parametrize the base $\IP^1$ of $\IF_n$.

	We can obtain distinguished curve classes, whose intersection matrix involves the negative affine Cartan matrix of the Lie algebra $\mg$, for a type II variety by following the torus invariant curves of type I geometries through the flop. Aside from the class of the contractible curve itself, the curves whose classes are modified by the flop are,  in the notation of section \ref{ss:flops_II}, $V(\sigma_{1,u}) \cdot (-K)$ and $V(\sigma_{2,u}) \cdot (-K)$. To reproduce \eqref{eq:affine_cartan}, the definition of the distinguished curves $C_{r_i}$ on type~II varieties must therefore be modified as follows
	\begin{equation}
		C_{r_0} = \DX_s \cdot \DX_u  + \delta_{u_s \in \{u_1,u_2\}} \DX_u \cdot \DX_v \,,
	\end{equation}
	\begin{equation}  
	C_{r_i} = \frac{1}{n_i^{comp}} (\DX_{r_i} \cdot \DX_u + \delta_{u_{r_i} \in \{u_1,u_2\}} \DX_u \cdot \DX_v)  \,.
	\end{equation}

	\paragraph{Example: $(E_7)_3$} 
	The intersection numbers of the distinguished curves $C_\bullet$ introduced in subsection \ref{sss:uv_in_ISR} with the divisors $\DX_{x_i}$ for an $(E_7)_3$ variety of type II are recorded in table \ref{table:mori_vectors_E7_3_not_uv}.
	\begin{table}[!]
		\begin{center}
			\begin{tabular}{c|c|c|c|c|c|c|c|c|c|c|c|c|c|c} 
				& $D^X_{r_1}$ & $D^X_{r_2}$ & $D^X_{r_3}$ & $D^X_{r_4}$ & $D^X_{r_5}$ & $D^X_{r_6}$ & $D^X_{r_7}$ & $D^X_{s}$ & $D^X_{t}$ & $D^X_{u}$ & $D^X_{v}$ & $D^X_{x}$ & $D^X_{y}$ & $D^X_{z}$ \\ \hline
				$C_{r_1}$ & $-3$ & $1$ & $0$ & $0$ & $0$ & $0$ & $0$ & $0$ & $0$ & $1$ & $1$ & $0$ & $0$ & $0$ \\
				$C_{r_2}$ & $1$ & $-2$ & $1$ & $0$ & $0$ & $0$ & $0$ & $0$ & $0$ & $0$ & $0$ & $0$ & $0$ & $0$ \\
				$C_{r_3}$ & $0$ & $1$ & $-2$ & $1$ & $0$ & $0$ & $1$ & $0$ & $0$ & $0$ & $0$ & $0$ & $0$ & $0$ \\
				$C_{r_4}$ & $0$ & $0$ & $1$ & $-2$ & $1$ & $0$ & $0$ & $0$ & $0$ & $0$ & $0$ & $0$ & $0$ & $0$ \\
				$C_{r_5}$ & $0$ & $0$ & $0$ & $1$ & $-2$ & $1$ & $0$ & $0$ & $0$ & $0$ & $0$ & $1$ & $0$ & $0$ \\
				$C_{r_6}$ & $0$ & $0$ & $0$ & $0$ & $1$ & $-2$ & $0$ & $0$ & $0$ & $0$ & $0$ & $0$ & $1$ & $0$ \\
				$C_{r_7}$ & $0$ & $0$ & $1$ & $0$ & $0$ & $0$ & $-2$ & $0$ & $0$ & $0$ & $0$ & $0$ & $1$ & $0$ \\
				$C_{r_0}$ & $0$ & $0$ & $0$ & $0$ & $0$ & $0$ & $0$ & $-3$ & $0$ & $1$ & $1$ & $0$ & $0$ & $1$ \\
				$C_{F}$ & $0$ & $0$ & $0$ & $0$ & $0$ & $0$ & $0$ & $1$ & $1$ & $0$ & $0$ & $0$ & $0$ & $-2$ \\
				$C_{B}$ & $0$ & $0$ & $0$ & $0$ & $0$ & $0$ & $0$ & $-3$ & $0$ & $1$ & $1$ & $0$ & $0$ & $1$ \\
			\end{tabular}
			\caption{Intersection numbers for $(E_7)_3$ varieties of type~II.} \label{table:mori_vectors_E7_3_not_uv}
		\end{center}
	\end{table}
	Note that the upper left hand corner of this table does not quite reproduce the negative Cartan matrix of $E_7$. The Mori vector of the curve $\tCcontr = \DX_u \cdot \DX_v$ is
	
	\begin{center}
		$\begin{tabu}{l|c|c|c|c|c|c|c|c|c|c|c|c|c|c} 
		& \DX_{r_1} & \DX_{r_2} & \DX_{r_3} & \DX_{r_4} & \DX_{r_5} & \DX_{r_6} & \DX_{r_7} & \DX_{s} & \DX_{t} & \DX_{u} & \DX_{v} & \DX_{x} & \DX_{y} & \DX_{z} \\
		\hline
		\tCcontr & 1 & 0 & 0 & 0 & 0 & 0 & 0 & 1 & 0 & -1 & -1 & 0 & 0 & 0
		\end{tabu}$
	\end{center}
	
	from which we read off that $\{u_1, u_2\} = \{u_{r_1},u_s\}$, i.e. this variety is related to a variety of type I by contracting the curve $\Ccontr = V(\sigma_{s,r_1})$ of the latter. From the Mori vectors of the distinguished curves of the type I variety, we can read off the decomposition
	\be
	[\Ccontr] = [C_B] - [C_{r_0}] \,.
	\ee
	Replacing $C_{r_1}$ and $C_{r_0}$ in table \ref{table:mori_vectors_E7_3_not_uv} by $C_{r_1} + \tCcontr$, $C_{r_0} + \tCcontr$ yields the negative Cartan matrix of $E_7$. 
		
	\paragraph{Example: $(A_2)_3$} Among all the geometries we consider, this is the only one for which all maximal projective subdivisions of the lattice polytope $\Psing$ have $u *v \not \in I_{\mathrm Stanley-Reisner}$, i.e. $(A_2)_3$ does not have an element of type~I.\footnote{Indeed, the existence of this geometry is what motivated us to include varieties of type~II in our discussion.} To understand why this is the case, we can consider all top-dimensional cones containing both 1-cones $\rho_u$ and $\rho_v$ as faces. Among the 12 maximal projective subdivisions of $\Psing$ (dropping 1-cones which lie on facets of $\Psing^\circ$, cf. the discussion in section \ref{ss:h11}), two cases occur.
	
	For the first, two pairs of such cones exist in the fan $\Sigma$, separated by the walls $\sigma_{u v r_2}$ and $\sigma_{u v y}$, with wall relations
	\be
	u_{\rho_u} + u_{\rho_v} - u_{\rho_{s}} - u_{\rho_{r_1}} - u_{\rho_{r_2}} = 0 
	\ee
	and
	\be
	u_{\rho_u} + u_{\rho_v} - u_{\rho_{s}} - 2u_{\rho_{r_1}} - u_{\rho_{y}} = 0 \,.
	\ee
	Via the $GL(4)$ map
	\begin{eqnarray} 
		u_{\rho_u} &\rightarrow& e_1 + e_2 + e_3 +e_4\,, \\
		u_{\rho_v} &\rightarrow& - e_1 \,, \nn\\
		u_{\rho_s} &\rightarrow& e_2 \,, \nn\\
		u_{\rho_{r_1}} &\rightarrow& e_3 \,, \nn\\ 
		u_{\rho_{r_2}} &\rightarrow& e_4 \,.\nn 
	\end{eqnarray}
	we can e.g. identify the local geometry of the curve $V(\sigma_{uvr_2})$  as
	\be
	\cO(-1) \oplus \cO(-1) \oplus \cO (-1) \rightarrow \IP^1\,.
	\ee
	To eliminate all cones which contain both the 1-cone $\rho_u$ and the 1-cone $\rho_v$ as faces, we can simultaneously flip both curves $\sigma_{u v r_2}$ and $\sigma_{u v y}$ by making the replacement
	\be
	\sigma_{u v r_2 r_1}, \sigma_{u v r_2 s}, \sigma_{u v y r_1}, \sigma_{u v y s} \rightarrow \sigma_{r_1 s r_2 u},\sigma_{r_1 s r_2 v},\sigma_{r_1 s y u},\sigma_{r_1 s y v}
	\ee
	in $\Sigma$. However, the geometry after this transition no longer has nef anti-canonical bundle. These flips are hence not $(-K)$-flops, which is why the resulting fan does not arise as a maximal projective subdivision of the polytope $\Psing$. 
	
	The second case is similar. Here, the 2-cone $\sigma_{u v}$ is the face of three top-dimensional cones obtained by combining the rays $\rho_u$ and $\rho_v$ with all pairs among $\{\rho_s, \rho_{r_1}, \rho_{r_2} \}$. A flipping transition maps this geometry to one where the wall $\sigma_{s r_1 s_2}$ separates the two cones $\sigma_{s r_1 r_2 u}$ and $\sigma_{s r_1 r_2 v}$. Again, this is not a flop transition as the anti-canonical line bundle of the resulting geometry is no longer nef. 
	
	We thus have to make do with only having type II geometries in $(A_2)_3$.
	
	For all such varieties, the intersection numbers of the distinguished curves $C_\bullet$ introduced in subsection \ref{sss:uv_in_ISR} with the divisors $\DX_{x_i}$ are recorded in table \ref{table:mori_vectors_A2_3}. As for all type~II geometries, we observe that the upper left corner of this table does not quite reproduce the Cartan matrix associated to the singularity.
	\begin{table}[h!]
		\begin{center}
			$\begin{tabu}{l|c|c|c|c|c|c|c|c|c} 
			& \DX_{r_1} & \DX_{r_2} & \DX_{s} & \DX_{t} & \DX_{u} & \DX_{v} & \DX_{x} & \DX_{y} & \DX_{z} \\
			\hline
			C_{r_1} & -3 & 0 & 0 & 0 & 1 & 1 & 1 & 3 & 0 \\
			C_{r_2} & 0 & -3 & 0 & 0 & 1 & 1 & 1 & 0 & 0 \\
			C_{r_0} & 0 & 0 & -3 & 0 & 1 & 1 & 0 & 0 & 1 \\
			C_{F} & 0 & 0 & 1 & 1 & 0 & 0 & 0 & 0 & -2 \\
			C_{B} & 0 & 0 & -3 & 0 & 1 & 1 & 0 & 0 & 1	
			\end{tabu}$
			\caption{Intersection numbers for the $(A_2)_3$ variety. \label{table:mori_vectors_A2_3}.} 
		\end{center}
	\end{table}
	Motivated by geometries for which a flop transition exists connecting type~I and type~II geometries, we compute the Mori vector of the curve $D_u^X \cdot D_v^X$,
	\begin{center}
	$\begin{tabu}{l|c|c|c|c|c|c|c|c|c} 
	& \DX_{r_1} & \DX_{r_2} & \DX_{s} & \DX_{t} & \DX_{u} & \DX_{v} & \DX_{x} & \DX_{y} & \DX_{z} \\
	\hline
	D_u^X \cdot D_v^X & 1 & 1 & 1 & 0 & -1 & -1 & 0 & 0 & 0 \\
	\end{tabu}$
	\end{center}	
	Replacing $C_\bullet$ for $\bullet \in \{r_1, r_2, r_0\}$ by $C_\bullet + D_u^X \cdot D_v^X$, we obtain an intersection matrix with the negative Cartan matrix of affine $A_2$ appearing in the upper left corner.

	\subsection{Constructing Higgsing trees torically}  \label{ss:constructing_Higgsing_trees}
	Using the formalism developed in this section, we can set up a simple algorithm to determine which singularities can occur torically over a given Hirzebruch base $\IF_n$. Special attention must be paid to cases in which the occurring singularity depends not only on the power of $s$ in the individual coefficients in the Tate form of the elliptic fibration, but also on factorization conditions involving relations between multiple coefficients.\footnote{This is intimately related to the occurrence of resolution divisors which are not rationally fibered, discussed in the opening paragraphs of section \ref{ss:distinguished_curves}.} The authors of \cite{Bershadsky:1996nh} retain the Kodaira name of the (surface) singularity in these cases, but add a superscript $ns, s$ for when factoring does not occur (the `{\bf n}on-{\bf s}plit' case, with lower rank singularity), and when it does (the '{\bf s}plit' case, with higher rank singularity); in the sole case of the $I^*_0$ singularity, the polynomial in question is of order 3, and also an intermediate factoring condition, indexed by $ss$ (for {\bf s}emi-{\bf s}plit), is necessary. When factorization is possible, a variable redefinition can lead to higher order vanishing of appropriate coefficients of the generic section, and thus to a higher rank singularity.
	
	The first step in the algorithm is to construct the polytope $\Psing$ as introduced in subsection \ref{ss:enhancing}. If $\Psing \cap M \neq \Msing$, the singularity cannot be constructed torically over the given base. If $\Psing \cap M = \Msing$ but $h^{1,1}(X) - 3$ does not equal the rank of any of the Lie algebras associated to the singularity, the geometry does not match the pattern discussed in this paper and merits further study, see \cite{Morrison:2012np,Haghighat:2014vxa}. This criterion excludes in particular $n=9,10,11$ from our study, as well as $B_k$ and $D_k$ singularities, $k>6$, for $n=0,1,2,3$. Otherwise, the question of factorization can be settled by computing $h^{1,1}(X)$, which reveals the rank of the Lie algebra associated to the singularity. The conclusion can then be checked by computing the intersection matrix between distinguished divisors and curves, as outlined in subsection \ref{ss:distinguished_curves}. If the correction term
	\be \label{eq:reducible_intersection}
	\sum_{\Theta^\circ \in F_2(\Pp)} l^*(\Theta^\circ) l^*(\Theta) 
	\ee
	to $h^{1,1}(X)$ vanishes, the intersections of the toric divisors of $Y$ with $X$ are irreducible, and the generators of the Picard group of $X$ are in 1-1 correspondence with those of $Y$. In this case, the upper left corner of the intersection matrix as presented e.g. in table \ref{table:mori_vectors_E6_n} directly yields the negative Cartan matrix of the associated Lie algebra. When the correction term is non-vanishing, the matrix computed following the steps outlined in section \ref{ss:distinguished_curves} can be obtained from the negative Cartan matrix of the Lie algebra associated to the singularity by summing rows and columns corresponding to the irreducible components of the reducible intersections. 
	
	In cases when the correction term is non-vanishing, one can be tempted to modify the geometry to obtain a better toric embedding of $X$. As we see from \eqref{eq:reducible_intersection}, the only toric divisors of $Y$ which may have reducible intersections with $X$ are those that correspond to cone generators which lie in codimension 2 faces $\Theta^\circ$ of $\Pp$. The intersection of such divisors with $X$ are reducible if the corresponding dual face $\Theta$, a 1-face of $P$, contains interior points. Chipping away at these 1-faces by removing an endpoint of the 1-face from $\Msing$\footnote{Generically, only one choice leads to a compact polytope.} often leads to an admissible variety with reduced correction term \eqref{eq:reducible_intersection}. However, the variety  thus obtained may no longer coincide with $X$. 
	
	\paragraph{Example: $(G_2)_4$} 
	
	Following the above procedure naively suggests that a $G_2$ singularity can be imposed over $\IF_4$, but the resulting hypersurface $X$ with singularities resolved has $h^{1,1}(X) = 7$. This indicates that in fact, the imposed singularity must have rank 4, suggesting its identification as $D_4$ (i.e. $I^{* ns}_0 \rightarrow I^{* s}_0$). We can check this conclusion by computing the intersection matrix of the distinguished curves $C_{\bullet}$ of the geometry with the toric invariant divisors, which we give in table \ref{table:mori_vectors_G2_4}.
	\begin{table}[h!]
		\begin{center}
			$\begin{tabu}{l|c|c|c|c|c|c|c|c|c} 
			& \DX_{r_1} & \DX_{r_2} & \DX_{s} & \DX_{t} & \DX_{u} & \DX_{v} & \DX_{x} & \DX_{y} & \DX_{z} \\
			\hline
			C_{r_1} &-2& 1 & 0 & 0 & 0 & 0 & 0 & 1 & 0 \\
			C_{r_2} &3 & -2 & 1 & 0 & 0 & 0 & 1 & 0 & 0 \\
			C_{r_0} & 0 & 1 & -2 & 0 & 0 & 0 & 0 & 0 & 1 \\
			C_{F} & 0 & 0 & 1 & 1 & 0 & 0 & 0 & 0 & -2 \\
			C_{B} & 0 & 0 & -4 & 0 & 1 & 1 & 0 & 0 & 2
			\end{tabu}$
			\caption{Intersection numbers for a $(G_2)_4$ variety. \label{table:mori_vectors_G2_4}.} 
		\end{center}
	\end{table}
	The analysis is complicated by the fact that for this example, some divisors of the ambient space $Y$ intersect the hypersurface $X$ reducibly; the value of the correction term \eqref{eq:reducible_intersection} is 2. In particular, the negative Cartan matrix appearing in the upper left corner of the table is that of $G_2$. By decomposing $\DX_{r_1}$ in terms of irreducible components 
	\be
	\DX_{r_1} =\DX_{r_1,1} + \DX_{r_1,2} + \DX_{r_1,3} \,,
	\ee
	we can see that the intersection numbers given in table \ref{table:mori_vectors_G2_4} are indeed consistent with the hypersurface $X$ resulting from the resolution of a $D_4$ singularity:
	\be
	C_{D_4} = 
	\begin{pmatrix}
		2 & -1 & 0 & 0 \\
		-1 & 2 & -1 & -1 \\
		0 & -1 & 2 & 0 \\
		0 & -1 & 0 & 2 \\
	\end{pmatrix}
	\rightarrow 
	\begin{pmatrix}
		2 & -3 & 2 & 2 \\
		-1 & 2 & -1 & -1 \\
	\end{pmatrix}
	\rightarrow 
	\begin{pmatrix}
		6 & -3   \\
		-3 & 2  \\
	\end{pmatrix}
	\rightarrow 
	\begin{pmatrix}
		2 & -1   \\
		-3 & 2  \\
	\end{pmatrix}
	= C_{G_2}
	\ee
	Here, the first, third and forth row and column are associated to the three reducible components of $\DX_{r_1}$.
	
	We can try to reduce the value 2 of the correction term \eqref{eq:reducible_intersection} by applying the excision procedure outlined above twice. The first application does not modify $X$ while reducing the correction term by 1. One can obtain the same toric data by directly imposing a $D_4$ singularity over $\IF_4$ to obtain $\Psing$. A second application however changes the geometry to a $B_4$ singularity over $\IF_4$.\footnote{This contradicts a statement in \cite{Haghighat:2014vxa, DelZotto:2017mee}. It appears however that the genus 0 Gromov-Witten invariants of $(D_4)_4$ and $(B_4)_4$, in as far as these enter into the considerations of these two references, coincide.  We are currently investigating this phenomenon.}
	
	The procedure outlined in this subsection allows us to construct the varieties underlying all gauge groups and matter contents occurring in the Higgsing trees over $\IF_n$ \cite{Bershadsky:1996nh, DelZotto:2018tcj} with vanishing correction term \eqref{eq:reducible_intersection}, with the following exceptions:
	\begin{itemize}
		\item Attempting to impose the $I_{2k}^{ns}$ singularities to obtain the symplectic groups over $\IF_2$ yields $I_{2k}^{s}$ singularities instead, leading to the gauge group $A_{2k-1}$.
		\item $I_{2k-2}^s$ singularities leading to gauge group $D_{2k+2}$ require imposing the same degree of vanishing on the coefficients $a_i$ introduced in section \ref{ss:anti_canonical_hypersurface} as for the singularities $I_{2k-2}^{ns}$ leading to gauge group $B_{2k+1}$, complemented with an additional factorization constraint on a polynomial built from these coefficients \cite{Bershadsky:1996nh}. The only $D_{2k+2}$ singularity we can impose following the above algorithm is $D_4$ over $\IF_4$, but one divisor of the ambient space intersects the hypersurface reducibly, see the discussion above.
	\end{itemize}

	\section{Computing genus 0 Gromov-Witten invariants via mirror symmetry} \label{s:mirror_sym}
	Mirror symmetry maps the problem of computing the genus 0 Gromov-Witten invariants $n_{\bk}$ 
	of a Calabi-Yau manifold $X$ to that of computing the appropriate periods of the top form $[\Omega] \in H^{(3,0)}(X')$ of the holomorphic Dolbeault cohomology ring on the mirror Calabi-Yau manifold $X'$.
	
	The generating function for the genus 0 Gromov-Witten invariants of $X$ is called the prepotential $F(\bt)$. It is computed \cite{Witten:1991zz,Bershadsky:1993cx} by the topological string A-model on $X$; the variable dependence $(\bt)$ is on flat coordinates $\bt$ on the complexified K\"ahler structure moduli space $\Mkaehler(X)$ of $X$. It takes the universal form
	\be \label{eq:A_prepotential}
	F(\bt) = \frac{c_3(X)}{(2\pi i)^3} \zeta(3) + \sum_i \frac{c_2(X) \cdot D_i}{24}t_i + \frac{1}{3!} \sum_{i,j,k} d_{ijk} t_i t_j t_k + \sum_{\bk, n} \frac{n_{\bk}}{n^3} e^{2 \pi i n \bk \cdot \bt}  \,,
	\ee
	where $D_i$ are divisor classes associated to the coordinate $t_i$, and $d_{ijk} = D_i \cdot D_j \cdot D_k$. The coefficients of the polynomial (perturbative) terms in $\bt$ depend on topological invariants of $X$, whereas the coefficients of the exponential (non-perturbative) terms are the enumerative invariants that we are after.
	
	To express $F(\bt)$ in terms of  the periods of an appropriately normalized representative $\Omega$ of  $H^{(3,0)}(X')$ of the mirror manifold $X'$, one must choose a symplectic basis $\{\alpha_I, \beta_I\}$ of $H_3(X',\IZ)$ (discussed below in subsection \ref{ss:appropriate_basis_of_periods}), and write
	\be
	X_I = \int_{\alpha_I} \Omega \,, \quad  	F_I = \int_{\beta_I} \Omega \,.  
	\ee
	We will refer to the $X_I$ as the $A$-periods of $\Omega$, and the $F_I$ as the $B$-periods. A theorem of Bryant and Griffiths \cite{Bryant:1983} implies that when the basis of cycle classes is chosen appropriately, the $B$-periods $F_I$  are fully determined in terms of the $A$-periods $X_I$.  As $\Omega$ is determined by its periods, we can write $\Omega(\bX)$. With regard to the complex structure determined by the periods $\bX^0$ (i.e. such that $[\Omega(\bX^0) ] \in H^{(3,0)}(X')$), a local argument implies
	\be
	[ \partial_{X_I} \Bigr|_{\bX = \bX^0}\Omega(\bX) ] \in H^{(3,0)}(X') \oplus H^{(2,1)}(X') \,.
	\ee
	Hence,
	\be \label{eq:Griffiths_transversality}
	0 = \int \Omega(\bX^0) \wedge \partial_{X_I} \Bigr|_{\bX = \bX^0}\Omega(\bX) \,.
	\ee
	By the Riemann bilinear identities, the RHS of  \eqref{eq:Griffiths_transversality} can be expressed as
	\be 
	0 = \sum_J X_J^0 \partial_{X_I}  \Bigr|_{\bX=  \bX^0} F_J(\bX) - F_I(\bX^0) \,,
	\ee
	whence
	\be
	2 F_I(\bX^0) =	\partial_{X_I}\Bigr|_{\bX=  \bX^0}  \sum_J X_J   F_J(\bX) \,.
	\ee
	It follows that the period $F_I(\bX)$ can be obtained as the $X_I$ derivative
	\be \label{eq:B_periods_as_derivatives}
	F_I(\bX) = \partial_{X_I} F(\bX)
	\ee
	of one quantity $F(\bX)$, defined as
	\be \label{eq:B_prepotential}
	F(\bX) = \frac{1}{2}  \sum_J X_J   F_J(\bX) \,.
	\ee
	This is the definition of the prepotential based on complex structure data of $X'$. The statement of mirror symmetry is that upon an appropriate choice of basis of $H_3(X',\IZ)$ discussed below, the variable identification
	\be \label{eq:mirror_map}
	t_i = \frac{X_i}{X_0} 
	\ee
	 allows us to equate
	\be
	F(\bX) = X_0^2 \, F(1, \frac{X_i}{X_0}) = X_0^2 F(\bt) \,.
	\ee 
	$F(\bt)$ here denotes the $A$-model prepotential \eqref{eq:A_prepotential}, $F(\bX)$ the period expression \eqref{eq:B_prepotential}, whose homogeneity of degree 2 property (which follows from its definition) we have used.

	In the following subsection, we will fill in the details required to explicitly perform the computation of $F(\bX)$ in the class of anti-canonical hypersurfaces of toric varieties. In a nutshell, the steps required are the following:
	\begin{enumerate}
		\item Write down a representative for the class of $\Omega$ as a function of appropriate coordinates on $\Mcplx(X')$.
		\item Find a complete system of differential equations (the Picard-Fuchs system) satisfied by these periods. While it is possible to compute periods by identifying cycles in $H_3(X', \IZ)$ and performing integrals, it is much more convenient to compute them as solutions of the Picard-Fuchs system.
		\item Identify the linear combination of solutions corresponding to the appropriate symplectic basis of $H_3(X')$. The fastest route to this identification is by imposing the perturbative part of $F(\bt)$ given in terms of the topological invariants of $X$ as written in \eqref{eq:A_prepotential}. Note that having recourse to (easily computable) data of $X$ is for computational convenience only.
	\end{enumerate}

	We then address how to apply the formalism to the task at hand, identifying matter in $F$-theory compactifications on elliptic fibrations, in the ensuing sections \ref{ss:complex_structure_coordinates}, \ref{ss:curve_invariants} and \ref{ss:GW_and_flops}.
	
	\subsection{Review: an algorithm for computing the Gromov-Witten invariants of anti-canonical hypersurfaces of toric varieties} \label{ss:mirror_symmetry_algorithm}

	The theory of mirror symmetry on Calabi-Yau hypersurfaces of toric varieties is very well developed \cite{Batyrev:1994hm, Hosono:1993qy}.\footnote{The theory of complete intersections is similar in many aspects  \cite{Batyrev:1994pg, Hosono:1994ax}, but we will not be discussing it here.} In this subsection, we will review all aspects required to understand the algorithm for the computation of the genus $0$ Gromov-Witten invariants for this class of Calabi-Yau threefolds. Aside from the original references, most of the material reviewed here can be found in the book \cite{CoxKatz}.
	
	{\it A point on notation:}  The mirror family to the anti-canonical hypersurface $X$ sitting inside the toric variety $Y_\Sigma$, where $\Sigma$ is a maximal projective subdivision of a lattice polytope $\LP$, is given by anti-canonical hypersurfaces $X'$ sitting inside the toric variety $Y'_{\Sigma'}$, where $\Sigma'$ is a maximal projective subdivision  of a lattice polytope $\LP^\circ$, the polar polytope to $\LP$. It is standard to designate the lattice of one-parameter subgroups of the torus $T_N \subset Y$ by the letter $N$, and its character lattice by the letter $M$. For the mirror hypersurface, we have $T_M \subset Y'$, and $M$ and $N$ are exchanged. We feel that this exchange on casual reading can be a source of confusion. We hope to alleviate this by introducing the notation $M^\circ = M^\vee = N$, $N^\circ = N^\vee = M$.
	
	\subsubsection{A representative of $\Omega$ in appropriate coordinates on the complex structure moduli space $\Mcplx(X')$}
	When $X'$ is an anti-canonical hypersurface of a toric variety $Y'$, we can obtain a representative $\Omega$ of  a generator of $H^{3,0}(X')$ as the residue of the extension of the form 
	\be  \label{eq:general_form_omega}
	\omega = \frac{1}{f} \, \frac{dt_1}{t_1} \wedge \ldots \wedge \frac{dt_4}{t_4} \,,
	\ee
	defined on the torus $T \subset Y'$ \cite{Batyrev:1993}. Here, $t_i$ are natural coordinates on $T \cong (\IC^*)^4$ (not to be confused with coordinates on the moduli space $\Mkaehler(X)$), and in the notation introduced in \eqref{eq:Laurent_polynomials_to_polytope}, $f \in L(\LP^\circ)$. $f$ yields the hypersurface $X'$ as its zero locus, $f=0$. It has the form
	\be \label{eq:expansion_of_f}
	f = \sum_i \lambda_i \bt^{m_i} \,, \quad \bt^{m_i} = \prod t_j^{(m_i)_j} \,,
	\ee
	with the sum over the index set parametrizing the $r+1$ points $m_i \in \LP^\circ \cap M^\circ$. For later convenience, we will assign the origin the index 0, $m_0 = \mathbf{0}$. 
	
	Before computing the periods of $\Omega$, we need to express $\omega$ in terms of good coordinates on the complex structure moduli space $\Mcplx(X')$. The coefficients $\{\lambda_i\}$ occurring in \eqref{eq:expansion_of_f} parametrize $\Mcplx(X')$ redundantly:	
	\begin{itemize}
		\item $T$ acts on itself and thus on $f$, leading to isomorphic hypersurfaces, via
				\be
					\bnu \cdot f = \sum_i \lambda_i (\bnu \cdot \bt)^{m_i}  \,, \quad \bnu \in T \,.
				\ee
		\item Rescaling of $f$ by a non-zero constant $c\in \IC^*$ does not change its zero locus.
	\end{itemize}
	To eliminate this $T \times \IC^*$ redundancy, we can consider a basis of the lattice $\Lambda$ of relations amongst the $r$ vectors $m_i$, $i \neq 0$.  As we are considering four dimensional ambient spaces, upon a choice of generating set for $M^\circ$, we can identify the $m_i$ with elements of $\IZ^4$; there will hence be $r-4$ such relations. We will label these as $l^\alpha$, $\alpha = 1, \ldots, r-4$, such that 
	\be \label{eq:lattice_of_relations}
	\sum_{i=1}^r (l^\alpha)_i m_i = 0 \,, \quad \alpha = 1, \ldots r-4 \,.
	\ee
	The coordinates
	\be
	z_\alpha = \lambda_0^{- \sum_{i=1}^r (l^\alpha)_i} \prod_{i=1}^{r} \lambda_i^{(l^\alpha)_i} 
	\ee
	are then invariant under the $T \times \IC^*$ action. Note that upon introducing the set of points
	\be
	\Xi = 1 \times ( \LP^\circ \cap M^\circ ) 
	\ee
	in $\IZ^5$, the basis of relations $l^\alpha$ of points $\LP^\circ \cap M^\circ$ is naturally extended to a basis of relations $L^\alpha$ of the points $\Xi$, in terms of which the coordinates $z_\alpha$ are expressed as
	\be \label{eq:cplx_structure_coord_via_relations}
	z_\alpha = \prod_{i=0}^r \lambda_i^{(L^\alpha)_i}. 
	\ee
	Either way, the $z_\alpha$ coordinatize the quotient
	\be \label{eq:M_cplx_toric}
	L(\LP^\circ) / T \times \IC^*  = \IP(L(\LP^\circ)) / T \,,
	\ee
	which is a first approximation to the complex structure moduli space $\Mcplx(X')$.
	
	The choice of coordinates on the space \eqref{eq:M_cplx_toric} thus maps to a choice of basis $\{ L^\alpha \}$ on the space of relations among the points of $\Xi$. We will discuss this choice further in sections \ref{sss:complex_str_coordinates_prequel} and \ref{ss:complex_structure_coordinates}.
	
	\subsubsection{The Picard-Fuchs system}
	Rather than calculating the periods of $\Omega$ directly by identifying appropriate cycles in $X'$ and integrating, it is computationally more convenient to derive a set of differential operators which annihilate these periods. The complete set of such operators (i.e. such that each element of their common kernel is a linear combination of periods) spans the so-called Picard-Fuchs ideal. The problem of computing periods is thus mapped to finding the corresponding set of differential equations, determining  their general family of solutions,  and identifying the linear combinations of solutions corresponding to periods with regard to an appropriate basis of $H_3(X', \IZ)$. 
	
	Given the explicit expression \eqref{eq:general_form_omega} for $\omega$, it is not difficult to derive elements of the Picard-Fuchs ideal. To this end, for any relation $L$ among the points $\Xi$,  define the differential operator $\Box_{L}$ via
	\be
	\Box_{L} = \prod_{L_i > 0} \theta_i^{L_i} - \prod_{L_i < 0} \theta_i^{-L_i} \,,
	\ee
	with $\theta_i$ the logarithmic derivative
	\be
	\theta_i = \frac{1}{\lambda_i} \frac{\partial}{\partial \lambda_i} \,.
	\ee	
	It is then a simple calculation to check that 
	\be
	\Box_L \omega = 0 \,.
	\ee
	Multiplying $\omega$ by $\lambda_0$ to render it invariant under the $\IC^*$ action, we obtain a set of Picard-Fuchs operators
	\be \label{eq:PF_system}
	\Box_{L^\alpha} \frac{1}{\lambda_0}  \,, \quad \alpha = 1, \ldots r-4.
	\ee
	This set however generically does not generate the complete Picard-Fuchs ideal. Methods for obtaining the missing differential operators in the case of toric hypersurfaces are discussed in \cite{Hosono:1993qy}.
	
	\subsubsection{A distinguished basis of periods at a MUM point}
	Determining the Gromov-Witten invariants of $X$ requires computing the periods of $\Omega$ at a point in $\Mcplx(X')$ that is mirror to the large radius point of $X$. Matching the expected structure of the periods suggests that this should be a point of maximally unipotent monodromy (a MUM point) \cite{Morrison:1993b}: at such a point, all indices of the Picard-Fuchs system are equal (and in fact vanish).  Introducing local coordinates $z_i$ on $\Mcplx(X')$ such that a boundary point $p$ of $\Mcplx(X')$, $p \in \overline{\cM}_{cplx}(X') - \Mcplx(X')$, is given by the vanishing of these coordinates, $p$ is a MUM point if exactly one period is analytic here, and $h^{1,1}(X)$ periods have logarithmic growth in $z_i$. One can include as part of the definition of the MUM point that integer linear combinations of the logarithmic periods exist which each have logarithmic growth with regard to exactly one coordinate $z_i$. 
	
	In the toric case, a basis of periods at a MUM point can be computed as follows. Power series solutions to the Picard-Fuchs system around the point $\bz = \mathbf{0}$ are given by\footnote{To be precise, the coordinates $\bz$ used here differ from those introduced in \eqref{eq:cplx_structure_coord_via_relations} by signs $(-1)^{(L^\alpha)_0}$.}
	\be \label{eq:power_series_solution}
	\Pi_{power} (\bz; \brho) = \sum_{\bn \in \IN_0^{h^{1,1}(X)}} \frac{\Gamma(1-\sum_\alpha (n_\alpha + \rho_\alpha) (L^\alpha)_0)}{\prod_{i>0}\Gamma(1+\sum_\alpha (n_\alpha + \rho_\alpha) (L^\alpha)_i)} \frac{\prod_{i>0}\Gamma(1+\sum_\alpha  \rho_\alpha (L^\alpha)_i)}{\Gamma(1-\sum_\alpha  \rho_\alpha (L^\alpha)_0)} \bz^{\bn + \brho} \,,
	\ee
	where $\brho$ is any of the $2 + 2h^{1,1}(X)$ (not necessarily distinct) indices of the system. At a point at which all indices vanish, there exists a unique holomorphic solution $\Pi_0$, obtained by setting $\brho = \boldsymbol{0}$ in \eqref{eq:power_series_solution}. $\Pi_0$ can then be completed to a basis of solutions via the Frobenius method:
	\ba \label{eq:periods_frobenius}
	\Pi_0 &=& \Pi_{power} (\bz; 0) \,, \\
	\Pi_\alpha &=& \partial_{\rho_{\alpha}} \Bigr|_{\brho = \bzero} \Pi_{power} (\bz; \brho) \,, \\
	 \Pi_{\alpha,\beta}  &=&  \partial_{\rho_{\alpha}} \partial_{\rho_{\beta}} \Bigr|_{\brho = \bzero} \Pi_{power} (\bz; \brho) \,, \\
	 \Pi_{\alpha,\beta,\gamma} &=& \partial_{\rho_{\alpha}} \partial_{\rho_{\beta}}  \partial_{\rho_{\gamma}} \Bigr|_{\brho = \bzero} \Pi_{power} (\bz; \brho) \,.
	\ea
	A complete set of solutions of the Picard-Fuchs system at such a point therefore consists of one holomorphic, $h^{1,1}(X)$ logarithmic, $h^{1,1}(X)$ doubly logarithmic, and one triply logarithmic solution.  
	
	Extracting Gromov-Witten invariants from the Picard-Fuch system requires identifying appropriate linear combinations of these solutions, corresponding to the choice of an adapted symplectic basis of $H_3(X',\IZ)$. To address this task, we need to take a closer look at the choice of variables $\bz$ on $\Mcplx(X')$.
	
	\subsubsection{How to choose coordinates on complex structure moduli space -- the prequel} \label{sss:complex_str_coordinates_prequel}
	
	The choice of variables $z_\alpha$, and thus the point in complex structure moduli space that $\bz = \bzero$ designates, clearly depends on the choice of basis $\{l_\alpha\}$ for the lattice of relations $\Lambda$ of the points $\LP^\circ \cap M^\circ$ introduced in \eqref{eq:lattice_of_relations}. Note that as $\LP^\circ$ is reflexive, its only interior point is the origin. All other elements of the intersection are thus vertices of $\LP^\circ$, hence elements of $\Sigma(1)$. Now recall from subsection \ref{ss:mori_cone} that on $X$, every curve $C$ gives rise to a relation amongst the elements of $\Sigma(1)$, as encoded in the Mori vector of the curve. The generators of the Mori cone yield a basis of all such relations. We have thus identified generators of $\Lambda$ whose duals generate the K\"ahler cone of $X$. As mirror symmetry between $X$ and $X'$ requires relating $\Mkaehler(X)$ to $\Mcplx(X')$, it is a natural conjecture \cite{Aspinwall:1993rj} that the sought after basis of $\Lambda$ determining distinguished coordinates $\bz$ on $\Mcplx(X')$ should be given by these generators.
	
	In practice, as discussed in subsection \ref{ss:mori_cone}, we generally do not have a basis of the Mori cone of $X$ at our disposal. We will discuss the repercussions of this fact in the context of the class of examples we are considering in section \ref{ss:complex_structure_coordinates}, after we have completed our review of how to compute the prepotential on anti-canonical hypersurfaces of toric varieties, assuming an appropriate choice of basis of $\Lambda$ has been found.
	
	\subsubsection{Identifying the linear combinations of solutions which coincide with periods of $\Omega$ in a symplectic basis of $H_3(X',\IZ)$} \label{ss:appropriate_basis_of_periods}
	
	We can identify the periods $\Pi_0$ and $\Pi_\alpha$ introduced in \eqref{eq:power_series_solution} and \eqref{eq:periods_frobenius} with the $A$-periods $X_0$ and $X_i$ of $\Omega$. By invoking mirror symmetry, this can be argued for by studying the leading behavior of contributions to the mirror volume form $\int \Omega \wedge \overline{\Omega}$ in the variables $t_i$. Having identified $X_0$ and $X_i$ allows us to compute the coordinates $t_i$ via \eqref{eq:mirror_map}. Linear combinations of the periods $\Pi_{\alpha, \beta}$ must then describe the $B$-periods $F_i$ dual to the logarithmic periods $X_i$, $i=1, \ldots, h^{1,1}(X)$. To find the appropriate linear combinations, we invoke mirror symmetry, which identifies the symplectic product of periods \eqref{eq:B_prepotential} with the prepotential \eqref{eq:A_prepotential}, upon imposition of the mirror map \eqref{eq:mirror_map}. Given \eqref{eq:B_periods_as_derivatives}, the perturbative contribution to $F(\bt)$ which is cubic in $t_i$ supplies sufficient information to fix the periods $F_i$ in terms of $\Pi_{\alpha, \beta}$. These in turn can be integrated to obtain the non-perturbative piece of $F(\bt)$, the generating function for Gromov-Witten invariants.

	\subsection{How to choose coordinates on complex structure moduli space} \label{ss:complex_structure_coordinates}
	
	As discussed in section \ref{ss:mori_cone}, any projective subdivision of the lattice polytope $\LP$ gives rise to a fan $\Sigma$ and a projective variety $Y_\Sigma$ with associated  K\"ahler cone $\Nef(Y_\Sigma)$ and the associated dual cone, the Mori cone $\MC(Y_\Sigma)$ of $Y_\Sigma$. The corresponding cones of the hypersurface $X$ sitting inside $Y_\Sigma$ are generically larger, $\Nef(Y_\Sigma) \subset \Nef(X)$ and smaller $\MC(X) \subset \Nef(Y_\Sigma)$, respectively. We have found experimentally that the Gromov-Witten invariants of $X$ can be calculated using any smooth cone $\cC(X) \subset \Nef(X)$, as long as for each generator of $\cC(X)$, the sum of its entries is non-negative, i.e. a putative associated curve does not violate the nef condition on the anti-canonical class of $X$.
	
	The most naive choice for $\cC(X)$ is $\Nef(Y_\Sigma)$ for any projective subdivision $\Sigma$ of $\LP$ such that $X \subset Y_\Sigma$. The following situations may occur:
	\begin{enumerate}
		\item If $\NefY$ is smooth, the algorithm reviewed in section \ref{ss:mirror_symmetry_algorithm} based on $\NefY$ will yield the Gromov-Witten invariants of $X$. Depending on how well $\NefY$ approximates $\Nef(X)$, many of the invariants computed in the basis provided by the generators of $\NefY$ will vanish.
		\item \label{simplicial}$\Nef(Y_\Sigma)$ is not smooth, but it is top dimensional (i.e. has the dimension of $N^1(X)$) and simplicial. Appropriately refining $\Nef(Y_\Sigma)$ leads to smooth subcones contained in $\Nef(X)$. Any such subcone leads to the correct Gromov-Witten invariants of $X$, provided that for each generator of the subcone, the sum of its entries is non-negative. 
		\item $\Nef(Y_\Sigma)$ is top dimensional but not simplicial. In this case, we can consider simplicial subdivisions of $\Nef(Y_\Sigma)$ as a starting point for \ref{simplicial}.
		\item $\Nef(Y_\Sigma)$ is not top dimensional. This occurs when $\MC(X)$ is not strictly convex, i.e. when pairs of curves exist whose Mori vectors map to each other upon multiplication by $-1$. Such subdivisions $\Sigma$ do not provide a convenient approximation to $\Nef(X)$ for the computation of the Gromov-Witten invariants of $X$. 
	\end{enumerate}
	
	A better approximation to $\Nef(X)$ than $\NefY$ for $X \subset Y_\Sigma$ is the dual of the toric Mori cone \eqref{eq:toric_Mori_cone} which we introduced in section \ref{ss:mori_cone}. In most geometries that we have studied, this cone is smooth. When it is not, the steps outlined above must be pursued to obtain from it a cone on which the mirror symmetry algorithm can be based.

	\subsection{The Gromov-Witten invariants of interest} \label{ss:curve_invariants}
	The interpretation of Gromov-Witten invariants is simplest when all of the curves in a given class are isolated. In this case, the invariant yields the number of such curves. Another simple case \cite{Candelas:1994hw} is when the curves in a given class are parametrized by a smooth variety $B$ of dimension $b$. Then the invariant is given by $c_b(\Omega_B^1)$, the appropriate Chern number of the holomorphic cotangent bundle of $B$. These are the two cases that occur in our considerations together with a third hybrid case: the moduli space of the curves in a given class is disconnected, containing both a family and isolated curves. In this case, the invariant is the sum of the invariants of all components.

	\subsection{Gromov-Witten invariants of birationally equivalent varieties} \label{ss:GW_and_flops}
	Following early results assuming genericity \cite{Witten:1993ed, Morrison:1993}, the flop invariance of the genus 0 topological string partition function was proved in \cite{Li:2001}.\footnote{Note that \cite{Witten:1993ed} shows a stronger result for a particular example: that the genus $0$ topological string free energy coincides for the two birationally equivalent manifolds upon analytic continuation in the K\"ahler parameter of the flopped curve. Here, we are merely considering the Gromov-Witten invariants outside the locus in which the two varieties differ.} At the level of Gromov-Witten invariants, this implies that curves lying in the intersection of the Mori cones (identified via their image to the common singular manifold with the exceptional locus removed)  of two varieties related by a flop have the same Gromov-Witten invariants. We can easily check this for all of our examples.

	\section{Identifying matter: the formalism applied} \label{s:formalism_applied}
	
	We are at long last ready to apply the formalism developed above to determine the gauge algebra $\mg$ and the matter content $\bigoplus_i \mR_i$, with $\mR_i$ denoting representations of $\mg$, of an F-theory compactification on the elliptically fibered anti-canonical hypersurface $X$.

	\subsection{Embedding the root lattice $\Lroot(\mg)$ in $N_1(X)$}
	Using our results from section \ref{ss:distinguished_curves}, we can identify the distinguished curves in $X$ within the toric Mori cone (or, when the latter is not smooth, within the dual of the smooth refinement of the toric K\"ahler cone) with regard to which we compute the Gromov-Witten invariants of $X$. In the notation of \ref{ss:distinguished_curves}, our analysis will be based on curves with no component in the classes of $C_B$, $C_F$, and $C_{r_0}$.
	
	As discussed in section \ref{s:F-matter}, curves which give rise to fields residing in vector multiplets come in $\IP^1$ families. These are hence identifiable via the Gromov-Witten invariants $c_1(\Omega_{\IP^1}^1) = -2$. The curves in $X$ with vanishing $[C_B]$, $[C_F]$, and $[C_{r_0}]$ components and Gromov-Witten invariant $-2$ will furnish a basis of the root lattice of $\mg$ (up to a subtlety we shall discuss in section \ref{ss:roots_and_weights}). Comparing the intersection matrix of these curves with the divisors $\DX_{r_i}$ to the Cartan matrices of simple Lie algebras allows us to identify $\mg$, and to define a linear embedding $\phi$ of the root lattice $\Lroot(\mg)$ of $\mg$ into $N_1(X)$, with image inside the Mori cone of $X$:
	\be
	\phi : \Lroot(\mg) \rightarrow N_1(X) \,.
	\ee

	\subsection{Identifying matter} \label{ss:identifying_matter}
	
	All remaining curves with vanishing $[C_B]$, $[C_F]$, and $[C_{r_0}]$ components are isolated, hence give rise to fields residing in charged matter hypermultiplets. By section \ref{ss:curve_invariants}, the Gromov-Witten invariants associated to the class of such curves count their number (unless non-isolated curves lie in the same class; this is the hybrid case invoked in section \ref{ss:curve_invariants} and discussed further in section \ref{ss:roots_and_weights}). The Mori vectors of such curves lie in the image of $\phi$ extended over $\IQ$. Their inverse image under $\phi_{\IQ}$ lies inside the weight lattice $\Lweight(\mg)$ of $\mg$. 
	
	Recall that $ \Lweight \subset (\Lroot)_{\IQ}$, i.e. weights $\lambda$ expanded in a basis of simple roots will generically exhibit rational coefficients. Nevertheless, the image under $\phi_{\IQ}$ of those weights that belong to representations $\mR_i$ furnished by $X$ lies in $N_1(X)$.

	\subsubsection{Complex vs. self-conjugate representations} \label{ss:complex_vs_self_conjugate}
	Given the Gromov-Witten invariant associated to a curve class giving rise to fields residing in a hypermultiplet transforming in the representation $\mR$, determining the number of such hypermultiplets depends on whether the representation $\mR$ is complex or (pseudo-)real.
	
	Let $S(\mR)$ be the set of curves that give rise to the scalars in the representation $\mR$. The collection $S(\mR)$ will generically contain both holomorphic and anti-holomorphic curves. Only the holomorphic elements are counted by Gromov-Witten invariants. 
	
	If the representation $\mR$ is complex, then $\lambda \in \Pi(\mR) \rightarrow - \lambda \in \Pi(\overline{\mR})$. Either $\lambda$ or $-\lambda$ is represented by a holomorphic curve class, but not both. Thus, only some of the weights of $\mR$ will be identified via Gromov-Witten invariants. Nevertheless, the analysis outlined in the introduction to this subsection will find non-zero invariants for all classes $\phi_{\IQ}(\Pi(\mR))$. This is required by CPT symmetry: the representation content of a hypermultiplet associated to a complex representation is $\mR \oplus \overline{\mR}$. Thus, holomorphic and anti-holomorphic curves combined must furnish this reducible representation. It follows that the holomorphic curves in $S(\mR) \cup S(\overline{\mR})$ combined must yield precisely the elements of $\Pi(\mR)$. 
	
	If $\mR$ is self-conjugate, then $\lambda \in \Pi(\mR) \leftrightarrow - \lambda \in \Pi(\mR)$. Again, either $\lambda$ or $-\lambda$ is represented by a holomorphic curve class, but not both.
	The analysis outlined above applied to this situation will therefore only find part of $\Pi(\mR)$ represented by classes in $N_1$ with non-zero Gromov-Witten invariants. Note that when $\mR$ is self-conjugate, CPT symmetry does not dictate the doubling of degrees of freedom for hypermultiplets: half-hypermultiplets are permitted.
	
	To summarize, these considerations entail the following for the bookkeeping of matter content: if $\mR$ is complex, an isolated curve and its complex conjugate give rise to a field and its conjugate in a hypermultiplet, the corresponding Gromov-Witten invariant hence allows us to read off the number of hypermultiplets. If $\mR$ is self-conjugate, an isolated curve and its complex conjugate give rise to different fields in the same half-hypermultiplet, the Gromov-Witten invariant hence counts the number of half-hypermultiplets.
	
	The distribution of holomorphic vs. anti-holomorphic curves in $S(\mR)$ changes under flops. It thus differs among the elements in $(\mg)_n$.

	\subsubsection{When roots and weights coincide} \label{ss:roots_and_weights}
	The deduction of field content from Gromov-Witten invariants requires additional care when some weights and roots coincide. This is only possible if the highest weight of the representation in question is an element of the root lattice $\Lroot$. E.g., all weights of the $\mathbf{7}$ representation of $G_2$, the $\mathbf{26}$ representation of $F_4$, and the vector representation of $B_n$ are also roots.
	
	\paragraph{Example: $(G_2)_3$} The image of only half of the simple roots arises in $N_1(X)$ with Gromov-Witten invariant -2. No other classes with vanishing $[C_B]$, $[C_F]$, and $[C_{r_0}]$ components have non-vanishing Gromov-Witten invariant. The interpretation is that a hypermultiplet in the real $\mathbf{7}$ representation is present: the holomorphic curve classes in $\phi_{\IQ}(\Pi(\mathbf{7}))$ contribute $+2$ to the respective Gromov-Witten invariants ($\mathbf{7}$ is self-conjugate, hence a contribution of $+2$ to the Gromov-Witten invariant implies 2 half-hypermultiplets). As all of the weights of the $\mathbf{7}$ representation coincide with roots of $G_2$, the net effect of the presence of such curves is to cancel the Gromov-Witten invariants associated to these roots.
	
	\subsubsection{Matter curves and the toric Mori cone} \label{ss:toric_mori_cone_generators_interpretation}
	Representations $\mR$ for which $\Pi(\mR) \not \subset \Lroot$ can leave an imprint on the generators of the toric Mori cone. When the toric Mori cone is smooth, we find that its generators can be expressed in terms of the distinguished curve classes introduced in section \ref{ss:distinguished_curves} as follows: three are linear combinations of the classes $[C_B]$, $[C_F]$, and $[C_{r_0}]$, and rank($\mg$) are linear combinations of the classes of the curves $C_{r_i}$. These latter linear combinations correspond either to simple roots of $\mg$ or to weights in a representation $\mR$ (or either $\mR$ or $\overline{\mR}$, for complex representations) of $\mg$. Unlike the naive expectation but in agreement with the discussion in section \ref{ss:complex_vs_self_conjugate}, the weights that occur are not the highest weight of $\mR$ (or $\overline{\mR}$), hence the image under $\phi_{\IQ}$ of the weights of $\mR$ (or $\overline{\mR}$) does not lie in the toric Mori cone. However, for all weights $\lambda \in \Pi(\mR)$ not mapped into the toric Mori cone, $-\lambda$ is. Which weights occur in the toric Mori cone depends on which of the birationally equivalent $(\mg)_n$ varieties we consider. 
	
	For varieties with smooth toric Mori cones, the presence of matter curves associated to representations $\mR$ for which $\Pi(\mR) \not \subset \Lroot$ can hence be inferred from a  generator whose preimage under $\phi_{\IQ}$ lies in $\Pi(\mR)$. In the case of $(E_6)_n$ and $(E_7)_n$, we can even determine the multiplicity with which $\mR$ occurs: at least one of the generators in question occurs as an irreducible component of the intersection of a torus invariant surface of the ambient space with $X$. The multiplicity of this reducible intersection coincides with the multiplicity of $\mR$. It would be interesting to study the systematics underlying this observation further.

	\paragraph{Example: $(E_6)_3$}
	We consider a variety $X$ of type~I in $(E_6)_3$. In table \ref{table:mc_in_terms_of_dc}, we give the generators of the toric Mori cone in terms of the classes of distinguished curves whose Mori vectors are listed in table \ref{table:mori_vectors_E6_n}. We note that $\phi_{\IQ}^{-1}$ maps $C_1$, $C_2$, $C_3$ to simple roots of $E_6$,   $C_4$ and $C_6$ to weights of the $\boldsymbol{27}$ representation, and  $C_5$ to a weight of the $\boldsymbol{\overline{27}}$ representation.  
	\begin{table}[h!]
		\begin{center}
			$\begin{tabu}{l|c|c|c|c|c|c|c|c|c|c|c|c|c} 
			& C_{r_1} & C_{r_2} & C_{r_3} & C_{r_4} & C_{r_5} & C_{r_6} & C_{r_0} & C_F & C_B \\ \hline
			C_1 & 0 & 0 & 1 & 0 & 0 & 0 & 0 & 0 & 0 \\
			C_2 & 0 & 0 & 0 & 1 & 0 & 0 & 0 & 0 & 0 \\
			C_3 & 0 & 0 & 0 & 0 & 0 & 1 & 0 & 0 & 0 \\
			C_4 & -\frac{1}{3} & \frac{1}{3} &  0 & -\frac{1}{3} & \frac{1}{3} & 0 & 0 & 0 & 0 \\
			C_5 & \frac{1}{3} & -\frac{1}{3} & 0 & \frac{1}{3} & \frac{2}{3} & 0 & 0 & 0 & 0 \\
			C_6 & \frac{2}{3} & \frac{1}{3} & 0 & -\frac{1}{3} & -\frac{2}{3} & 0 & 0 & 0 & 0 \\
			C_7 & 0 & 0 & 0 & 0 & 0 & 0 & 1 & 0 & 0 \\
			C_8 & 0 & 0 & 0 & 0 & 0 & 0 & 0 & 1 & 0\\
			C_9 & 0 & 0 & 0 & 0 & 0 & 0 & -1 & 0 & 1 	
			\end{tabu}$
			\caption{Generators of the toric Mori cone for an $(E_6)_3$ variety of type~I expressed in terms of the Mori vectors of the distinguished curves listed in table \ref{table:mori_vectors_E6_n}.} \label{table:mc_in_terms_of_dc}
		\end{center}
	\end{table}
	Furthermore, $3\,C_4 = [\DX_{r_2} \cdot \DX_{r_5}]$, and $3\,C_5 = [\DX_{r_1}\cdot \DX_{r_6}]$, 3 being the number of hypermultiplets in the $\boldsymbol{27}$ representation which arise upon compactification on $X$.
	
	\section*{Acknowledgements}
	We are grateful to Michele Del Zotto, Albrecht Klemm, Guglielmo Lockhart and Timo Weigand for discussions.
	
	\appendix
	\section{Assorted data on $(\mg)_n$ varieties}\label{s:data}
	In this appendix, we record some toric data of the geometries we have studied in this paper, notably the generators of the toric Mori cones. To keep this section within a reasonable length, we discuss only the first few members of the $A$-, $B$-, and $D$-series, and after the first example, $(A_2)_n$, refrain from listing data of varieties of type~II (except for the case $(A_2)_3$, where the only element is of this type).
	
	\subsection{$A$-series}
	
	$\boldsymbol{A_2}$
	
	{\bf Occurs over Hirzebruch bases $\IF_n$, $n=0, \ldots, 3$.}
	
	{\bf Additional 1-cones:} 
	\be
		\begin{array}{ccccc}
				u_{\rho_{r_1}}= & (-1 & -1 & 0 & -1) \\
				u_{\rho_{r_2}}= & (-1 & -2 & 0 & -1)
				 
		\end{array}
	\ee
	
	{\bf Matter content:}  $6(3-n)$ hypermultiplets in the complex representation $\boldsymbol{3}$.
	
	For $n=0, \ldots, 2$, $(A_2)_n$ contains exactly one variety of type~I. The corresponding toric Mori cones coincide and are smooth. Their generators are given in table \ref{table_A:mc_in_terms_of_dc_A_2}. The curve $C_2$ corresponds to a weights of the representation $\boldsymbol{3}$.
	\begin{table}[h!]
		\begin{center}
			\scalebox{0.7}{$\begin{tabu}{l|c|c|c|c|c} 
				& C_{r_1} & C_{r_2} & C_{r_0} & C_F & C_B \\ \hline
				C_1 & 0 & 1 & 0 & 0 & 0 \\
				C_2 & \frac{1}{3} & -\frac{1}{3} & 0 & 0 & 0 \\
				C_3 & 0 & 0 & 1 & 0 & 0 \\
				C_4 & 0 & 0 & 0 & 1 & 0 \\
				C_5 & 0 & 0 & 0 & 0 & 1
				\end{tabu}$} 
			\caption{The generators of the toric Mori cone of the $(A_2)_n$ varieties of type I.} \label{table_A:mc_in_terms_of_dc_A_2}
		\end{center}
	\end{table}
	
	In addition, $(A_2)_n$ for $n=1,3$ also contains a variety of type~II. For $n=3$, this is the only variety contained in $(A_2)_n$. The corresponding toric Mori cones are smooth. Their generators, expressed in terms of the Mori vectors of the distinguished curves introduce in section \ref{ss:distinguished_curves} for the case that $u*v \not \in \ISR(Y)$ are given in table \ref{table_A:mc_in_terms_of_dc_A_2_not_uv}.
	\begin{table}[h!]
		\begin{center}
			\scalebox{0.7}{$\begin{tabu}{l|c|c|c|c|c} 
				& C_{r_1} & C_{r_2} & C_{r_0} & C_F & C_B \\ \hline
				C_1 & 0 & 1 & 0 & 0 & 0 \\
				C_2 & \frac{1}{3} & -\frac{1}{3} & 0 & 0 & 0 \\
				C_3 & 0 & 0 & 1 & 0 & -1 \\
				C_4 & 0 & 0 & 0 & 1 & -1 \\
				C_5 & 0 & 0 & 0 & 0 & 1
				\end{tabu}$} \quad \quad
			\scalebox{0.7}{$\begin{tabu}{l|c|c|c|c|c} 
				& C_{r_1} & C_{r_2} & C_{r_0} & C_F & C_B \\ \hline
				C_1 & 0 & 1 & -1 & 0 & 1 \\
				C_2 & \frac{1}{3} & -\frac{1}{3} & 0 & 0 & 0 \\
				C_3 & 0 & 0 & 1 & 0 & -1 \\
				C_4 & 0 & 0 & 0 & 1 & 0 \\
				C_5 & 0 & 0 & 0 & 0 & 1
				\end{tabu}$} 
			\caption{The generators of the toric Mori cone ($n=1$ on the left, $n=2$ on the right) of the $(A_2)_n$ varieties of type~II.} \label{table_A:mc_in_terms_of_dc_A_2_not_uv}
		\end{center}
	\end{table}

	\vspace{1cm}
	$\boldsymbol{A_3}$
	
	{\bf Occurs over Hirzebruch bases $\IF_n$, $n=0, \ldots, 2$.}
	
	{\bf Additional 1-cones:} 
	\be
		\begin{array}{ccccc}
			u_{\rho_{r_1}}= & (-1 & -1 & 0 & -1) \\
			u_{\rho_{r_2}}= & (0 & -1 & 0 & -1) \\
			u_{\rho_{r_3}}= & (-1 & -2 & 0 & -1) 
		\end{array}
	\ee
	
	{\bf Matter content:}  
	\begin{itemize}
	\item $(A_3)_1$ : $12$ hypermultiplets in the complex representation $\boldsymbol{4}$, 2 half hypermultiplets in the self-conjugate representation $\boldsymbol{6}$.
	\item $(A_3)_2$:  $8$ hypermultiplets in the complex representation $\boldsymbol{4}$.
	\end{itemize}
	
	For $n=0, 1, 2$, $(A_3)_n$ contains exactly one variety of type~I. The corresponding toric Mori cones coincide and are smooth. Their generators are given in table \ref{table_A:mc_in_terms_of_dc_A_3}. Curves corresponding to weights of the representation $\boldsymbol{4}$ are highlighted in red, and those corresponding to the representation $\boldsymbol{6}$ in green.
	\begin{table}[h!]
		\begin{center}
			\scalebox{0.7}{$\begin{tabu}{l|c|c|c|c|c|c|c} 
				& C_{r_1} & C_{r_2} & C_{r_3} & C_{r_0} & C_F & C_B \\ \hline
				C_1 & 0 & 0 & 1 & 0 & 0 & 0 \\
				\red{C_2} & -\frac{1}{4} & \frac{1}{2} & \frac{1}{4} & 0 & 0 & 0 \\
				\green{C_3} & \frac{1}{2} & 0 & -\frac{1}{2} & 0 & 0 & 0 \\
				C_4 & 0 & 0 & 0 & 1 & 0 & 0 \\
				C_5 & 0 & 0 & 0 & 0 & 1 & 0 \\
				C_6 & 0 & 0 & 0 & 0 & 0 & 1
				\end{tabu}$} 
			\caption{The generators of the toric Mori cone of the $(A_3)_n$ varieties of type~I.} \label{table_A:mc_in_terms_of_dc_A_3}
		\end{center}
	\end{table}
	
	In addition, $(A_3)_1$ also contains a type~II variety.

	\subsection{$B$-series}
	$\boldsymbol{B_3}$
	
	{\bf Occurs over Hirzebruch bases $\IF_n$, $n=0, \ldots, 3$.}
	
	{\bf Additional 1-cones:} 
	\be
	\begin{array}{ccccc}
		u_{\rho_{r_1}}= & (0 & -1 & 0 & -1) \\
		u_{\rho_{r_2}}= & (-2 & -3 & 0 & -2) \\
		u_{\rho_{r_3}}= & (-1 & -1 & 0 & -1)
		\end{array}
	\ee
	
	{\bf Matter content:}  
	\begin{itemize}
		\item $2*(3-n)$ half-hypermultiplets in the self-conjugate representation $\boldsymbol{7}$.
		\item $2*2(4-n)$ half-hypermultiplets in the self-conjugate representation $\boldsymbol{8}$.
	\end{itemize}
	
	For all $n=0, \ldots, 3$, $(B_3)_n$ contains exactly one variety of type~I. The corresponding toric Mori cones  are smooth. They have six generators, five of which are independent of $n$, given in table \ref{table_A:mc_in_terms_of_dc_B_3}.  The curve $C_3$ corresponds to a weight in the representation $\boldsymbol{8}$.
	\begin{table}[h!]
		\begin{center}
			\scalebox{0.7}{$\begin{tabu}{l|c|c|c|c|c|c|c} 
				& C_{r_1} & C_{r_2} & C_{r_3} & C_{r_0} & C_F & C_B \\ \hline
				C_1 & 0 & 1 & 0 & 0 & 0 & 0 \\
				C_2 & 0 & 0 & 1 & 0 & 0 & 0 \\
				C_3 & \frac{1}{2} & 0 & -\frac{1}{2} & 0 & 0 & 0 \\
				C_4 & 0 & 0 & 0 & 1 & 0 & 0 \\
				C_5 & 0 & 0 & 0 & 0 & 1 & 0 \\
				\end{tabu}$} \quad
			\scalebox{0.7}{$\begin{tabu}{l|c|c|c|c|c|c|c} 
				& C_{r_1} & C_{r_2} & C_{r_3} & C_{r_0} & C_F & C_B \\ \hline
				(C_6)_{0,1,2} & 0 & 0 & 0 & 0 & 0 & 1 \\
				(C_6)_3 &0 & 0 & 0 & -1 & 0 & 1
				\end{tabu}$}
			\caption{The five $n$-independent generator of the toric Mori cone of the varieties in $(B_3)_n$ of type~I given on the left, and the last $n$-dependent generator given on the right.} \label{table_A:mc_in_terms_of_dc_B_3}
		\end{center}
	\end{table}
	
	In addition, $(B_3)_n$ for $n=1,3$ also contains a variety of type~II.
	
	Note that all the weights of the representation $\boldsymbol{7}$ are also roots. The associated Gromov-Witten invariants at base $\IF_n$ are thus $2*(3-n)-2 = 4-2n$. In particular, at $n=2$, the contributions from roots and weights cancel.
	
	\vspace{1cm}
	\newpage
	$\boldsymbol{B_4}$
	
	\nopagebreak
	{\bf Occurs over Hirzebruch bases $\IF_n$, $n=0, \ldots, 4$.}
	
	{\bf Additional 1-cones:} 
	\be
	\begin{array}{ccccc}
		u_{\rho_{r_1}}= & (-1 & -1 & 0 & -1) \\
		u_{\rho_{r_2}}= & (-2 & -3 & 0 & -2) \\
		u_{\rho_{r_3}}= & (-1 & -2 & 0 & -2) \\
		u_{\rho_{r_4}}= & (0 & -1 & 0 & -1) 
	
	\end{array}
	\ee
	
	{\bf Matter content:}  
	\begin{itemize}
		\item $2*(5-n)$ half-hypermultiplets in the self-conjugate representation $\boldsymbol{9}$.
		\item $2*(4-n)$ half-hypermultiplets in the self-conjugate representation $\boldsymbol{16}$.
	\end{itemize}
	
	For all $n=0, \ldots, 4$, $(B_4)_n$ contains exactly one variety of type~I. The corresponding toric Mori cones  are smooth. They have seven generators, six of which are independent of $n$, given in table \ref{table_A:mc_in_terms_of_dc_B_4}. The curve $C_4$ corresponds to a weight in the representation $\boldsymbol{16}$.
	
	\begin{table}[h!]
		\begin{center}
			\scalebox{0.7}{$\begin{tabu}{l|c|c|c|c|c|c|c} 
				& C_{r_1} & C_{r_2} & C_{r_3} & C_{r_4} & C_{r_0} & C_F & C_B \\ \hline
				C_1 & 1 & 0 & 0 & 0 & 0 & 0 & 0 \\
				C_2 &0 & 1 & 0 & 0 & 0 & 0 & 0 \\
				C_3 &0 & 0 & 0 & 1 & 0 & 0 & 0 \\
				C_4 &	-\frac{1}{2} & 0 & \frac{1}{2} & 0 & 0 & 0 & 0 \\
				C_5 &0 & 0 & 0 & 0 & 1 & 0 & 0 \\
				C_6 &0 & 0 & 0 & 0 & 0 & 1 & 0
				\end{tabu}$} \quad
			\scalebox{0.7}{$\begin{tabu}{l|c|c|c|c|c|c|c} 
				& C_{r_1} & C_{r_2} & C_{r_3} & C_{r_4} & C_{r_0} & C_F & C_B \\ \hline
				(C_7)_{0,1,2} & 0 & 0 & 0 & 0 & 0 & 0 & 1 \\
				(C_7)_3 & 0 & 0 & 0 & 0 & -1 & 0 & 1 \\
				(C_7)_4 & 0 & 0 & 0 & 0 & -2 & 0 & 1
				\end{tabu}$}
			\caption{The six $n$-independent generator of the toric Mori cone of the $(B_4)_n$ varieties of type~I given on the left, and the last $n$-dependent generator given on the right.} \label{table_A:mc_in_terms_of_dc_B_4}
		\end{center}
	\end{table}

	In addition, $(B_4)_n$ for $n=1,3$ also contains a variety of type~II.
	
	Note that all the weights of the representation $\boldsymbol{9}$ are also roots. The associated Gromov-Witten invariants at base $\IF_n$ are thus $2*(5-n)-2 = 2(4-n)$. These Gromov-Witten invariants hence coincide with those associated to the $\boldsymbol{16}$ representation. At $n=4$, the contributions associated to the weights of the representation $\boldsymbol{9}$ cancel against that of roots.
	
	\vspace{1cm}
	$\boldsymbol{B_5}$
	
	{\bf Occurs over Hirzebruch bases $\IF_n$, $n=0, \ldots, 4$.}
	
	{\bf Additional 1-cones:} 
	\be
	\begin{array}{ccccc}
		u_{\rho_{r_1}}= & (-1 & -1 & 0 & -1) \\
		u_{\rho_{r_2}}= & (-2 & -3 & 0 & -2) \\
		u_{\rho_{r_3}}= & (-1 & -2 & 0 & -2) \\
		u_{\rho_{r_4}}= & (0 & -1 & 0 & -2) \\
		u_{\rho_{r_5}}= & (0 & 0 & 0 & -1)
	\end{array}
	\ee
	
	{\bf Matter content:}  
	\begin{itemize}
		\item $2*(7-n)$ half-hypermultiplets in the self-conjugate representation $\boldsymbol{11}$.
		\item $(4-n)$ half-hypermultiplets in the self-conjugate representation $\boldsymbol{32}$.
	\end{itemize}
	
	For $n=0, \ldots, 3$, $(B_5)_n$ contains four varieties of type~I. The corresponding toric Mori cones  are smooth. They have eight generators, as given in table \ref{table_A:mc_in_terms_of_dc_B_5}. Seven of these are independent of $n$. The curves in table \ref{table_A:mc_in_terms_of_dc_B_5} highlighted in green map to weights in the representation $\boldsymbol{32}$.
	\begin{table}
		\begin{center}
			\scalebox{0.7}{\begin{tabular}{c|cccccccc}
				& $C_{r_1}$ & $C_{r_2}$ & $C_{r_3}$ & $C_{r_4}$ & $C_{r_5}$ & $C_{r_0}$ & $C_F$ & $C_B$ \\ \hline
				$C_1$ & $1$ & $0$ & $0$ & $0$ & $0$ & $0$ & $0$ & $0$ \\
				$C_2$ & $0$ & $1$ & $0$ & $0$ & $0$ & $0$ & $0$ & $0$ \\
				$C_3$ & $0$ & $0$ & $0$ & $1$ & $0$ & $0$ & $0$ & $0$ \\
				$C_4$ & $0$ & $0$ & $0$ & $0$ & $1$ & $0$ & $0$ & $0$ \\
				$\green{C_5}$ & $-\frac{1}{2}$ & $0$ & $\frac{1}{2}$ & $0$ & $-\frac{1}{2}$ & $0$ & $0$ & $0$ \\
				$C_6$ & $0$ & $0$ & $0$ & $0$ & $0$ & $1$ & $0$ & $0$ \\
				$C_7$ & $0$ & $0$ & $0$ & $0$ & $0$ & $0$ & $1$ & $0$ \\
				\end{tabular}} \quad \quad \quad
			\scalebox{0.7}{
			\begin{tabular}{c|cccccccc}
				& $C_{r_1}$ & $C_{r_2}$ & $C_{r_3}$ & $C_{r_4}$ & $C_{r_5}$ & $C_{r_0}$ & $C_F$ & $C_B$ \\ \hline
				$C_1$ & $0$ & $1$ & $0$ & $0$ & $0$ & $0$ & $0$ & $0$ \\
				$C_2$ & $0$ & $0$ & $0$ & $1$ & $0$ & $0$ & $0$ & $0$ \\
				$\green{C_3}$ & $\frac{1}{2}$ & $0$ & $-\frac{1}{2}$ & $0$ & $\frac{1}{2}$ & $0$ & $0$ & $0$ \\
				$\green{C_4}$ & $\frac{1}{2}$ & $0$ & $\frac{1}{2}$ & $0$ & $-\frac{1}{2}$ & $0$ & $0$ & $0$ \\
				$\green{C_5}$ & $-\frac{1}{2}$ & $0$ & $\frac{1}{2}$ & $0$ & $\frac{1}{2}$ & $0$ & $0$ & $0$ \\
				$C_6$ & $0$ & $0$ & $0$ & $0$ & $0$ & $1$ & $0$ & $0$ \\
				$C_7$ & $0$ & $0$ & $0$ & $0$ & $0$ & $0$ & $1$ & $0$ \\
			\end{tabular}} \\  \vspace{0.5cm}
		\scalebox{0.7}{
			\begin{tabular}{c|cccccccc}
				& $C_{r_1}$ & $C_{r_2}$ & $C_{r_3}$ & $C_{r_4}$ & $C_{r_5}$ & $C_{r_0}$ & $C_F$ & $C_B$ \\ \hline
				$C_1$ & $1$ & $0$ & $0$ & $0$ & $0$ & $0$ & $0$ & $0$ \\
				$C_2$ & $0$ & $0$ & $1$ & $0$ & $0$ & $0$ & $0$ & $0$ \\
				$C_3$ & $0$ & $0$ & $0$ & $1$ & $0$ & $0$ & $0$ & $0$ \\
				$\green{C_4}$ & $\frac{1}{2}$ & $1$ & $\frac{1}{2}$ & $0$ & $-\frac{1}{2}$ & $0$ & $0$ & $0$ \\
				$\green{C_5}$ & $-\frac{1}{2}$ & $0$ & $-\frac{1}{2}$ & $0$ & $\frac{1}{2}$ & $0$ & $0$ & $0$ \\
				$C_6$ & $0$ & $0$ & $0$ & $0$ & $0$ & $1$ & $0$ & $0$ \\
				$C_7$ & $0$ & $0$ & $0$ & $0$ & $0$ & $0$ & $1$ & $0$ \\
			\end{tabular}} \quad \quad \quad
		\scalebox{0.7}{\begin{tabular}{c|cccccccc}
				& $C_{r_1}$ & $C_{r_2}$ & $C_{r_3}$ & $C_{r_4}$ & $C_{r_5}$ & $C_{r_0}$ & $C_F$ & $C_B$ \\ \hline
				$C_1$ & $0$ & $1$ & $0$ & $0$ & $0$ & $0$ & $0$ & $0$ \\
				$C_2$ & $0$ & $0$ & $1$ & $0$ & $0$ & $0$ & $0$ & $0$ \\
				$C_3$ & $0$ & $0$ & $0$ & $0$ & $1$ & $0$ & $0$ & $0$ \\
				$\green{C_4}$ & $\frac{1}{2}$ & $0$ & $-\frac{1}{2}$ & $0$ & $-\frac{1}{2}$ & $0$ & $0$ & $0$ \\
				$C_5$ & $-1$ & $0$ & $1$ & $1$ & $1$ & $0$ & $0$ & $0$ \\
				$C_6$ & $0$ & $0$ & $0$ & $0$ & $0$ & $1$ & $0$ & $0$ \\
				$C_7$ & $0$ & $0$ & $0$ & $0$ & $0$ & $0$ & $1$ & $0$ \\
			\end{tabular}}   \\ \vspace{0.5cm}
		\scalebox{0.7}{\begin{tabular}{c|cccccccc}
				& $C_{r_1}$ & $C_{r_2}$ & $C_{r_3}$ & $C_{r_4}$ & $C_{r_5}$ & $C_{r_0}$ & $C_F$ & $C_B$ \\ \hline
				$(C_8)_{0,1,2}$ & $0$ & $0$ & $0$ & $0$ & $0$ & $0$ & $0$ & $1$ \\
				$(C_8)_{3}$ & $0$ & $0$ & $0$ & $0$ & $0$ & $-1$ & $0$ & $1$ \\
			\end{tabular}}
			\caption{Generators of the toric Mori cones for the four $(B_5)_n$ varieties of type~I for $n=0, \ldots, 3$, expressed in terms of the Mori vectors of the distinguished curves introduced in section \ref{ss:mori_cone}. Only the generator $C_{8}$ depends on the base surface $\IF_n$; its expression in terms of the distinguished curves coincides for three of the four $(B_5)_n$ varieties of type~I. The corresponding classes are denoted as $(C_{8})_n$ in the last table. \label{table_A:mc_in_terms_of_dc_B_5}}
		\end{center}
	\end{table}

	In addition, $(B_5)_n$ for $n=1,3$ also contains four varieties of type~II.
	
	The class $(B_5)_4$ contains a single variety. Its toric Mori cone, given in table \ref{table_A:mc_in_terms_of_dc_B_5_4}, is not simplicial.
	\begin{table}
		\begin{center}
			\scalebox{0.7}{\begin{tabular}{c|cccccccc}
					& $C_{r_1}$ & $C_{r_2}$ & $C_{r_3}$ & $C_{r_4}$ & $C_{r_5}$ & $C_{r_0}$ & $C_F$ & $C_B$ \\ \hline
					$C_1$ & $1$ & $0$ & $0$ & $0$ & $0$ & $0$ & $0$ & $0$ \\
					$C_2$ & $0$ & $1$ & $0$ & $0$ & $0$ & $0$ & $0$ & $0$ \\
					$C_3$ & $0$ & $0$ & $1$ & $0$ & $0$ & $0$ & $0$ & $0$ \\
					$C_4$ & $0$ & $0$ & $0$ & $1$ & $0$ & $0$ & $0$ & $0$ \\
					$C_5$ & $0$ & $0$ & $0$ & $0$ & $1$ & $0$ & $0$ & $0$ \\
					$C_6$ & $\frac{1}{2}$ & $1$ & $\frac{1}{2}$ & $0$ & $-\frac{1}{2}$ & $0$ & $0$ & $0$ \\
					$C_7$ & $-1$ & $0$ & $1$ & $1$ & $1$ & $0$ & $0$ & $0$ \\
					$C_8$ & $0$ & $0$ & $0$ & $0$ & $0$ & $1$ & $0$ & $0$ \\
					$C_9$ & $0$ & $0$ & $0$ & $0$ & $0$ & $0$ & $1$ & $0$ \\
					$C_{10}$ & $0$ & $0$ & $0$ & $0$ & $0$ & $-2$ & $0$ & $1$ \\
			\end{tabular}}
			\caption{Generators of the non-simplicial toric Mori cone for the unique $(B_5)_4$ variety. \label{table_A:mc_in_terms_of_dc_B_5_4}}
		\end{center}
	\end{table}
	
	Note that all the weights of the representation $\boldsymbol{11}$ are also roots. The associated Gromov-Witten invariants at base $\IF_n$ are thus $2*(7-n)-2 = 2*(6-n)$.
	
	\vspace{1cm}

	\subsection{$D$-series}
	$\boldsymbol{D_5}$
	
	{\bf Occurs over Hirzebruch bases $\IF_n$, $n=0, \ldots, 4$.}
	
	{\bf Additional 1-cones:} 
	\be
	\begin{array}{ccccc}
		u_{\rho_{r_1}}= & (-1 & -1 & 0 & -1) \\
		u_{\rho_{r_2}}= & (-2 & -3 & 0 & -2) \\
		u_{\rho_{r_3}}= & (-1 & -2 & 0 & -2) \\
		u_{\rho_{r_4}}= & (0 & -1 & 0 & -1) \\
		u_{\rho_{r_5}}= & (0 & 0 & 0 & -1) 
	\end{array}
	\ee
	
	{\bf Matter content:}  
	\begin{itemize}
		\item $2*(6-n)$ half-hypermultiplets in the self-conjugate representation $\boldsymbol{10}$.
		\item $(4-n)$ hypermultiplets in the complex representation $\boldsymbol{16}$.
	\end{itemize}
	
	For $n=0, \ldots, 3$, $(D_5)_n$ contains three varieties of type~I. The corresponding toric Mori cones are smooth. They have eight generators, given in table  \ref{table_A:mc_in_terms_of_dc_D_5_3}. Curves corresponding to weights of the representation $\mathbf{10}$ are highlighted in blue, those of the representation $\mathbf{16}$ in green, and those of the representation $\mathbf{\overline{16}}$ in red.
	\begin{table}[h!]
		\begin{center}			
			\scalebox{0.7}{$\begin{tabu}{l|c|c|c|c|c|c|c|c} 
				& C_{r_1} & C_{r_2} & C_{r_3} & C_{r_4} & C_{r_5} & C_{r_0} & C_F & C_B \\ \hline
				C_1 & 0 & 1 & 0 & 0 & 0 & 0 & 0 & 0 \\
				C_ 2& 0 & 0 & 0 & 1 & 0 & 0 & 0 & 0 \\
				\red{C_3} & -\frac{1}{2} & 0 & \frac{1}{2} & -\frac{1}{4} & \frac{1}{4} & 0 & 0 & 0 \\
				\green{C_4} & \frac{1}{2} & 0 & \frac{1}{2} & \frac{1}{4} & -\frac{1}{4} & 0 & 0 & 0 \\
				\red{C_5} & \frac{1}{2} & 0 & -\frac{1}{2} & -\frac{1}{4} & \frac{1}{4} & 0 & 0 & 0 \\
				C_6 & 0 & 0 & 0 & 0 & 0 & 1 & 0 & 0 \\
				C_7 & 0 & 0 & 0 & 0 & 0 & 0 & 1 & 0 
				\end{tabu}$}  \quad \quad \quad
			\scalebox{0.7}{$\begin{tabu}{l|c|c|c|c|c|c|c|c} 
				& C_{r_1} & C_{r_2} & C_{r_3} & C_{r_4} & C_{r_5} & C_{r_0} & C_F & C_B \\ \hline
				C_1 & 1 & 0 & 0 & 0 & 0 & 0 & 0 & 0 \\
				C_2 &0 & 0 & 1 & 0 & 0 & 0 & 0 & 0 \\
				C_3 &0 & 0 & 0 & 1 & 0 & 0 & 0 & 0 \\
				\green{C_4} & \frac{1}{2} & 1 & \frac{1}{2} & \frac{1}{4} & -\frac{1}{4} & 0 & 0 & 0 \\
				\red{C_5} & -\frac{1}{2} & 0 & -\frac{1}{2} & -\frac{1}{4} & \frac{1}{4} & 0 & 0 & 0 \\
				C_6 &0 & 0 & 0 & 0 & 0 & 1 & 0 & 0 \\
				C_7 &0 & 0 & 0 & 0 & 0 & 0 & 1 & 0
				\end{tabu}$}  \\ 		 \vspace{0.5cm}
			\scalebox{0.7}{$\begin{tabu}{l|c|c|c|c|c|c|c|c} 
				& C_{r_1} & C_{r_2} & C_{r_3} & C_{r_4} & C_{r_5} & C_{r_0} & C_F & C_B \\ \hline
				C_1 & 1 & 0 & 0 & 0 & 0 & 0 & 0 & 0 \\
				C_2 & 0 & 1 & 0 & 0 & 0 & 0 & 0 & 0 \\
				C_3 & 0 & 0 & 0 & 1 & 0 & 0 & 0 & 0 \\
				\green{C_4} & -\frac{1}{2} & 0 & \frac{1}{2} & \frac{1}{4} & -\frac{1}{4} & 0 & 0 & 0 \\
				\blue{C_5} & 0 & 0 & 0 & -\frac{1}{2} & \frac{1}{2} & 0 & 0 & 0 \\
				C_6 & 0 & 0 & 0 & 0 & 0 & 1 & 0 & 0 \\
				C_7 & 0 & 0 & 0 & 0 & 0 & 0 & 1 & 0
				\end{tabu}$} 
			\quad
			\scalebox{0.7}{$\begin{tabu}{l|c|c|c|c|c|c|c|c} 
				& C_{r_1} & C_{r_2} & C_{r_3} & C_{r_4} & C_{r_5} & C_{r_0} & C_F & C_B \\ \hline
				(C_8)_{0,1,2} & 0 & 0 & 0 & 0 & 0 & 0 & 0 & 1 \\
				(C_8)_3 & 0 & 0 &  0 & 0 & 0 & -1 & 0 & 1
				\end{tabu}$}
			\caption{Generators of the toric Mori cones for the three $(D_5)_n$ varieties of type~I for $n=0, \ldots, 3$, expressed in terms of the Mori vectors of the distinguished curves introduce in section \ref{ss:mori_cone}. Only the generator $C_{8}$ depends on the base surface $\IF_n$, but its expression in terms of the distinguished curves coincides for all three $(D_5)_n$ varieties of type~I. The corresponding classes are denoted as $(C_{8})_n$ in the last table. \label{table_A:mc_in_terms_of_dc_D_5_3}}
		\end{center}
	\end{table}

	In addition, $(D_5)_n$ for $n=1,3$ also contain three varieties of type~II.
	
	The class $(D_5)_4$ contains a single variety.  Its toric Mori cone, given in table \ref{table_A:mc_in_terms_of_dc_D_5_4}, is not simplicial. 
	\begin{table}[h!]
		\begin{center}			
			\scalebox{0.7}{$\begin{tabu}{l|c|c|c|c|c|c|c|c} 
				& C_{r_1} & C_{r_2} & C_{r_3} & C_{r_4} & C_{r_5} & C_{r_0} & C_F & C_B \\ \hline
		C_1 & 0 & 1 & 0 & 0 & 0 & 0 & 0 & 0 \\
		C_1 & 0 & 0 & 1 & 0 & 0 & 0 & 0 & 0 \\
		C_3 & 0 & 0 & 0 & 1 & 0 & 0 & 0 & 0 \\
		C_4 & 0 & 0 & 0 & 0 & 1 & 0 & 0 & 0 \\
		C_5 & \frac{1}{4} & \frac{1}{2} & 0 & -\frac{1}{4} & -\frac{1}{2} & 0 & 0 & 0 \\
		C_6 & -\frac{1}{4} & \frac{1}{2} & 1 & \frac{1}{4} & \frac{1}{2} & 0 & 0 & 0 \\
		C_7 & \frac{1}{2} & 0 & 0 & -\frac{1}{2} & 0 & 0 & 0 & 0 \\
		C_8 & 0 & 0 & 0 & 0 & 0 & 1 & 0 & 0 \\
		C_9 & 0 & 0 & 0 & 0 & 0 & 0 & 1 & 0 \\
		C_{10} & 0 & 0 & 0 & 0 & 0 & -2 & 0 & 1
		\end{tabu}$}
		\caption{Generators of the non-simplicial toric Mori cone for the unique $(D_5)_4$ variety. \label{table_A:mc_in_terms_of_dc_D_5_4}}
		\end{center}
	\end{table}

	\vspace{1cm}
	\newpage
	$\boldsymbol{D_7}$
	
	{\bf Occurs over Hirzebruch base $\IF_4$.}
	
	{\bf Additional 1-cones:} 
	\be
	\begin{array}{ccccc}
		u_{\rho_{r_1}}= & (-1 & -1 & 0 & -1) \\
		u_{\rho_{r_2}}= & (-2 & -3 & 0 & -2) \\
		u_{\rho_{r_3}}= & (-1 & -2 & 0 & -2) \\
		u_{\rho_{r_4}}= & (0 & -1 & 0 & -2) \\
		u_{\rho_{r_5}}= & (1 & 0 & 0 & -2) \\
		u_{\rho_{r_6}}= & (1 & 0 & 0 & -1) \\
		u_{\rho_{r_7}}= & (1 & 1 & 0 & -1)
	\end{array}
	\ee
	
	{\bf Matter content:}  
	\begin{itemize}
		\item $2*(6)$ half-hypermultiplets in the self-conjugate representation $\boldsymbol{14}$.
	\end{itemize}
	
	The class $(D_7)_4$ contains a single variety.  Its toric Mori cone, given in table \ref{table_A:mc_in_terms_of_dc_D_7_4}, is not simplicial. 
	\begin{table}[h!]
		\begin{center}			
			\scalebox{0.7}{\begin{tabular}{c|cccccccccc}
					& $C_{r_1}$ & $C_{r_2}$ & $C_{r_3}$ & $C_{r_4}$ & $C_{r_5}$ & $C_{r_6}$ & $C_{r_7}$ & $C_{r_0}$ & $C_F$ & $C_B$ \\ \hline
					$C_{1}$ & $1$ & $0$ & $0$ & $0$ & $0$ & $0$ & $0$ & $0$ & $0$ & $0$ \\
					$C_{2}$ & $0$ & $1$ & $0$ & $0$ & $0$ & $0$ & $0$ & $0$ & $0$ & $0$ \\
					$C_{3}$ & $0$ & $0$ & $1$ & $0$ & $0$ & $0$ & $0$ & $0$ & $0$ & $0$ \\
					$C_{4}$ & $0$ & $0$ & $0$ & $1$ & $0$ & $0$ & $0$ & $0$ & $0$ & $0$ \\
					$C_{5}$ & $0$ & $0$ & $0$ & $0$ & $1$ & $0$ & $0$ & $0$ & $0$ & $0$ \\
					$C_{6}$ & $0$ & $0$ & $0$ & $0$ & $0$ & $1$ & $0$ & $0$ & $0$ & $0$ \\
					$C_{7}$ & $\frac{1}{2}$ & $1$ & $\frac{3}{2}$ & $1$ & $\frac{1}{2}$ & $\frac{1}{4}$ & $-\frac{1}{4}$ & $0$ & $0$ & $0$ \\
					$C_{8}$ & $0$ & $0$ & $0$ & $0$ & $0$ & $-\frac{1}{2}$ & $\frac{1}{2}$ & $0$ & $0$ & $0$ \\
					$C_{9}$ & $-\frac{1}{2}$ & $0$ & $\frac{1}{2}$ & $1$ & $\frac{1}{2}$ & $-\frac{1}{4}$ & $\frac{1}{4}$ & $0$ & $0$ & $0$ \\
					$C_{10}$ & $-1$ & $0$ & $1$ & $1$ & $1$ & $\frac{1}{2}$ & $\frac{1}{2}$ & $0$ & $0$ & $0$ \\
					$C_{11}$ & $1$ & $2$ & $2$ & $2$ & $1$ & $\frac{1}{2}$ & $-\frac{1}{2}$ & $1$ & $0$ & $0$ \\
					$C_{12}$ & $\frac{3}{2}$ & $3$ & $\frac{5}{2}$ & $2$ & $\frac{3}{2}$ & $\frac{3}{4}$ & $-\frac{3}{4}$ & $2$ & $1$ & $0$ \\
					$C_{13}$ & $0$ & $0$ & $0$ & $0$ & $0$ & $0$ & $0$ & $1$ & $0$ & $0$ \\
					$C_{14}$ & $0$ & $0$ & $0$ & $0$ & $0$ & $0$ & $0$ & $0$ & $1$ & $0$ \\
					$C_{15}$ & $0$ & $0$ & $0$ & $0$ & $0$ & $0$ & $0$ & $-2$ & $0$ & $1$ \\
			\end{tabular}}
			\caption{Generators of the non-simplicial toric Mori cone for the unique $(D_7)_4$ variety. \label{table_A:mc_in_terms_of_dc_D_7_4}}
		\end{center}
	\end{table}

	\newpage
	\subsection{$E_6$}
	{\bf Occurs over Hirzebruch bases $\IF_n$, $n=0, \ldots, 6$.}
	
	{\bf Additional 1-cones:} 
	\be
	\begin{array}{ccccc}
		u_{\rho_{r_1}}= & (0 & 0 & 0 & -1) \\
		u_{\rho_{r_2}}= & (-1 & -1 & 0 & -2) \\
		u_{\rho_{r_3}}= & (-2 & -3 & 0 & -3) \\
		u_{\rho_{r_4}}= & (-1 & -2 & 0 & -2) \\
		u_{\rho_{r_5}}= & (0 & -1 & 0 & -1) \\
		u_{\rho_{r_6}}= & (-2 & -3 & 0 & -2) 
	\end{array}
	\ee
	
	{\bf Matter content:}  $6-n$ hypermultiplets in the complex representation $\boldsymbol{27}$.
	
	The cases $n=0, \ldots, 5$ are similar: there are 4 varieties in $(E_6)_n$ of type~I. All corresponding toric Mori cones are smooth. They are given in table \ref{table_A:mc_in_terms_of_dc_E_6} in terms of the classes of distinguished curves given in section \ref{ss:distinguished_curves}. Weights of the representation $\boldsymbol{27}$ are highlighted in green, those of the representation $\boldsymbol{\overline{27}}$ in red.

	\begin{table}[h!]
		\begin{center}
			\scalebox{0.7}{$\begin{tabu}{l|c|c|c|c|c|c|c|c|c} 
				& C_{r_1} & C_{r_2} & C_{r_3} & C_{r_4} & C_{r_5} & C_{r_6} & C_{r_0} & C_F & C_B \\ \hline
				C_1 & 0 & 0 & 1 & 0 & 0 & 0 & 0 & 0 & 0 \\
				C_2 & 0 & 0 & 0 & 1 & 0 & 0 & 0 & 0 & 0 \\
				C_3 & 0 & 0 & 0 & 0 & 0 & 1 & 0 & 0 & 0 \\
				\green{C_4} &-\frac{1}{3} & \frac{1}{3} & 0 & -\frac{1}{3} & \frac{1}{3} & 0 & 0 & 0 & 0 \\
				\red{C_5} &\frac{1}{3} & -\frac{1}{3} & 0 & \frac{1}{3} & \frac{2}{3} & 0 & 0 & 0 & 0 \\
				\green{C_6} &\frac{2}{3} & \frac{1}{3} & 0 & -\frac{1}{3} & -\frac{2}{3} & 0 & 0 & 0 & 0 \\
				C_7 &0 & 0 & 0 & 0 & 0 & 0 & 1 & 0 & 0 \\
				C_8 &0 & 0 & 0 & 0 & 0 & 0 & 0 & 1 & 0
				\end{tabu}$} \quad
			\scalebox{0.7}{$\begin{tabu}{l|c|c|c|c|c|c|c|c|c} 
				& C_{r_1} & C_{r_2} & C_{r_3} & C_{r_4} & C_{r_5} & C_{r_6} & C_{r_0} & C_F & C_B \\ \hline
				C_1 &1 & 0 & 0 & 0 & 0 & 0 & 0 & 0 & 0 \\
				C_2 &0 & 0 & 1 & 0 & 0 & 0 & 0 & 0 & 0 \\
				C_3 &0 & 0 & 0 & 1 & 0 & 0 & 0 & 0 & 0 \\
				C_4 &0 & 0 & 0 & 0 & 1 & 0 & 0 & 0 & 0 \\
				C_5 &0 & 0 & 0 & 0 & 0 & 1 & 0 & 0 & 0 \\
				\green{C_6} & -\frac{1}{3} & \frac{1}{3} & 0 & -\frac{1}{3} & -\frac{2}{3} & 0 & 0 & 0 & 0 \\
				C_7 & 0 & 0 & 0 & 0 & 0 & 0 & 1 & 0 & 0 \\
				C_8 &0 & 0 & 0 & 0 & 0 & 0 & 0 & 1 & 0 
				\end{tabu}$} \\ \vspace{0.5cm}
			\scalebox{0.7}{$\begin{tabu}{l|c|c|c|c|c|c|c|c|c} 
			& C_{r_1} & C_{r_2} & C_{r_3} & C_{r_4} & C_{r_5} & C_{r_6} & C_{r_0} & C_F & C_B \\ \hline
			C_1 & 0 & 0 & 1 & 0 & 0 & 0 & 0 & 0 & 0 \\
			C_2 & 0 & 0 & 0 & 0 & 1 & 0 & 0 & 0 & 0 \\
			C_3 & 0 & 0 & 0 & 0 & 0 & 1 & 0 & 0 & 0 \\
			\red{C_4} & \frac{1}{3} & -\frac{1}{3} & 0 & \frac{1}{3} & -\frac{1}{3} & 0 & 0 & 0 & 0 \\
			\green{C_5} & -\frac{1}{3} & \frac{1}{3} & 0 & \frac{2}{3} & \frac{1}{3} & 0 & 0 & 0 & 0 \\
			\red{C_6} &\frac{1}{3} & \frac{2}{3} & 0 & -\frac{2}{3} & -\frac{1}{3} & 0 & 0 & 0 & 0 \\
			C_7 &0 & 0 & 0 & 0 & 0 & 0 & 1 & 0 & 0 \\
			C_8 &0 & 0 & 0 & 0 & 0 & 0 & 0 & 1 & 0
			\end{tabu}$} \quad
		\scalebox{0.7}{$\begin{tabu}{l|c|c|c|c|c|c|c|c|c} 
			& C_{r_1} & C_{r_2} & C_{r_3} & C_{r_4} & C_{r_5} & C_{r_6} & C_{r_0} & C_F & C_B \\ \hline
			C_1 &0 & 1 & 0 & 0 & 0 & 0 & 0 & 0 & 0 \\
			C_2 &0 & 0 & 0 & 1 & 0 & 0 & 0 & 0 & 0 \\
			C_3 &0 & 0 & 0 & 0 & 1 & 0 & 0 & 0 & 0 \\
			C_4 &0 & 0 & 0 & 0 & 0 & 1 & 0 & 0 & 0 \\
			\red{C_5} &\frac{1}{3} & -\frac{1}{3} & 0 & -\frac{2}{3} & -\frac{1}{3} & 0 & 0 & 0 & 0 \\
			\green{C_6} &-\frac{1}{3} & \frac{1}{3} & 1 & \frac{2}{3} & \frac{1}{3} & 0 & 0 & 0 & 0 \\
			C_7 &0 & 0 & 0 & 0 & 0 & 0 & 1 & 0 & 0 \\
			C_8 &0 & 0 & 0 & 0 & 0 & 0 & 0 & 1 & 0
			\end{tabu}$} \\ \vspace{0.5cm}
			\scalebox{0.7}{$\begin{tabu}{l|c|c|c|c|c|c|c|c|c} 
				& C_{r_1} & C_{r_2} & C_{r_3} & C_{r_4} & C_{r_5} & C_{r_6} & C_{r_0} & C_F & C_B \\ \hline
				(C_9)_{0,1,2} & 0 & 0 & 0 & 0 & 0 & 0 & 0 & 0 & 1 \\
				(C_9)_3 & 0 & 0 & 0 & 0 & 0 & 0 & -1 & 0 & 1 \\
				(C_9)_4 & 0 & 0 & 0 & 0 & 0 & 0 & -2 & 0 & 1 \\
				(C_9)_5 & 0 & 0 & 0 & 0 & 0 & -1 & -3 & 0 & 1	
				\end{tabu}$}
			\caption{Generators of the toric Mori cones for the four  $(E_6)_n$ varieties of type~I in terms of the Mori vectors of the distinguished curves listed in table \ref{table:mori_vectors_E6_n}. Only the generator $C_9$ depends on the base surface $\IF_n$, but its expression in terms of the distinguished curves coincides for all four $(E_6)_n$ varieties of type~I. The corresponding classes are denoted as $(C_9)_n$ in the last table.} \label{table_A:mc_in_terms_of_dc_E_6}
		\end{center}
	\end{table}

	In addition, $(E_6)_n$ for $n=1,3,5$ contains four varieties of type~II.

	The class $(E_6)_6$ contains a single variety, compactification on which gives rise to a theory without charged matter.  Its toric Mori cone, given in table \ref{table_A:mc_in_terms_of_dc_E_6_6}, is not simplicial. 
	\begin{table}[h!]
		\begin{center}
			\scalebox{0.7}{$\begin{tabu}{l|c|c|c|c|c|c|c|c|c} 
				& C_{r_1} & C_{r_2} & C_{r_3} & C_{r_4} & C_{r_5} & C_{r_6} & C_{r_0} & C_F & C_B \\ \hline
				C_1 & 1 & 0 & 0 & 0 & 0 & 0 & 0 & 0 & 0 \\
				C_2 &0 & 1 & 0 & 0 & 0 & 0 & 0 & 0 & 0 \\
				C_3 &0 & 0 & 1 & 0 & 0 & 0 & 0 & 0 & 0 \\
				C_4 &0 & 0 & 0 & 1 & 0 & 0 & 0 & 0 & 0 \\
				C_5 &0 & 0 & 0 & 0 & 1 & 0 & 0 & 0 & 0 \\
				C_6 &0 & 0 & 0 & 0 & 0 & 1 & 0 & 0 & 0 \\
				C_7 &-\frac{1}{3} & \frac{1}{3} & 1 & \frac{2}{3} & \frac{1}{3} & 0 & 0 & 0 & 0 \\
				C_8 &\frac{2}{3} & \frac{1}{3} & 0 & -\frac{1}{3} & -\frac{2}{3} & 0 & 0 & 0 & 0 \\
				C_9 &\frac{1}{3} & \frac{2}{3} & 0 & -\frac{2}{3} & -\frac{1}{3} & 0 & 0 & 0 & 0 \\
				C_{10} &0 & 0 & 0 & 0 & 0 & 0 & 1 & 0 & 0 \\
				C_{11} &0 & 0 & 0 & 0 & 0 & 0 & 0 & 1 & 0 \\
				C_{12} &0 & 0 & 0 & 0 & 0 & -2 & -4 & 0 & 1
				\end{tabu}$}
		\caption{Generators of the non-simplicial toric Mori cone for the unique $(E_6)_6$ variety.} \label{table_A:mc_in_terms_of_dc_E_6_6}
		\end{center}
	\end{table}
	 \newpage
	\subsection{$E_7$}
	{\bf Occurs over Hirzebruch bases $\IF_n$, $n=0, \ldots, 8$.}
	
	{\bf Additional 1-cones:} 
	\be
	\begin{array}{ccccc}
			u_{\rho_{r_1}}= & (-2 & -3 & 0 & -2) \\
			u_{\rho_{r_2}}= & (-2 & -3 & 0 & -3) \\
			u_{\rho_{r_3}}= & (-2 & -3 & 0 & -4) \\
			u_{\rho_{r_4}}= & (-1 & -2 & 0 & -3) \\
			u_{\rho_{r_5}}= & (0 & -1 & 0 & -2) \\
			u_{\rho_{r_6}}= & (0 & 0 & 0 & -1)\\
			u_{\rho_{r_7}}= & (-1 & -1 & 0 & -2) 
		\end{array}
	\ee
	
	{\bf Matter content:}  $8-n$, $n=0, \dots, 8$, half-hypermultiplets in the self-conjugate representation $\boldsymbol{56}$.
	
	For $n=0, \ldots, 8$, $(E_7)_n$ contains 4 varieties of type~I. All corresponding toric Mori cones except for $n=8$ are smooth. The generators for $n=0,\ldots,7$ are given in table \ref{table_A:mc_in_terms_of_dc_E_7_3_to_7} in terms of the classes of distinguished curves given in section \ref{ss:distinguished_curves}. Curves corresponding to  weights of the representation $\boldsymbol{56}$ are highlighted in green.

	\begin{table}[h!]
		\begin{center}
			\scalebox{0.7}{$\begin{tabu}{l|c|c|c|c|c|c|c|c|c|c} 
				& C_{r_1} & C_{r_2} & C_{r_3} & C_{r_4} & C_{r_5} & C_{r_6} & C_{r_7} & C_{r_0} & C_F & C_B \\ \hline
				C_1 & 1 & 0 & 0 & 0 & 0 & 0 & 0 & 0 & 0 & 0 \\
				C_2 & 0 & 1 & 0 & 0 & 0 & 0 & 0 & 0 & 0 & 0 \\
				C_3 & 0 & 0 & 1 & 0 & 0 & 0 & 0 & 0 & 0 & 0 \\
				C_4 & 0 & 0 & 0 & 0 & 1 & 0 & 0 & 0 & 0 & 0 \\
				\green{C_5} & 0 & 0 & 0 & \frac{1}{2} & 0 & -\frac{1}{2} & \frac{1}{2} & 0 & 0 & 0 \\
				\green{C_6} & 0 & 0 & 0 & \frac{1}{2} & 0 & \frac{1}{2} & -\frac{1}{2} & 0 & 0 & 0 \\
				\green{C_7} & 0 & 0 & 0 & -\frac{1}{2} & 0 & \frac{1}{2} & \frac{1}{2} & 0 & 0 & 0 \\
				C_8 & 0 & 0 & 0 & 0 & 0 & 0 & 0 & 1 & 0 & 0 \\
				C_9 & 	0 & 0 & 0 & 0 & 0 & 0 & 0 & 0 & 1 & 0 
				\end{tabu}$} \quad
			\scalebox{0.7}{$\begin{tabu}{l|c|c|c|c|c|c|c|c|c|c} 
				& C_{r_1} & C_{r_2} & C_{r_3} & C_{r_4} & C_{r_5} & C_{r_6} & C_{r_7} & C_{r_0} & C_F & C_B \\ \hline
				C_1 &1 & 0 & 0 & 0 & 0 & 0 & 0 & 0 & 0 & 0 \\
				C_1 &0 & 1 & 0 & 0 & 0 & 0 & 0 & 0 & 0 & 0 \\
				C_3 &0 & 0 & 1 & 0 & 0 & 0 & 0 & 0 & 0 & 0 \\
				C_4 &0 & 0 & 0 & 0 & 1 & 0 & 0 & 0 & 0 & 0 \\
				C_5 &0 & 0 & 0 & 0 & 0 & 1 & 0 & 0 & 0 & 0 \\
				C_6 &0 & 0 & 0 & 0 & 0 & 0 & 1 & 0 & 0 & 0 \\
				\green{C_7} &0 & 0 & 0 & \frac{1}{2} & 0 & -\frac{1}{2} & -\frac{1}{2} & 0 & 0 & 0 \\
				C_8 &0 & 0 & 0 & 0 & 0 & 0 & 0 & 1 & 0 & 0 \\
				C_9 &0 & 0 & 0 & 0 & 0 & 0 & 0 & 0 & 1 & 0
				\end{tabu}$}	\\ \vspace{0.5cm}
			\scalebox{0.7}{$\begin{tabu}{l|c|c|c|c|c|c|c|c|c|c} 
				& C_{r_1} & C_{r_2} & C_{r_3} & C_{r_4} & C_{r_5} & C_{r_6} & C_{r_7} & C_{r_0} & C_F & C_B \\ \hline
				C_1 & 1 & 0 & 0 & 0 & 0 & 0 & 0 & 0 & 0 & 0 \\
				C_2 & 0 & 1 & 0 & 0 & 0 & 0 & 0 & 0 & 0 & 0 \\
				C_3 & 0 & 0 & 1 & 0 & 0 & 0 & 0 & 0 & 0 & 0 \\
				C_4 & 0 & 0 & 0 & 1 & 0 & 0 & 0 & 0 & 0 & 0 \\
				C_5 & 0 & 0 & 0 & 0 & 0 & 1 & 0 & 0 & 0 & 0 \\
				\green{C_6} & 0 & 0 & 0 & -\frac{1}{2} & 0 & -\frac{1}{2} & \frac{1}{2} & 0 & 0 & 0 \\
				\green{C_7} & 0 & 0 & 0 & \frac{1}{2} & 1 & \frac{1}{2} & -\frac{1}{2} & 0 & 0 & 0 \\
				C_8 & 0 & 0 & 0 & 0 & 0 & 0 & 0 & 1 & 0 & 0 \\
				C_9 & 0 & 0 & 0 & 0 & 0 & 0 & 0 & 0 & 1 & 0
				\end{tabu}$} \quad
			\scalebox{0.7}{$\begin{tabu}{l|c|c|c|c|c|c|c|c|c|c} 
				& C_{r_1} & C_{r_2} & C_{r_3} & C_{r_4} & C_{r_5} & C_{r_6} & C_{r_7} & C_{r_0} & C_F & C_B \\ \hline
				C_1 & 1 & 0 & 0 & 0 & 0 & 0 & 0 & 0 & 0 & 0 \\
				C_2 & 0 & 1 & 0 & 0 & 0 & 0 & 0 & 0 & 0 & 0 \\
				C_3 & 0 & 0 & 0 & 1 & 0 & 0 & 0 & 0 & 0 & 0 \\
				C_4 & 0 & 0 & 0 & 0 & 1 & 0 & 0 & 0 & 0 & 0 \\
				C_5 & 0 & 0 & 0 & 0 & 0 & 0 & 1 & 0 & 0 & 0 \\
				\green{C_6} & 0 & 0 & 0 & -\frac{1}{2} & 0 & \frac{1}{2} & -\frac{1}{2} & 0 & 0 & 0 \\
				\green{C_7} & 0 & 0 & 1 & \frac{1}{2} & 0 & -\frac{1}{2} & \frac{1}{2} & 0 & 0 & 0 \\
				C_8 & 0 & 0 & 0 & 0 & 0 & 0 & 0 & 1 & 0 & 0 \\
				C_9 & 0 & 0 & 0 & 0 & 0 & 0 & 0 & 0 & 1 & 0
				\end{tabu}$} \\ \vspace{0.5cm}
			\scalebox{0.7}{$\begin{tabu}{l|c|c|c|c|c|c|c|c|c|c} 
				& C_{r_1} & C_{r_2} & C_{r_3} & C_{r_4} & C_{r_5} & C_{r_6} & C_{r_7} & C_{r_0} & C_F & C_B \\ \hline
				(C_{10})_{0,1,2} & 0 & 0 & 0 & 0 & 0 & 0 & 0 & 0 & 0 & 1 \\
				(C_{10})_3 & 0 & 0 & 0 & 0 & 0 & 0 & 0 & -1 & 0 & 1 \\
				(C_{10})_4 & 0 & 0 & 0 & 0 & 0 & 0 & 0 & -2 & 0 & 1 \\
				(C_{10})_5 & -1 & 0 & 0 & 0 & 0 & 0 & 0 & -3 & 0 & 1 \\
				(C_{10})_6 & -2 & 0 & 0 & 0 & 0 & 0 & 0 & -4 & 0 & 1 \\
				(C_{10})_7 & -3 & -1 & 0 & 0 & 0 & 0 & 0 & -5 & 0 & 1
				\end{tabu}$}
			\caption{Generators of the toric Mori cones for the four $(E_7)_n$ varieties of type~I for $n=0, \ldots, 7$, expressed in terms of the Mori vectors of the distinguished curves introduce in section \ref{ss:mori_cone}. Only the generator $C_{10}$ depends on the base surface $\IF_n$, but its expression in terms of the distinguished curves coincides for all four $(E_7)_n$ varieties of type~I. The corresponding classes are denoted as $(C_{10})_n$ in the last table.} \label{table_A:mc_in_terms_of_dc_E_7_3_to_7}
		\end{center}
	\end{table}
	
	In addition, $(E_7)_n$ for $n=1,3,5,7$ contains four varieties of type~II.
	
	The class $(E_7)_8$ contains a single variety, compactification on which gives rise to a theory without charged matter. Its toric Mori cone, given in table \ref{table_A:mc_in_terms_of_dc_E_7_8}, is not simplicial. 
	\begin{table}[h]
		\begin{center}
			\scalebox{0.7}{$\begin{tabu}{l|c|c|c|c|c|c|c|c|c|c} 
				& C_{r_1} & C_{r_2} & C_{r_3} & C_{r_4} & C_{r_5} & C_{r_6} & C_{r_7} & C_{r_0} & C_F & C_B \\ \hline
				C_1 & 1 & 0 & 0 & 0 & 0 & 0 & 0 & 0 & 0 & 0 \\
				C_2 & 0 & 1 & 0 & 0 & 0 & 0 & 0 & 0 & 0 & 0 \\
				C_3 & 0 & 0 & 1 & 0 & 0 & 0 & 0 & 0 & 0 & 0 \\
				C_4 & 0 & 0 & 0 & 1 & 0 & 0 & 0 & 0 & 0 & 0 \\
				C_5 & 0 & 0 & 0 & 0 & 1 & 0 & 0 & 0 & 0 & 0 \\
				C_6 & 0 & 0 & 0 & 0 & 0 & 1 & 0 & 0 & 0 & 0 \\
				C_7 & 0 & 0 & 0 & 0 & 0 & 0 & 1 & 0 & 0 & 0 \\
				C_8 & 0 & 0 & 0 & \frac{1}{2} & 1 & \frac{1}{2} & -\frac{1}{2} & 0 & 0 & 0 \\
				C_9 & 0 & 0 & 1 & \frac{1}{2} & 0 & -\frac{1}{2} & \frac{1}{2} & 0 & 0 & 0 \\
				C_{10} & 0 & 0 & 0 & 0 & 0 & 0 & 0 & 1 & 0 & 0 \\
				C_{11} & 0 & 0 & 0 & 0 & 0 & 0 & 0 & 0 & 1 & 0 \\
				C_{12} & -4 & -2 & 0 & 0 & 0 & 0 & 0 & -6 & 0 & 1
				\end{tabu}$}
			\caption{Generators of the non-simplicial toric Mori cone for the unique $(E_7)_8$ variety.} \label{table_A:mc_in_terms_of_dc_E_7_8}
		\end{center}
	\end{table}

	\subsection{$E_8$}
	{\bf Occurs over Hirzebruch base $\IF_{12}$}
	
	{\bf Additional 1-cones:} 
	\be
	\begin{array}{ccccc}
		u_{\rho_{r_1}}= & 	(0 & -1 & 0 & -2) \\
		u_{\rho_{r_2}}= & (-1 & -2 & 0 & -4) \\
		u_{\rho_{r_3}}= & (-2 & -3 & 0 & -6) \\
		u_{\rho_{r_4}}= & (-2 & -3 & 0 & -5) \\
		u_{\rho_{r_5}}= & (-2 & -3 & 0 & -4) \\
		u_{\rho_{r_6}}= & (-2 & -3 & 0 & -3) \\
		u_{\rho_{r_7}}= & (-2 & -3 & 0 & -2) \\
		u_{\rho_{r_8}}= & (-1 & -1 & 0 & -3)
	\end{array}
	\ee
	
	{\bf Matter content:}  none.
	
	$(E_8)_{12}$ contains exactly one variety. It is of type~I. Its toric Mori vector is smooth. Its generators are given in table \ref{table_A:mc_in_terms_of_dc_E_8} in terms of the classes of distinguished curves given in section \ref{ss:distinguished_curves}. 
	\begin{table}[h!]
		\begin{center}
			\scalebox{0.7}{$\begin{tabu}{l|c|c|c|c|c|c|c|c|c|c|c} 
				& C_{r_1} & C_{r_2} & C_{r_3} & C_{r_4} & C_{r_5} & C_{r_6} & C_{r_7} & C_{r_8} & C_{r_0} & C_F & C_B \\ \hline
				C_1 & 1 & 0 & 0 & 0 & 0 & 0 & 0 & 0 & 0 & 0 & 0 \\
				C_2 & 0 & 1 & 0 & 0 & 0 & 0 & 0 & 0 & 0 & 0 & 0 \\
				C_3 & 0 & 0 & 1 & 0 & 0 & 0 & 0 & 0 & 0 & 0 & 0 \\
				C_4 & 0 & 0 & 0 & 1 & 0 & 0 & 0 & 0 & 0 & 0 & 0 \\
				C_5 & 0 & 0 & 0 & 0 & 1 & 0 & 0 & 0 & 0 & 0 & 0 \\
				C_6 & 0 & 0 & 0 & 0 & 0 & 1 & 0 & 0 & 0 & 0 & 0 \\
				C_7 & 0 & 0 & 0 & 0 & 0 & 0 & 1 & 0 & 0 & 0 & 0 \\
				C_8 & 0 & 0 & 0 & 0 & 0 & 0 & 0 & 1 & 0 & 0 & 0 \\
				C_9 & 0 & 0 & 0 & 0 & 0 & 0 & 0 & 0 & 1 & 0 & 0 \\
				C_{10} & 0 & 0 & 0 & 0 & 0 & 0 & 0 & 0 & 0 & 1 & 0 \\
				C_{11} & 0 & 0 & 0 & -2 & -4 & -6 & -8 & 0 & -10 & 0 & 1
				\end{tabu}$} 		
			\caption{Generators of the toric Mori cone of the unique $(E_8)_{12}$ variety.} \label{table_A:mc_in_terms_of_dc_E_8}
		\end{center}
	\end{table}
	
	\newpage
	\subsection{$F_4$}
	{\bf Occurs over Hirzebruch bases $\IF_n$, $n=0, \ldots, 5$.}
	
	{\bf Additional 1-cones:} 
	\be
	\begin{array}{ccccc}
		u_{\rho_{r_1}}= & (-2 & -3 & 0 & -2) \\
		u_{\rho_{r_2}}= & (-2 & -3 & 0 & -3) \\
		u_{\rho_{r_3}}= & (-1 & -2 & 0 & -2) \\
		u_{\rho_{r_4}}= & (0 & -1 & 0 & -1)
	\end{array}
	\ee
	
	{\bf Matter content:}  $2*(5-n)$ half-hypermultiplets in the self-conjugate representation $\boldsymbol{26}$.
	
	For all $n=0, \ldots, 5$, $(F_4)_n$ contains exactly one variety of type~I. The corresponding toric Mori cones  are smooth. They have seven generators. The classes of six of these are given by $[C_{r_1}], \ldots, [C_{r_4}], [C_{r_0}], [C_F]$. The class of the last generator depends on $n$, and is given in table \ref{table_A:mc_in_terms_of_dc_F_4}.
	\begin{table}[h!]
		\begin{center}
			\scalebox{0.7}{$\begin{tabu}{l|c|c|c|c|c|c|c|c|c} 
				& C_{r_1} & C_{r_2} & C_{r_3} & C_{r_4} & C_{r_0} & C_F & C_B \\ \hline
				(C_7)_1 & 0 & 0 & 0 & 0 & 0 & 0 & 1 \\
				(C_7)_2 & 0 & 0 & 0 & 0 & 0 & 0 & 1 \\
				(C_7)_3 & 0 & 0 & 0 & 0 & -1 & 0 & 1 \\
				(C_7)_4 & 0 & 0 & 0 & 0 & -2 & 0 & 1 \\
				(C_7)_5 & -1 & 0 & 0 & 0 & -3 & 0 & 1
				\end{tabu}$}
			\caption{The $n$-dependent generator of the toric Mori cone of the $(F_4)_n$ varieties of type~I.} \label{table_A:mc_in_terms_of_dc_F_4}
		\end{center}
	\end{table}

	In addition, $(F_4)_n$ for $n=1,3,5$ also contains a variety of type~II.
	
	Note that all weights of the $\boldsymbol{26}$ representation are also roots. The corresponding Gromov-Witten invariants at base $\IF_n$ are thus $2 * (5-n) - 2$.

	\subsection{$G_2$}
	{\bf Occurs over Hirzebruch bases $\IF_n$, $n=0, \ldots, 3$.}
	
	{\bf Additional 1-cones:} 
	\be
	\begin{array}{ccccc}
		u_{\rho_{r_1}}= & (-1 & -1 & 0 & -1) \\
		u_{\rho_{r_1}}= & (-2 & -3 & 0 & -2) 
	\end{array}
	\ee
	
	{\bf Matter content:}  $2*(7-2n)$ half-hypermultiplets in the self-conjugate representation $\boldsymbol{7}$.
	
	For all $n=0, \ldots, 3$, $(G_2)_n$ contains exactly one variety of type~I. The corresponding toric Mori cones  are smooth. They have five generators. The classes of four of these are given by $[C_{r_1}], [C_{r_2}], [C_{r_0}], [C_F]$. The class of the last generator depends on $n$, and is given in table \ref{table_A:mc_in_terms_of_dc_G_2}.
	\begin{table}[h!]
		\begin{center}
			\scalebox{0.7}{$\begin{tabu}{l|c|c|c|c|c|c|c|c|c} 
				& C_{r_1} & C_{r_2} & C_{r_0} & C_F & C_B \\ \hline
				(C_5)_1 & 0 & 0 & 0 & 0 & 1 \\
				(C_5)_2 & 0 & 0 & 0 & 0 & 1 \\
				(C_5)_3 & 0 & 0 & -1 & 0 & 1
				\end{tabu}$}
			\caption{The $n$-dependent generator of the toric Mori cone of the $(G_2)_n$ variety of type~I.} \label{table_A:mc_in_terms_of_dc_G_2}
		\end{center}
	\end{table}

	In addition, $(G_2)_n$ for $n=1,3$ also contains a variety of type~II.
	
	Note that all weights of the $\boldsymbol{7}$ are also roots. The corresponding Gromov-Witten invariants at base $\IF_n$ are thus $
	2*(7-2n)-2 = 12-4n$. In particular, at $n=3$, the contributions from roots and weights cancel.

	\bibliography{matter_from_top_biblio}	

\end{document}